\documentclass[fleqn,usenatbib]{mnras}

\usepackage{graphicx, caption, subcaption}

\usepackage{amssymb}

\usepackage[T1]{fontenc}
\DeclareRobustCommand{\VAN}[3]{#2}
\let\VANthebibliography\thebibliography
\def\thebibliography{\DeclareRobustCommand{\VAN}[3]{##3}\VANthebibliography}

\usepackage{graphicx}	% Including figure files
\usepackage{amsmath}	% Advanced maths commands
\usepackage{array}
\usepackage{newtxtext,newtxmath}
\title[Hot Neptunes in TESS]{A systematic validation of hot Neptunes in TESS data}

\author[C. Magliano et al.]{
Christian Magliano$^{1,2,3}$\thanks{E-mail: christian.magliano@unina.it},
Giovanni Covone$^{1,2,3}$,
Richa Dobal$^4$,
Luca Cacciapuoti$^5$,
Luca Tonietti$^6$,
\newauthor
Steven Giacalone$^{7}$,
Jose I. Vines$^{8}$,
Laura Inno$^{3,6}$,
James S. Jenkins$^{9,10}$,
Jack J. Lissauer$^{11}$,
Allyson Bieryla$^{12}$,
\newauthor
Fabrizio Oliva$^{13}$,
Isabella Pagano$^{14}$,
Veselin Kostov$^{15,16}$,
Carl Ziegler$^{17}$,
David R.Ciardi $^{18}$,
\newauthor
Erica J.Gonzales $^{19}$,
Courtney D. Dressing$^{7},$
Lars A. Buchhave$^{20}$,
Steve B. Howell$^{21}$,
Rachel A. Matson$^{22}$,
\newauthor
Elisabeth Matthews$^{23}$,
Alessandra Rotundi$^{6,13}$,
Douglas Alves$^{8}$,
Stefano Fiscale$^6$,
Riccardo M. Ienco$^1$,
\newauthor
Pablo Peña$^{9,10}$,
Francesco Gallo$^1$,
Maria T. Muscari Tomajoli$^{6}$.
\\
$^{1}$Dipartimento di Fisica "Ettore Pancini", Università di Napoli Federico II, Napoli, Italy\\
$^{2}$INFN, Sezione di Napoli, Complesso Universitario di Monte S. Angelo,
Via Cintia Edificio 6, 80126 Napoli, Italy \\
$^{3}$INAF - Osservatorio Astronomico di Capodimonte,
via Moiariello 16, 80131 Napoli, Italy \\
$^{4}$Indian Institute of Technology Gandhinagar, Gujarat, India \\
$^{5}$European Southern Observatory, Karl-Schwarzschild-Strasse 2 D-85748 Garching bei Munchen, Germany \\  
$^{6}$Science and Technology Department, Parthenope University of Naples, Naples, Italy \\
$^{7}$Department of Astronomy, University of California Berkeley, Berkeley, CA 94720, USA \\
$^{8}$Departamento de Astronom\'ia, Universidad de Chile, Casilla 36-D, Santiago, Chile\\
$^{9}$N\'ucleo de Astronom\'ia, Facultad de Ingenier\'ia y Ciencias, Universidad Diego Portales, Av. Ej\'ercito 441, Santiago, Chile\\
$^{10}$Centro de Astrof\'isica y Tecnolog\'ias Afines (CATA), Casilla 36-D, Santiago, Chile\\
$^{11}$Space Science $\&$ Astrobiology Division, MS 245-3, NASA Ames Research Center, Moffett Field, CA 94035, USA\\
$^{12}$Center for Astrophysics, Harvard $\&$ Smithsonian, 60 Garden Street,
Cambridge, MA 02138, USA\\
$^{13}$Istituto di Astrofisica e Planetologia Spaziali (IAPS/INAF), Rome, Italy\\
$^{14}$INAF - Osservatorio Astrofisico di Catania, Catania, Italy\\
$^{15}$NASA Goddard Space Flight Center, 8800 Greenbelt Road, Greenbelt, MD 20771, USA\\
$^{16}$GSFC Sellers Exoplanet Environments Collaboration, USA\\
$^{17}$Department of Physics, Engineering and Astronomy, Stephen F. Austin State University, 1936 North St, Nacogdoches, TX 75962, USA\\
$^{18}$NASA Exoplanet Science Institute, Caltech/IPAC, Mail Code 100-22, 1200 E. California Boulevard, Pasadena, CA 91125, USA\\
$^{19}$University of California, Santa Cruz, 1156 High Street, Santa Cruz, CA 95064, USA\\
$^{20}$DTU Space,  National Space Institute, Technical University of Denmark, Elektrovej 328, DK-2800 Kgs. Lyngby, Denmark\\
$^{21}$NASA Ames Research Center, Moffett Field, CA 94035, USA\\
$^{22}$U.S. Naval Observatory, Washington, D.C. 20392, USA\\
$^{23}$Départment d’astronomie de l’Université de Genève, Chemin Pegasi 51, 1290 Versoix, Switzerland\\
}

\date{Accepted XXX. Received YYY; in original form ZZZ}

\pubyear{2022}

\begin{document}
\label{firstpage}
\pagerange{\pageref{firstpage}--\pageref{lastpage}}
\maketitle

\begin{abstract}
%Despite the exponential rise in the number of confirmed exoplanets, extensive research has shown a depletion of Neptunian/sub-Jovian planets at periods shorter than approximately $4$ days, the \textit{Hot Neptune Desert}. In order to shed light on the physical origin of the observed paucity, theoretical models require a robust sample of these rare objects.
We statistically validated a sample of hot Neptune candidates applying a two-step vetting technique using \texttt{DAVE} and \texttt{TRICERATOPS}.
We performed a systematic validation of 250 transit-like events in the Transiting Exoplanet Survey Satellite (TESS) archive in the parameter region defined by $P\leq 4\,\text{d}$ and  $3\, R_{\oplus}\leq R \leq 5\,R_{\oplus}$. 
Through our analysis, we identified 18 hot Neptune-sized candidates, with a false positive probability $< 50\%$.
Nine of these planet candidates still need to be confirmed. For each of the nine targets we retrieved the stellar parameters using \texttt{ARIADNE} and derived constraints on the planetary parameters by fitting the lightcurves with the \texttt{juliet} package.
Within this sample of nine candidates, we statistically validated (i.e, with false positive probability $< 0.3\%$) two systems (TOI-277 b and TOI-1288 b)  by re-processing the candidates with \texttt{TRICERATOPS} along with follow-up observations. These new validated exoplanets expand the known hot Neptunes population and are high-priority targets for future radial velocities 
follow-up.
\end{abstract}


%%%%%%%%%%%%%%%%%%%%%%%%%%%%%%%%%%%%%%%%%%%%%%%%%%

\section{Introduction}
A variety of ground and space-based hunting missions have discovered over 5,000 planets orbiting other stars in the Galaxy \citep[e.g.,][]{Howell_2020FrASS}, and several more may be discovered outside the Milky Way in the future \citep{Covone2000, DiStefano_2021NatAs}. In particular, the \textit{Kepler} mission \citep{2010ApJ...713L..79K} has revolutionized our knowledge of exoplanets by allowing to conduct robust statistical studies of the exoplanets population \citep[e.g.,][]{2011ApJS..197....8L,2021AJ....161...16H} for the first time. 
One of the most unexpected and intriguing result is the discovery of the so called "Hot Neptune Desert": an apparent paucity of strongly irradiated planets with the sizes of Neptune which orbit their host star in $\lesssim 4$ days, see e.g. \cite{2013ApJ...763...12B} and \cite{2011ApJ...727L..44S}.
Indeed, most close-in known exoplanets are hot Jupiters that have enough mass to retain most of their atmosphere against the photoevaporation caused by the star, or small dry rocky exoplanets that have likely lost their atmosphere due to the intense host star radiation long ago \citep{2019AREPS..47...67O}.
It is still not clear whether the formation of planets is somehow prohibited inside the Desert or if they can start forming inside it but then rapidly migrate outwards or fall onto the host star \citep{2011ApJ...727L..44S}.
\cite{Mazeh_2016} investigated the Neptunian Desert conundrum in the period-mass and period-radius diagrams. By applying two different statistical techniques to a sample of confirmed exoplanets with known mass, they estimated the lower and upper boundaries within which the data show a dearth of objects.
Many authors tried to give physical explanations to the two observed boundaries which define the Desert.
The photoevaporation of the H/He atmospheres of an initially low-mass planet is consistent with the shape of the lower boundary in the radius-period plane \citep{2013ApJ...763...12B,2014ApJ...792....1L,2016NatCo...711201L,2018MNRAS.479.5012O}. 
Photoevaporation is the physical process by which exoplanets lose their H/He envelopes due to the UV/X-ray photons emitted by the host star. 
As a consequence, the planet experiences a downsizing. 
According to this scenario a Neptune-size planet, which formed inside or moved into the Desert, is rapidly shrunk to a super Earth. We also expect that the shape of the lower bound will be affected by the total lifetime UV/X-ray flux as well as the spectral type of the host star \citep{2019ApJ...876...22M}.
However, photoevaporation can not be the only process responsible for the upper limit of the Neptune Desert. Indeed, more massive planets ($M\gtrsim 0.5M_J$) are able to resist photoevaporation and numerical simulations \citep{2018MNRAS.476.5639I,2018MNRAS.479.5012O} showed that in the scenario where photoevaporation also creates the upper boundary the radius-period plane would be filled with sub-Jovian planets at very short period, in contrast with the observations. %given in Fig. \ref{fig:dataset_tot}.
Furthermore, by detecting the helium absorption signal of seven gas-giant planets near the upper edge, \cite{2022BAAS...54e.143V} found that their atmospheric lifetimes are much longer than $10$ Gyr.
According to \cite{2018MNRAS.479.5012O}, the upper boundary can be explained in the high-eccentricity migration scenario where, due to gravitational interactions, a planet is pushed into a highly-eccentric orbit which becomes tidally circularized as the planet proceeds towards its parent star. After orbital circularization, planets with masses greater than $1M_J$ can move even closer to their parent star thanks to stellar tidal decay.
Important observational constraints to decipher which scenario dominates (i.e., if these planets formed {\it in situ} or by migration) will come from the characterization of planetary bulk density estimates obtained with high accuracy radial velocity data along with precise radii measurements. In addition, the systematic characterization of planetary atmospheres that will be performed by the James Webb Space Telescope (JWST) and the Ariel mission \citep{2018ExA....46..135T} may shed light on whether these giant planets are gradually loosing their gas envelopes or they are depleted of volatile chemicals. The mixing of density estimates along with planetary atmosphere characterization will be a crucial step in understanding the origin of these worlds. Indeed, the {\it in situ} scenario formation is compatible with a planetary composition dominated by refractory compounds while a migration-scenario may lead to an abundance of frozen volatiles and amorphous ices.

The Transiting Exoplanet Survey Satellite (TESS) mission (\citealp{2015JATIS...1a4003R}) has found more than $5000$ candidate planets according to the NASA Exoplanet Archive\footnote{https://exoplanetarchive.ipac.caltech.edu/index.html} to date, making a remarkable contribution to detect planets within and around the Neptune Desert.
Most of the target stars in the TESS Input Catalog (TIC) are bright enough for follow-up spectroscopy observations to measure the mass of the planet, and thereby allowing to retrieve its mean density.
So far there has been little discussion about an overall analysis of Hot Neptune Desert within the TESS database.
However, there are several works based on TESS observations where authors have confirmed exoplanets lying in the hot Neptune Desert, see e.g. \cite{2020NatAs...4.1148J}. 

In this work we exploit the TESS archive and the follow-up observations to validate new hot Neptune candidates,
by applying a vetting procedure based on two recently developed tools, \texttt{DAVE} and \texttt{TRICERATOPS}. 
The outline of the paper is the following, in Sect. \ref{sec:sample} we focus our attention on a sub-region of the Neptune Desert in the TESS archive. In Sect. \ref{sec:vetting} we vet the whole catalog extracted in the sub-region by using the tools \texttt{DAVE} and \texttt{TRICERATOPS}, and then in Sect. \ref{sec:analysis} we perform transit fitting with \texttt{juliet} in order to obtain the planetary parameters while the adopted stellar parameters are retrieved by using the software package \texttt{ARIADNE}. 
In Sect. \ref{sec:followup} we  discuss follow-up observations of these TOIs carried by several facilities used by the TESS Follow-Up Observing Program Working Groups.
In Sect. \ref{sec:results} we analyze in detail each individual hot Neptune selected by our procedure.
In Sect. \ref{sec:discussion} we discuss the potential for atmospheric characterization of our selected sample. Finally we summarize our conclusions in Sect. \ref{sec:end}.

\section{The sample}
\label{sec:sample}

\begin{figure}
\centering
    \includegraphics[width=0.5\textwidth]{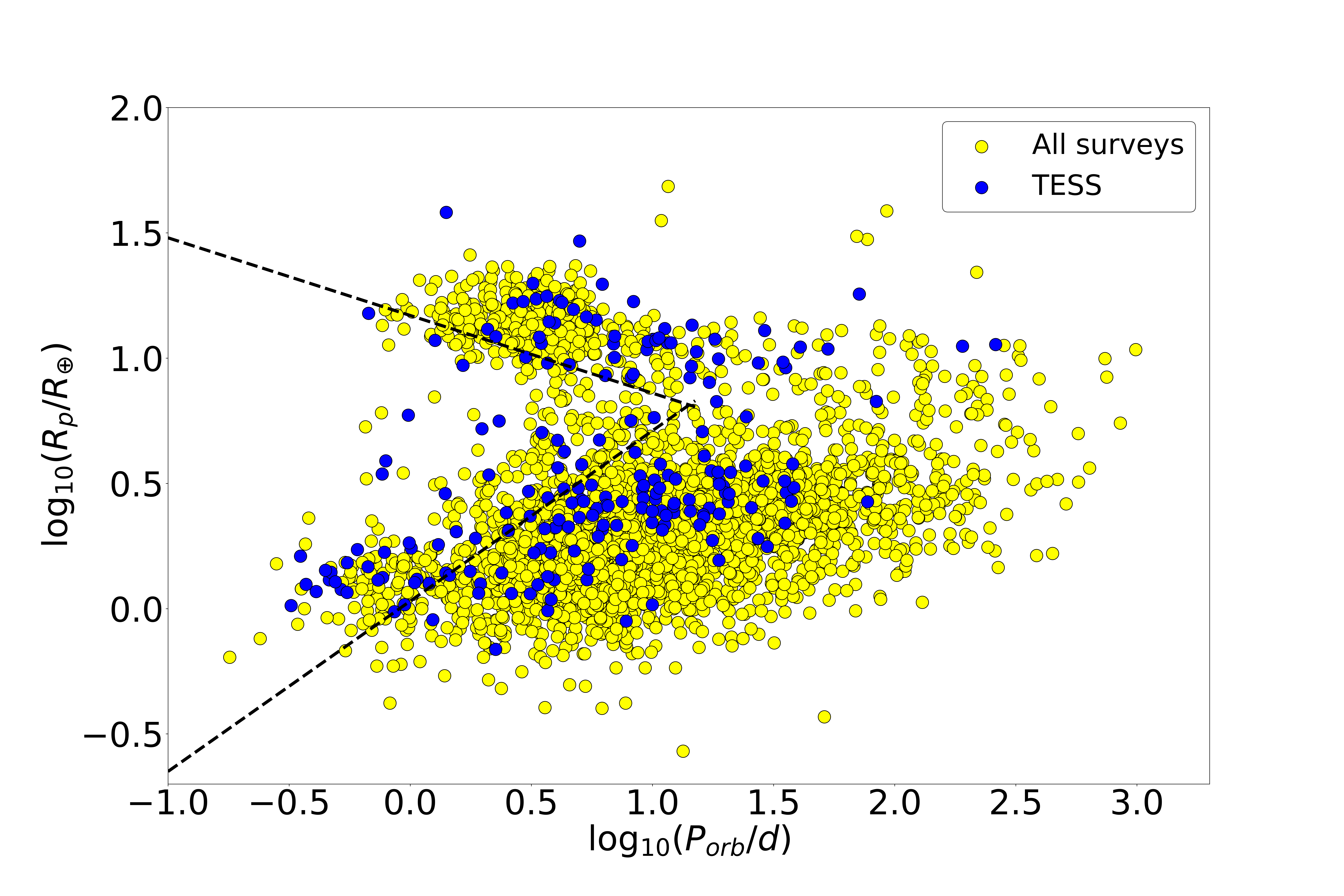}
    \caption{The distribution of the confirmed exoplanets observed with $P<10^{3}$ d and $R<100\,R_{\protect\oplus}$. The yellow dots are planets from all surveys, the blue dots represent the confirmed exoplanets discovered by TESS. The black dashed lines are the upper and lower boundaries of the Hot Neptunes Desert as calculated by \protect\cite{Mazeh_2016}.}
    \label{fig:dataset_tot}
\end{figure}

Figure \ref{fig:dataset_tot} shows the distribution of the confirmed exoplanets with $P<10^{3}$ days and $R<100 \, R_\oplus$ in the (P, R) plane, obtained from the NASA Exoplanet Archive. 
The dearth of Neptunian/sub-Jovian bodies orbiting very near their host star (i.e. $P\leq 4$ days) is clearly seen.
The two boundaries give rise to the well-know peculiar triangular shape, defining the borders of the so-called \textit{Hot Neptunes Desert} \citep{Mazeh_2016}.

Our aim  is to perform a homogeneous and statistically controlled validation of the sample of hot Neptunes candidates in the TESS data, as these will be valuable targets for follow-up spectroscopic observations aimed at understanding the origin of the hot Neptune Desert.

\begin{figure*}
     \centering
     \begin{subfigure}[b]{\columnwidth}
         \centering
         \includegraphics[width=\columnwidth]{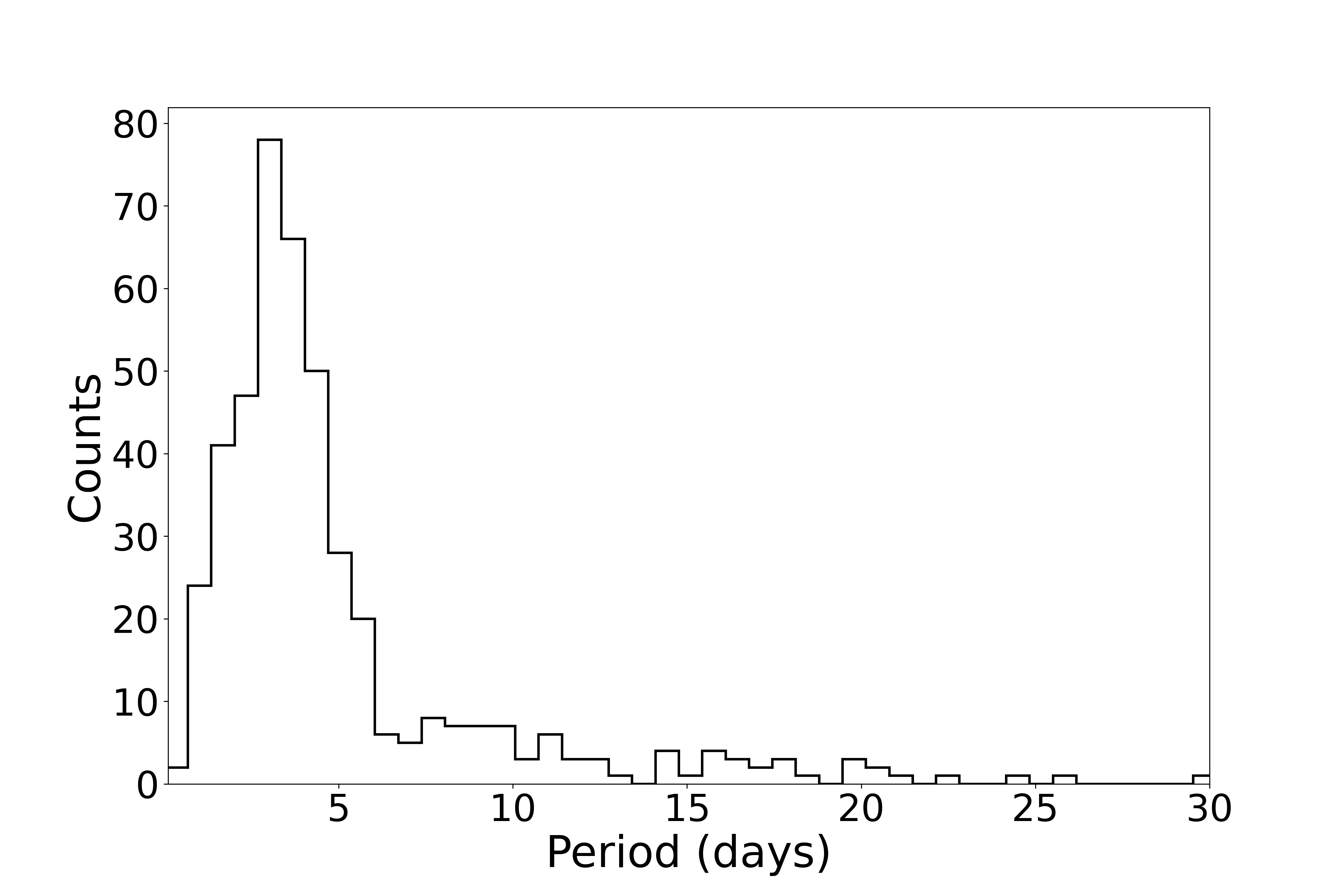}
         \caption{}
         \label{fig:histPtess}
     \end{subfigure}
     \hfill
    \begin{subfigure}[b]{\columnwidth}
         \centering
         \includegraphics[width=\columnwidth]{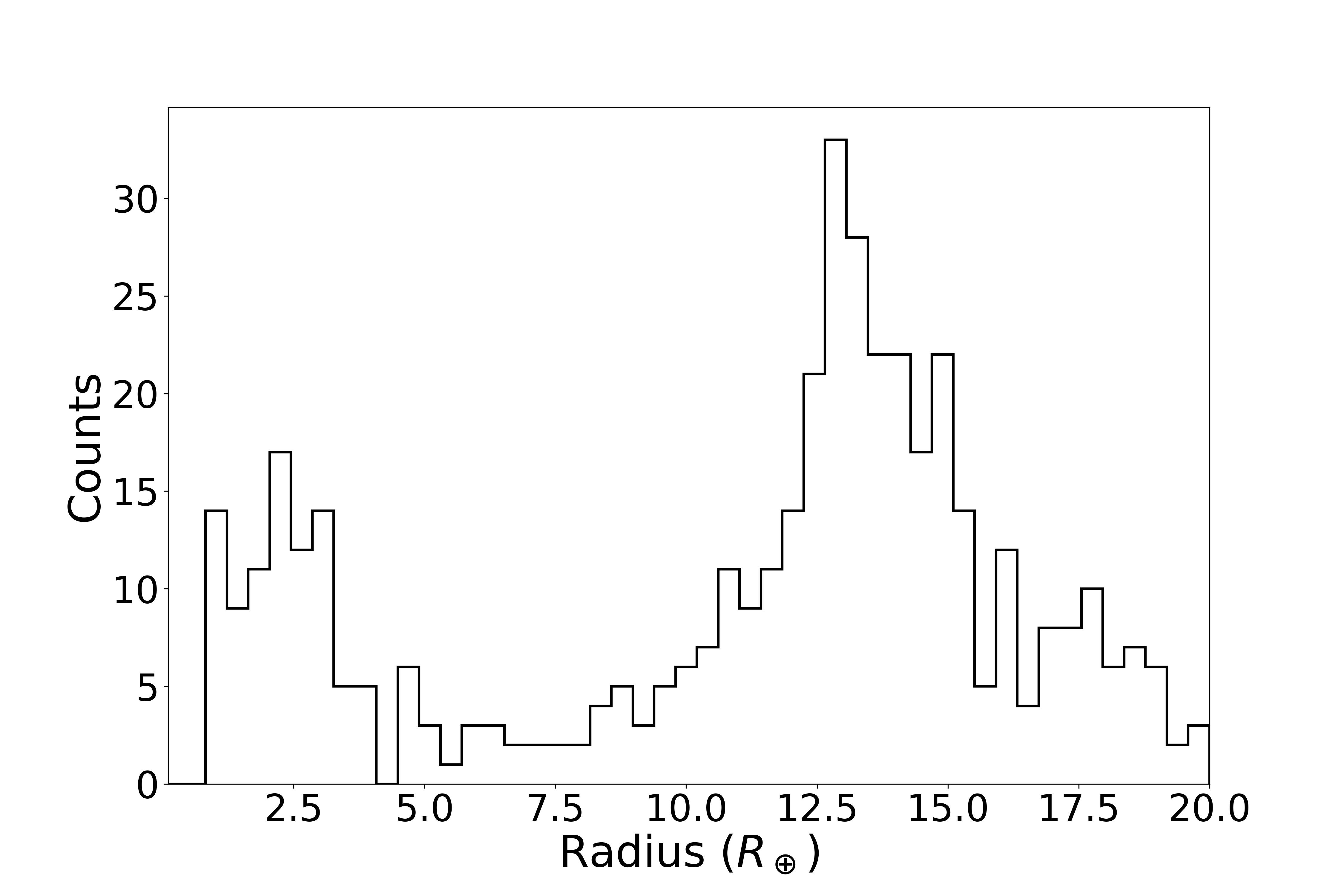}
         \caption{}
         \label{fig:histRtess}
     \end{subfigure}
        \caption{(a) Period distribution of confirmed exoplanets observed by TESS with $P<30$ days; (b) radii distribution of confirmed exoplanets observed by TESS with $R < 20 \, R\oplus$.}
        \label{fig:dataset_tess}
\end{figure*}

At January 2022, 453 objects observed by TESS were confirmed or were already known exoplanets, and their key features are publicly available in the ExoFOP archive\footnote{https://exofop.ipac.caltech.edu/tess/}. 
In Fig. \ref{fig:dataset_tess} we show the period and radii distributions of the confirmed and known planets observed by TESS.
The period distribution clearly shows the high rate of short-period exoplanets discovered by TESS. 
This is partly due to the way the TESS project has been designed. Given the typical baseline of 27.4 days for TESS sky sectors, TESS is more likely to find planets with period shorter than 13 days.
The bimodal radii distribution in the histogram in Fig. \ref{fig:dataset_tess} clearly shows a depletion of confirmed bodies with $3\,R_{\oplus}\lesssim R \lesssim 10\,R_{\oplus}$.
Only 24 out of 443 confirmed objects observed by TESS with $3\,R_{\oplus}\lesssim R \lesssim 10\,R_{\oplus}$ orbit their host stars in less than $4$ days.  
TESS observations confirm the dearth of Neptune/sub-Jovian size objects with very short period ($P\leq 4$ days) 
as pointed out by \cite{2013ApJ...763...12B} and \cite{2011ApJ...727L..44S} by using the \textit{Kepler} database.

In this region of the ($P, R$) space, ExoFOP archive contains more than $700$ TESS Objects of Interest (TOIs) still to be validated. This dataset was extremely large to be analyzed with the two-step vetting technique, so we restricted our analysis to a subregion of the Hot Neptune Desert. 
In particular we selected, within the ExoFOP archive, TOIs with $P\leq 4$ days and $3R_\oplus<R<5R_\oplus$, the ''Desert'' hereafter, where only nine objects have been confirmed as planets so far.
We found 250 TOIs populating the Desert, that we uniformly vetted by using \texttt{DAVE}.
The TOIs that successfully passed the \texttt{DAVE} test, were then processed using \texttt{TRICERATOPS} which quantifies the Bayesian probability that we are dealing with true planets. 
The details of the whole workflow are described in the next Section.

We checked whether the lack of confirmed planets in the Desert is statistically significant by performing a simple Monte Carlo (MC) test following the same procedure as in \cite{2013ApJ...763...12B}. 
In particular, we generated $10^4$ exoplanetary populations, each of 1000 planets uniformly distributed in the (P, R) space with $R\in [0.5, 10]R_\oplus$ and $P\in[0, P_{max}]$, where the  upper limit $P_{max}$ was varied for each different 
simulated sample within the range $[4,12]$ days, according to the period distribution shown in Fig. \ref{fig:dataset_tess}. 
In fact, the median period $\Tilde{P}$ of the confirmed TESS planet distribution is $\approx 3.8$ d with a standard deviation $\sigma_P\approx 8$ d. Therefore, we designated $[\Tilde{P}, \Tilde{P}+\sigma_P]$ as the range within which $P_{max}$ can take values in the MC simulations.
For a given sample and a given value of $P_{max}$, 
we calculated the occurrence rate of drawn planets that fall in the Desert, i.e., with $P<4$ days and $3R_\oplus < R < 5 R_\oplus$. 
At the end of the simulations we obtain a set of $10^4$ occurrence rates whose mean value depends on the value of $P_{max}$. 
As $P_{max}$ decreases, the averaged theoretical occurrence rate increases because we are gradually reducing the ($P, R$) region in which our simulated signals can fall, i.e. it is more probable for a randomly drawn exoplanet to fall within the Desert. In the worst case, i.e. $P_{max}=12$ days, the average occurrence rate obtained with simulation was $\approx 6\%$ against the observed one, $9/443 \approx 2\%$.

\section{The vetting procedure}
\label{sec:vetting}

Despite the theoretical simplicity of the transit method, a number of astrophysical sources can produce a large number of false positives.
There are many possible false positive scenarios: i) \textit{stellar binary}, that is two stars orbiting a common center of mass can mimic a planetary transit signal if they are different in size or if they have similar sizes but the eclipse is partial;
ii) \textit{blended binary star}, when the target blends the light of an ecplising binary such that the secondary
ecplise might be lost in the noise while the primary gets confused with a planetary transit; iii) other \textit{spurious single-star signals} originated from astrophysical sources (stellar spots, pulsation, rotation) or instrumentation artefacts (jitter noise and momentum dumps).

The vetting procedure is a first step in the long workflow which allows to confirm a transit-like signal as a genuine exoplanet. 
In this work we used two different tools to vet the 250 TOIs selected in Sect. \ref{sec:sample}. 
First, we used \texttt{DAVE} (Discovery and Vetting of Exoplanets, \citealt{kostov2019discovery}) to vet transit-like signals at both the pixel and lightcurve level by visual inspection. Then, after this skimming procedure, we used
\texttt{TRICERATOPS} \citep{2021AJ....161...24G} to analyze the selected candidates by calculating the Bayesian probability of  being astrophysical false positives. Below we describe in detail these two steps in our analysis. 

\subsection{\texttt{DAVE}}

The aim of the \texttt{DAVE} pipeline is to evaluate whether a transit-like signal might originate from a planetary candidate, or is a false positive due to different scenarios. \texttt{DAVE} was originally developed for vetting of exoplanet candidates using data from the K2 mission and later adapted to TESS (e.g. \citealt{kostov2019discovery}). While the pixel size, systematic effects and SNR differs between the two missions, the underlying procedure is effectively the same. Namely, each TOI is analyzed at the pixel level by photocenter analysis, and at the lightcurve level by a flux time-series analysis. This double-level analysis is carried out for each TESS sector in which the target has been observed.
The vetter can distinguish whether the transit is caused by a real planet candidate or a false positive by inspecting some \texttt{DAVE} outcomes. 

The \textit{centroids} analysis produces a difference image by subtracting the overall in-transit image from the out-of-transit image. Then it calculates the photocenter of the light distribution by fitting the TESS pixel response function to the image. This procedure is executed for each detected transit. The position of the centroid on the difference image is calculated by taking the average over all the events, the statistical significance of the measured offset is measured and provided to the user. This module allows the vetter to check whether the transit-like event did not originate from the target star: 
if \textit{centroids} gives a clear offset of the centroid with respect to the expected position of the target star then it is flagged as a False Positive (FP). However, it is possible that DAVE produces difference images that are difficult to interpret due to artefacts or low SNR flagged as bad quality images; in these cases, the photocenter analysis is pointless and the centroid unreliable hence, to be the most conservative, we pass the signal as a planet candidate.

The \textit{Modelshift} analysis generates a phase-folded lightcurve together with the best-fit trapezoid transit model. The purpose of this module is to inspect whether the source of the signal is an eclipsing binary system.
It also displays other important features of the transit such as the average primary signal, the second and the third most notable signals. Moreover, it also shows the averaged odd transits along with the even ones and it calculates the statistical significance of the odd-even difference. 
Another important feature that \textit{Modelshift} takes into account is the shape of the transit. 

Besides these two main modules, there are other supplementary metrics which allow a better comprehension for those targets harder to vet. In particular, of great importance is the Lomb-Scargle (LS) Periodogram \citep{1976Ap&SS..39..447L,1982ApJ...263..835S} which runs a LS periodogram on the transit-masked lightcurve. Then, it displays the phase-folded lightcurve with respect to the period with the highest peak in order to look for possible modulations of the lighcturve due to intrinsic and/or rotational variability. In general modulations can be caused by a variable background star inside the aperture mask used to extract the lightcurve, but if these are on the same period of the observed eclipse they may originated from ellipsoidal variations of a binary star system \citep{1993ApJ...419..344M,2011MNRAS.415.3921F,2017PASP..129g2001S}. When the lightcurve exhibits strong modulations, one needs to perform a detrending of the time series flux. 

In this work we followed the same procedure as discussed in \cite{2022MNRAS.513..102C}: 
for each TOI in the catalog, each member of the inspection team provides her/his personal ranking (or \textit{disposition}) of the TOI, according to the following prescriptions: 

\textit{i)} if the TOI shows no anomalies at both flux and pixel level then the signal is ranked as a Planetary Candidate (PC); 

\textit{ii)} if the TOI does not pass the vetting procedure then the signal is ranked as a FP. 
A FP disposition might be due to a clear centroid offset, a clear secondary eclipse at $0.5$ phase or a V-shaped transit along with a significant odd-even difference;

\textit{iii)} if the TOI shows a few red flags but not clear clues of a false positive scenario, then the signal is ranked as a probable False Positive, pFP. For example, we gave a pFP when the TESS lightcurve has a low SNR and at the same time we notice a potential secondary eclipse or the photocenter position is slightly shifted towards the first neighbour target's pixel, but not clear enough.

The vetter team was composed of CM, RD and LT who faced with three possible scenario:

\textit{i)} all three vetters ranked a signal as PC then the TOI was a PC; 

\textit{ii)} at least two vetters ranked a signal as pFP or FP, then the TOI was discarded from further analysis as FP; 

\textit{iii)} two vetters ranked a signal as a PC and the other one claimed a FP or pFP, then the TOI was further discussed as a group in order to decide one of the two above scenarios.

In order to be conservative and build a clean final catalog of Hot Neptune candidates, we chose to rule out also the systems with a pFP disposition,  while focusing our further analysis on PCs only. 

\begin{figure}
    \centering
    \includegraphics[width=0.5\textwidth]{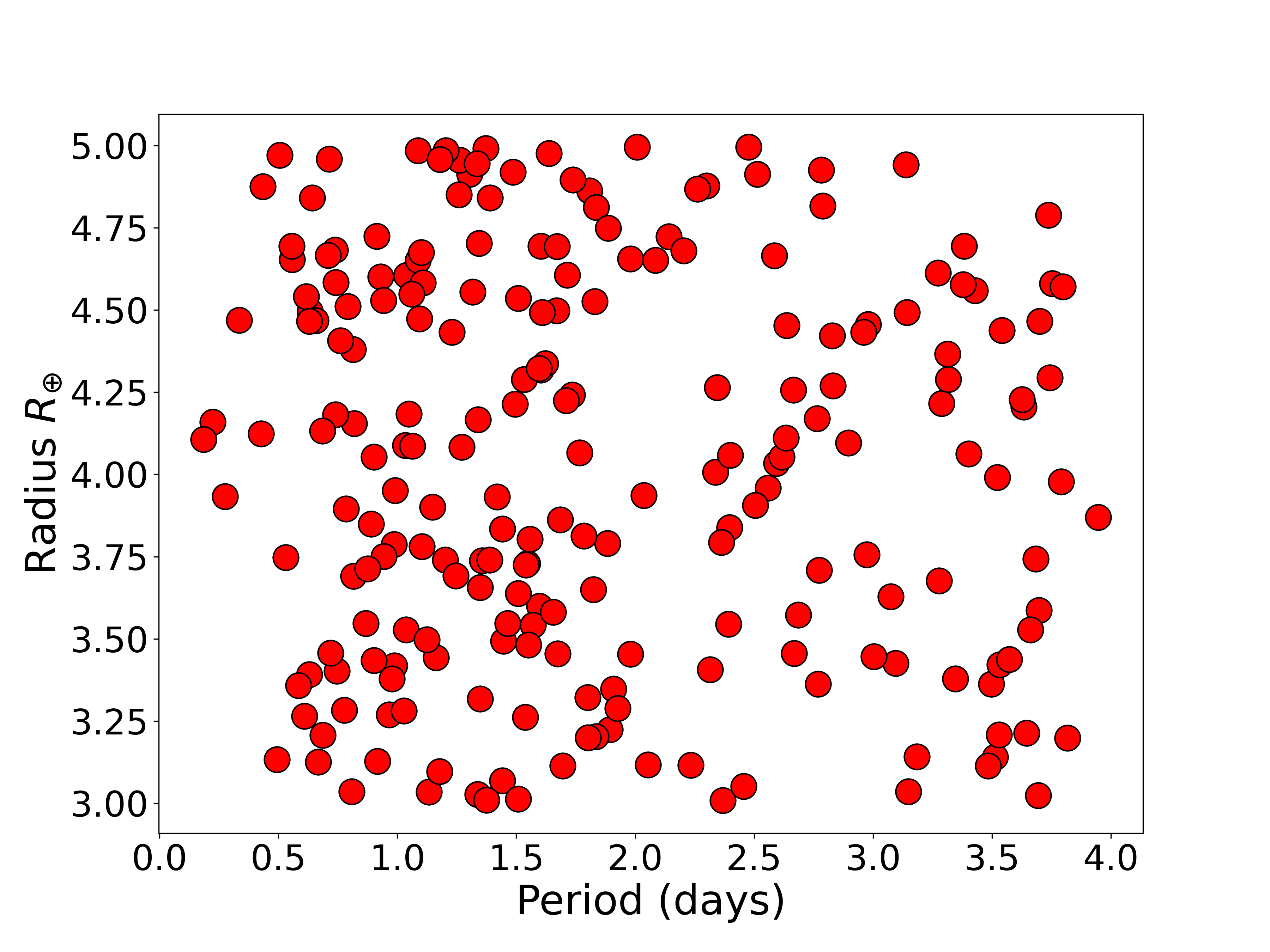}
    \caption{Distribution of the pFPs and FPs ranked by using DAVE in the ($P, R$) diagram.}
    \label{fig: FPs distribution}
\end{figure}

Out of the 250 TOIs, we ranked 62 signals as PC, that is about $25\%$ of the total. 
The occurrence rate for PC is very low if compared with the value $\approx 70\%$ obtained by \cite{2022MNRAS.513..102C}, due to the fact that in this work we considered a tiny region of the (P,R) space in which we expect to find a dearth of true planets. 
As a further check, we plot the distribution of the pFP and FP signals in the (P,R) diagram in Fig. \ref{fig: FPs distribution}. The distribution of the false positive candidates is quite uniform in this space, without any clear trend, ensuring that our classification based on DAVE is not affected by any bias. We only emphasize a slightly higher density of false positives at short periods ($P<2$ days), which is typical of binary star systems.

TESS has a focus-limited PSF because one TESS pixel corresponds to $21''$ in the sky. This design introduces an intrinsic limit in the \texttt{DAVE} dispositions based on TESS observation due to we can not handle with the field stars within the target pixel. 
We will discuss these details in Sect. \ref{sec:followup}. 
In this perspective \texttt{TRICERATOPS} tool comes in handy because of its capability of calculating the false positive probability taking into account sources less than $21''$ far away from the target.

\subsection{\texttt{TRICERATOPS}}

The \texttt{TRICERATOPS}\footnote{Web site: https://github.com/stevengiacalone/triceratops} pipeline calculates the Bayesian probability that a given lightcurve originates from a transiting planet or a wide range of false positive scenarios \citep{2021AJ....161...24G}.
For each target the pipeline allows to calculate the False Positive Probability (FPP), that is the probability that a planet candidate is a false positive, and the Nearby False Positive Probability (NFPP), 
namely the probability that a planet candidate is a false positive due to the presence of a resolved nearby star.

The classification of a lightcurve is based on the comparison between the obtained FPP and NFPP values and \textit{a priori} defined threshold values. 
\cite{2021AJ....161...24G} argued that when $\text{FPP}< 0.5$ and $\text{NFPP}<10^{-3}$, the observed  event is likely originated by a transiting planet.
Many authors consider a planet statistically validated when  $\text{FPP}<10^{-2}$ and $\text{NFPP}<10^{-3}$  (\citealp{2014ApJ...784...45R,2012ApJ...761....6M,2015ascl.soft03011M,2021AJ....161...24G}).
We stress that \texttt{TRICERATOPS} calculates the bayesian false positive probability based on planet occurrence rates priors which treat planet radius and orbital period as independent variables. However, in this work we are considering a region in the  ($P, R$) diagram where the planet occurrence rates are lower than one would expect looking at the abundances elsewhere in the diagram \citep{2019AJ....158..109H}. 
Consequently, all the probabilities calculated with \texttt{TRICERATOPS} will be underestimated by an unknown factor. In order to construct the cleanest possible sample  of high-priority targets, in this work for a planet candidate to be statistically validated we chose a stronger constraint than those aforementioned. 
We define a Hot Neptune candidate statistically validated to better than the 99.7th percentile if 
FPP$<3\cdot 10^{-3}$ and NFPP $<10^{-3}$ .
We stress that these are arbitrary threshold values depending on the desired degree of confidence chosen \textit{a priori}, so a target slight above the threshold still remains amenable for any follow-up observations. 
We also point out that \texttt{TRICERATOPS} works reliably on 2-minute cadence data, whereas planet candidates observed only at $30$- or $10$-minute cadence data cannot be confidently validated. For the sake of completeness we will also run \texttt{TRICERATOPS} on those candidates with only $10$ or $30$-min cadence data available, keeping in mind that even if they meet the above-mentioned threshold values, we will not consider them validated planets. Notwithstanding, these alleged candidates will still be amenable targets for follow-up measurements.

\begin{figure}
    \centering
    \includegraphics[width=0.5\textwidth]{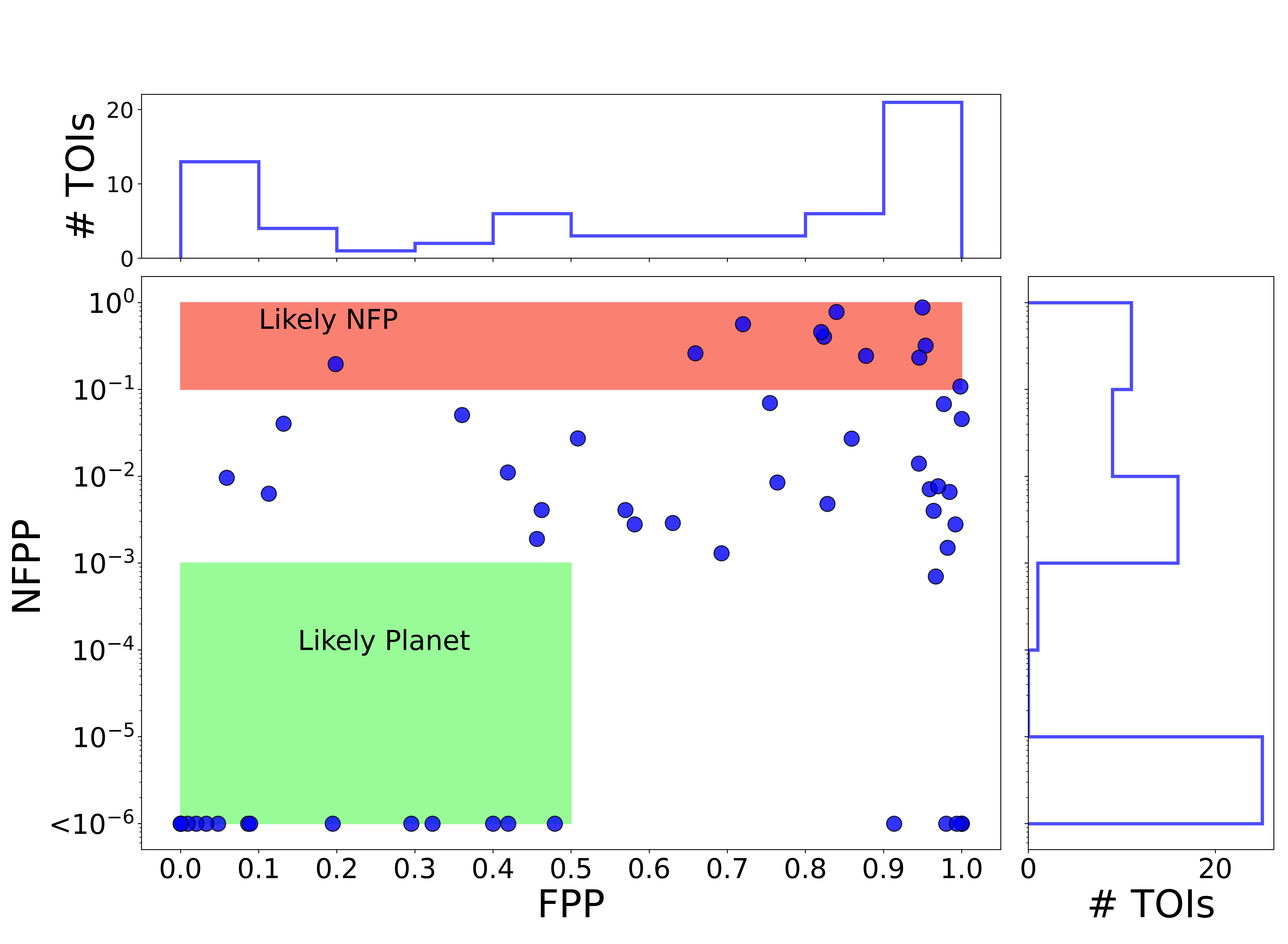}
    \caption{Distribution of the TRICERATOPS results within the (FPP, NFPP) plane,
    with the adopted threshold values as in Giacalone et al. (2021). 
    We consider TOIs with FPP $<0.5$ and NFPP $<10^{-3}$ as likely planets (green shaded region). 
    While, if a TOI has NFPP $>0.1$ then it is designated as a likely 
    false positive due to some nearby sources (red shaded region). } 
    \label{fig:triceratops}
\end{figure}

\texttt{TRICERATOPS} allows to determine the FPP and NFPP values by including also follow-up measurements, such as high-contrast images. However, we did not use any follow-up measurements at this step of the vetting procedure, as not all the 62 candidates had publicly available follow-up measurements. We ran \texttt{TRICERATOPS} for the 62 planet candidates previously classified with \texttt{DAVE}, using the same aperture mask employed for the visual-inspection. In order to take into account the intrinsic statistical fluctuations in the \texttt{TRICERATOPS} outcomes, for each TOI we run the pipeline 10 times taking the mean value for FPP and NFPP values.
We show the results of the \texttt{TRICERATOPS} analysis in Fig. \ref{fig:triceratops}.

As shown in Fig. \ref{fig:triceratops} and summarized in Table \ref{table:triceratops_res}, there are 18 TOIs  falling in the so called \textit{Likely Planet} region, in which there is a high probability 
that these signals are due to true planets. Instead, 10 TOIs lie inside the \textit{Likely NFP}, defined by NFPP $\geq 0.1$, composed by signals which have high chance to be false positive originating from the stars that fall in the TESS photometric aperture.
Eventually, 34 TOIs lie outside these two regions in the (FPP, NFPP).

For each TOI of the sample we checked its TESS Follow-up Observing Program Working Group\footnote{Web site: https://tess.mit.edu/followup} (TFOPWG) disposition based on available follow-up measurements. 
According to the publicly available TFOPWG dispositions, the sample of the 62 PCs
includes 46 PCs, 4 objects rejected as FPs throughout follow-up observations and 12 already confirmed planets. 
We found that 9 out of 12 confirmed exoplanets lie in the \textit{Likely Planet} region where, furthermore, no false positive occurs. In particular, within this region, there are 3 already known planets (KPs) from other surveys (GJ-436 b, GJ-3470 b and K2-55 b) and 6 confirmed planets by TFOPWG (TOI-1728 b, LTT 9779 b, TOI-849 b, TOI-132 b, TOI-1235 b, TOI-1260 b).
This comparison shows a close agreement between the results from our procedure and the TFOPWG analysis.
TFOPWG identified a false positive which falls in the \textit{Likely NFP} region while 3 confirmed exoplanets and 3 FPs that occupy the leftover area in the (FPP, NFPP) space.
We report in Table \ref{tab:triceratops_res} the summary of the results obtained with \texttt{TRICERATOPS} along with TFOPWG dispositions.

In the following we restrict our analysis to the 18 signals falling inside the \textit{Likely Planet} whose FPP and NFPP values are reported in Table \ref{table:10new}.
However, nine out of 18 TOIs inside the \textit{Likely Planet} region have already been confirmed as planet by the TFOPWG. Hence, we are left with nine planet candidates to be validated.

At the end of this vetting procedure we are left with nine new candidate Hot Neptunes. As reported in Table \ref{table:10new}, already at this stage in our analysis (that is, before taking into account any follow-up measurement) TIC 365733349.01 and TIC 439456714.01 
satisfy the requirement (i.e., FPP$ <3\cdot 10^{-3} $ and NFPP $ < 10^{-3}$) to be considered statistically validated planets as we will discuss in section \ref{sec:results} where we will repeat the \texttt{TRICERATOPS} test using the follow-up observations in order to have a uniformly statistically validated sample of hot Neptunes in TESS database.
The other seven candidates have good chance to be real planets but we require additional data to validate them (see Sect. \ref{sec:followup}) and rule out the possibility of unresolved companion or nearby star.
The reader can find the \texttt{DAVE} main outcomes for the nine TOIs in Appendix \ref{appendix: dave_results}.

\begin{table}
\caption{Summary of the \texttt{TRICERATOPS} outcomes along with the TFOPWG dispositions.}
\label{table:triceratops_res}
	\centering
	\label{tab:triceratops_res}
	\begin{tabular}{lccccr} 
		\hline
		 &Total&PC & CP & FP\\
		\hline
		 Likely Planet&18& $9$ & $9$ & $0$ \\
		 Likely NFP&10& $9$ & $0$ & $1$ \\
		 Other&34& $28$ & $3$ & $3$  \\
		 \hline
		 Total&62 & $46$ & $12$ &$4$\\
        \hline	
    \end{tabular}
\end{table}

\begin{table}
\caption{\texttt{TRICERATOPS} outcomes for the 18 likey hot Neptunes without follow-up observations along with the TFOPWG dispositions.}
\label{table:10new}
	\centering
	\label{tab:example_table}
	\begin{tabular}{lccccr} % four columns, alignment for each
		\hline
		TIC ID & FPP & NFPP & TFOPWG Disposition \\
		\hline
		439456714.01&$<3\cdot 10^{-3}$ & $<10^{-3}$&PC  \\
		365733349.01&  $<3\cdot 10^{-3}$ & $<10^{-3}$&PC \\
		138819293.01&$<3\cdot 10^{-3}$&$<10^{-3}$&KP (GJ-436 b)\\
        285048486.01&$<3\cdot 10^{-3}$&$<10^{-3}$&CP (TOI-1728 b)\\
        19028197.01&$<3\cdot 10^{-3}$&$<10^{-3}$&KP (GJ-3470b)\\
        183985250.01&$<3\cdot 10^{-3}$&$<10^{-3}$&CP (LTT 9779 b)\\
		63898957.01& $<0.5$ & $<10^{-3}$&PC \\
        73540072.01& $<0.5$  & $<10^{-3}$&PC \\
        372172128.01&  $<0.5$  & $<10^{-3}$&PC \\
        153078576.01& $<0.5$  & $<10^{-3}$&PC\\
        146523262.01&  $<0.5$  & $<10^{-3}$&PC\\
        358070912.01& $<0.5$  & $<10^{-3}$&PC\\
        124235800.01& $<0.5$  & $<10^{-3}$&PC\\
        33595516.01&$<0.5$&$<10^{-3}$&CP (TOI-849 b)\\
        437704321.01&$<0.5$&$<10^{-3}$&KP (K2-55 b)\\
        89020549.01&$<0.5$&$<10^{-3}$&CP (TOI-132 b)\\
        103633434.01&$<0.5$&$<10^{-3}$&CP (TOI-1235 b)\\
        355867695.01&$<0.5$&$<10^{-3}$&CP (TOI-1260 b)\\
        \hline
		
	\end{tabular}
	
\end{table}

\section{ANALYSIS}
\label{sec:analysis}

By following the vetting procedure discussed in Sect. \ref{sec:vetting} we selected a sample of 
nine new Hot Neptune candidates to be validated. In this Section we perform a transit fit using the \texttt{juliet} tool \citep{2019MNRAS.490.2262E}  based on stellar parameters inferred by using the software \texttt{ARIADNE} \citep{2022MNRAS.513.2719V}.

\subsection{Stellar Parameters}

We retrieved the stellar parameters of the nine host stars by performing a Spectral Energy Distribution (SED) modelling using \texttt{ARIADNE}. This tool has incorporated six atmosphere model grids, which were created by convolving the synthetic atmosphere SEDs with multiple bandpasses of different filters. The models used are --- \texttt{Phoenix v2} \citep{2013A&A...553A...6H}, BT-Settl, BT-Cond, \citep{2011ASPC..448...91A}, BT-NextGen \citep{Hauschildt1999, 2011ASPC..448...91A}, \cite{1993yCat.6039....0K}, and \cite{2003IAUS..210P.A20C} --- \texttt{ARIADNE} fits each selected model grid using the Dynamic Nested Sampling algorithm \citep{Skilling2004, Skilling2006, Higson2019} through \texttt{dynesty} \citep{Speagle2019}, and finally performs Bayesian Model Averaging over all utilized models to get the final set of parameters for a given star. Firstly, \texttt{ARIADNE} searches for broadband photometry using \texttt{astroquery} to access MAST and VizieR archives to query catalogs Tycho-2 \citep{2000A&A...355L..27H}, ASCC \citep{2001KFNT...17..409K}, 2MASS \citep{2006AJ....131.1163S}, GLIMPSE \citep{2009PASP..121..213C}, ALL-WISE \citep{2010AJ....140.1868W}, GALEX \citep{2011Ap&SS.335..161B}, APASS DR9 \citep{2014CoSka..43..518H}, SDSS DR12 \citep{2015ApJS..219...12A}, Strömgren Photometric Catalog \citep{2015A&A...580A..23P}, Pan-STARRS1 \citep{2016arXiv161205560C} and GAIA DR2 \citep{2016A&A...595A...2G,2018A&A...616A...1G}.
The models used for fitting were selected according to an initial effective temperature estimation from Gaia DR2 \citep{GaiaDR2}, this is because \cite{1993yCat.6039....0K} and \cite{2003IAUS..210P.A20C} are known to perform poorly on stars with T$_\text{eff} < 4000$~K.
For T$_\text{eff}$, $\log~g$ and [Fe/H] we used priors drawn from the TIC v8 catalog \citep{ticv8} values where available, and the default prior otherwise, while for the radius and distance we used \texttt{ARIADNE}'s default priors. Finally for the interstellar extinction A$_\text{v}$ we used a prior drawn from the Bayestar 3D dustmaps \citep{bayestar} when possible, otherwise we employed a uniform prior from 0 to the maximum line-of-sight extinction from the SFD dustmaps \citep{SFD1, SFD2}. A summary of the priors is shown in Table \ref{tab:ari_priors} and we report in Table \ref{tab:stellar_parameter} the outcomes of \texttt{ARIADNE}.

\begin{table*}
\caption{\texttt{ARIADNE} priors.}
\label{tab:ari_priors}
\centering
\begin{tabular}{lcccc}
\hline
TIC ID & T$_\text{eff}$ & $\log~g$ & [Fe/H] & A$_\text{v}$ \\ \hline
63898957 & $\mathcal{N}\left(5940, 100^2\right)$ & $\mathcal{N}\left(4.57, 0.1^2\right)$ & $\mathcal{N}\left(0.29, 0.1^2\right)$ & 0 (fixed)\\
439456714 & $\mathcal{N}\left(3866, 200^2\right)$ & $\mathcal{N}\left(4.58, 0.11^2\right)$ & $\mathcal{N}\left(-0.125, 0.234^2\right)$ & 0 (fixed) \\
365733349 & $\mathcal{N}\left(6180, 362^2\right)$ & $\mathcal{N}\left(4.62, 0.1^2\right)$ & $\mathcal{N}\left(-0.125, 0.234^2\right)$ & 0 (fixed) \\
73540072 & $\mathcal{N}\left(5145, 100^2\right)$ & $\mathcal{N}\left(4.34, 0.2^2\right)$ & $\mathcal{N}\left(0.31, 0.2^2\right)$ & 0 (fixed) \\
372172128 & $\mathcal{N}\left(5728, 130^2\right)$ & $\mathcal{N}\left(4.24, 0.1^2\right)$ & $\mathcal{N}\left(-0.35, 0.2^2\right)$ & $\mathcal{U}\left(0, 0.08\right)$ \\
153078576 & $\mathcal{N}\left(3596, 157^2\right)$ & $\mathcal{N}\left(4.79, 0.2^2\right)$ & $\mathcal{N}\left(-0.63, 0.5^2\right)$ & $\mathcal{U}\left(0, 0.043\right)$ \\
146523262 & $\mathcal{N}\left(6250, 200^2\right)$ & $\mathcal{N}\left(4.72, 0.1^2\right)$ & $\mathcal{N}\left(-0.63, 0.5^2\right)$ & 0 (fixed) \\
358070912 & $\mathcal{N}\left(5151, 124^2\right)$ & $\mathcal{N}\left(4.42, 0.1^2\right)$ & $\mathcal{N}\left(-0.35, 0.2^2\right)$ & $\mathcal{U}\left(0, 0.15\right)$ \\
124235800 & $\mathcal{N}\left(3830, 157^2\right)$ & $\mathcal{U}\left(3.5, 6\right)$ & $\mathcal{N}\left(-0.125, 0.234^2\right)$ & $\mathcal{U}\left(0, 0.2\right)$ \\ \hline
\end{tabular}
\end{table*}

\begin{table*}
\caption{Stellar parameters as obtained from \texttt{ARIADNE}.}
\label{table:stellar_parameter}
	\centering
	\label{tab:stellar_parameter}
	\begin{tabular}{lcccccccccr} % nine columns, alignment for each
		\hline
		TIC ID & TOI & Distance (pc) & $\text{T}_{\text{eff}} (K)$ & $\log g$&$R_{*}(R_\odot)$ & $M_{*}(M_\odot)$ & $\rho_{*}(kg/m^{3})$ & [Fe/H]\\
		\hline
		63898957 & 261 & $114.1_{-0.2}^{+0.8}$& $6076_{-21}^{+26}$&$4.57_{-0.05}^{+0.07}$&$1.150_{-0.013}^{+0.010}$&$1.78_{-0.09}^{+0.10}$&$1649\pm 104$&$0.13_{-0.03}^{+0.05}$\\[0.2cm]
		439456714 & 277 & $65.17_{-0.09}^{+0.18}$& $4031_{-23}^{+21}$&$4.58_{-0.13}^{+0.10} $ & $ 0.362_{-0.004}^{+0.008}$&$0.16\pm 0.01$&$4754\pm 380$ &$-0.01_{-0.09}^{+0.07}$ \\ [0.2cm]
		365733349 &1288&$115.06_{-0.15}^{+0.40}$& $5307_{-16}^{+18}$&$4.62\pm 0.05$&$0.974_{-0.008}^{+0.007}$&$1.46\pm 0.06$&$2227\pm 103$&$-0.04\pm 0.10$ \\[0.2cm]
        73540072 &1853& $166.8_{-0.5}^{+0.9}$&$5040_{-13}^{+22}$&$4.36_{-0.11}^{+0.09}$&$0.804_{-0.009}^{+0.007}$&$0.55_{-0.03}^{+0.04}$&$1491\pm 117$&$0.27_{-0.08}^{+0.07}$\\[0.2cm]
        372172128 &2196&$263.5_{-0.5}^{+3.0}$& $5655_{-51}^{+34}$&$4.28_{-0.08}^{+0.07}$&$1.044_{-0.011}^{+0.024}$&$0.75\pm 0.06$&$929\pm 89$&$0.11_{-0.09}^{+0.10}$\\[0.2cm]
        153078576 &2407&$92.23_{-0.12}^{+0.20}$& $3522_{-17}^{+22}$&$4.86_{-0.10}^{+0.09}$&$0.571_{-0.009}^{+0.006}$&$0.86\pm 0.06$&$6512\pm 514$&$-0.28_{-0.18}^{+0.09}$\\[0.2cm]
        146523262 &2465& $217.7_{-0.9}^{+1.7}$&$6146\pm 24$&$4.72\pm 0.05$&$1.200_{-0.011}^{+0.014}$&$2.74\pm 0.14$&$2235\pm 132$&$0.03_{-0.07}^{+0.04}$\\[0.2cm]
        358070912 &3261&$301.6_{-1.0}^{+1.5}$& $5126_{-28}^{+22}$&$4.48\pm 0.09$&$0.916_{-0.026}^{+0.021}$&$0.72\pm 0.04$&$1320\pm 124$&$-0.30_{-0.09}^{+0.08}$\\[0.2cm]
        124235800 &4898&$99.2
        _{-0.2}^{+0.3}$& $3823_{-22}^{+28}$&$5.08_{-0.19}^{+0.22}$&$0.697\pm 0.009$&$2.1\pm 0.2$&$8743\pm 899$& $-0.08 \pm 0.11$\\[0.2cm]
		\hline
    \end{tabular}
	
\end{table*}
\subsection{Transit fit}
\begin{figure*}
    \centering
    \includegraphics[width=\textwidth]{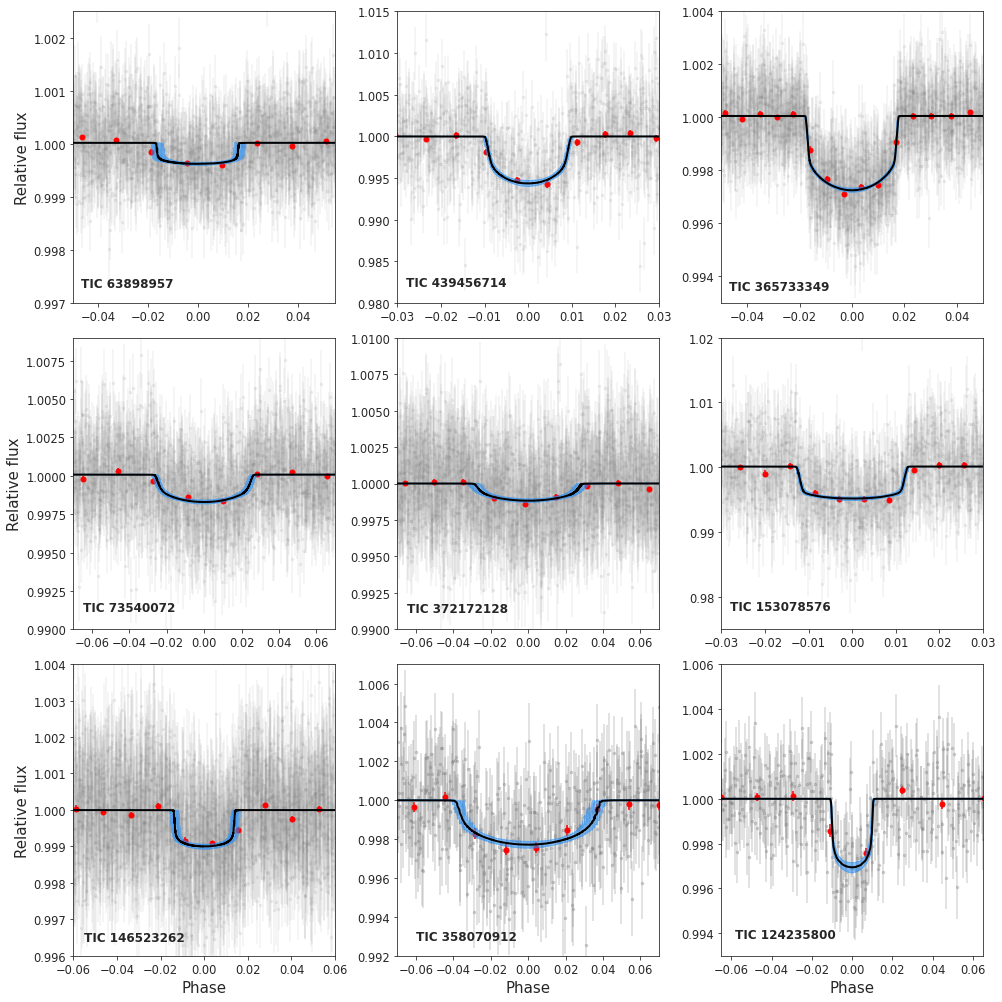}
    \caption{Phase-folded TESS lighcturves (grey points) for each of the nine likely Hot Neptunes of this work. Best-fit median model is shown by the solid black line along with the blue shaded region specifying the 68$\%$ posterior credibility interval for our best-fit model. The red dots with error bars represent the binned data. The corresponding best-fit parameters are summarized in Table \ref{tab:transit_fit}.}
    \label{fig:final_detrended_lc}
\end{figure*}

In this work we used the open source python package \texttt{juliet}\footnote{https://github.com/nespinoza/juliet} for the modelling of our $9$ likely Hot Neptunes. 
This tool exploits the \texttt{batman} package \citep{2015PASP..127.1161K} for transit fitting 
and allows to constrain the physical planetary parameters based on photometry and the  radial velocity
measurements when available.
The MultiNest sampler \citep{10.1111/j.1365-2966.2009.14548.x}, via the PyMultiNest wrapper \citep{2014A&A...564A.125B}, is employed to determine posterior parameters. 
The Presearch data conditioning simple aperture phtotometry lightcurves \citep{2012PASP..124..963K} are retrieved from the MAST archive{\footnote{http://archive.stsci.edu/tess/}}, where long term trends have been removed along with fewer systematic trends. 
Nevertheless, the lightcurves of the candidates TIC 124235800, 146523262, 358070912 and 37217218 still possessed some modulations in the lightcurve due to some artefacts and possible space telescope jitter. 
In order to take into account this noise, we used the Gaussian Process (GP) method with a Matern kernel using the celerite package, see \cite{Foreman_Mackey_2017}. The GP method provides a non-parametric technique to address the systematic trends or the noise are hard to model. 
After having detrended the modulations in the lightcurves with GP method, we defined priors for transit modelling. For orbital period $P$ and  mid-transit time $T_0$ based on ExoFOP values, we set a normal distribution $N(\mu$, $\sigma)$ where $\mu$ and $\sigma$ are respectively the mean and the standard deviation of the distribution. In regards to the setup of the impact parameter $b$ and the planet to stellar radius ratio $p=R_p/R_*$ we followed the sampling scheme as explained by \cite{2018RNAAS...2..209E}. In particular, we sampled two parameters $r1$ and $r2$, defined using a uniform prior between $0$ and $1$, which return all the physically meaningful values of  $p$ and $b$, with the condition $0<b<1+p$ to be satisfied. 
In addition, we used \texttt{ARIADNE}'s estimated stellar density, as a normal prior to precisely constrain the semi-major axis. 
Furthermore, we assumed a circular orbit by fixing the eccentricity value to zero.
Our model includes the quadratic law which fits the two limb darkening parameters $q1$ and $q2$ with a uniform prior between $0$ and $1$, as described by \cite{2013MNRAS.435.2152K}. We also fitted the jitter term, $\sigma_{w,i}$, to account for the white noises. However, when it was incapable of modelling more complicate cases we used the GP method as discussed above.
The dilution factor $DTess$, is the ratio of the out of transit flux of the target star to the total flux by other stars within the photometric aperture. $DTess$ is fixed to unity assuming no contamination by nearby sources.
We show the best-fit lightcurve models for each candidate in Figure \ref{fig:final_detrended_lc} along with the best-fit parameters in Table \ref{tab:transit_fit} as obtained by using \texttt{juliet}.

\begin{figure*}
    \centering
    \includegraphics[width=\textwidth]{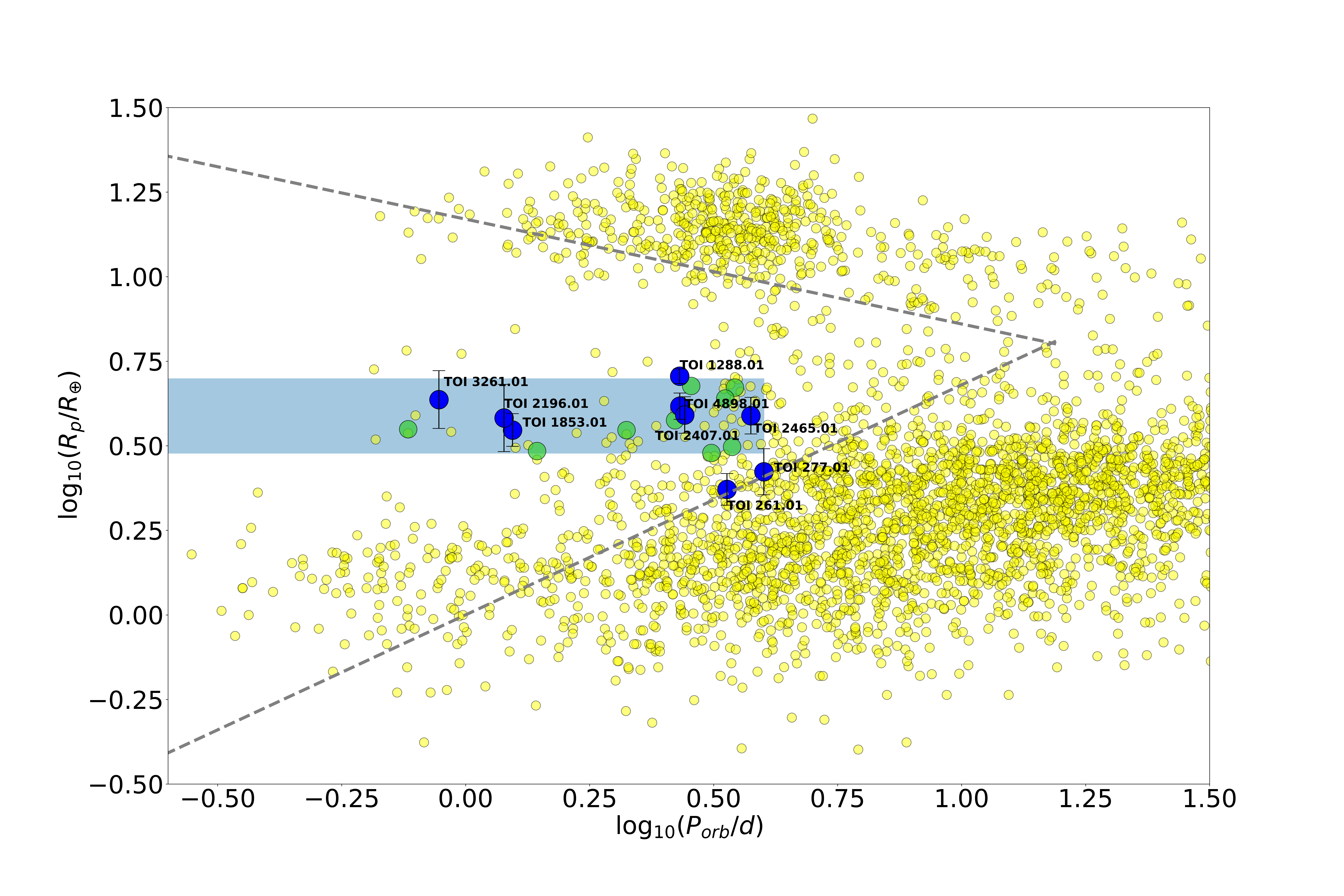}
    \caption{The nine new candidate Hot Neptunes in the (P,R) plane obtained in this work represented as blue dots. The yellow dots are confirmed exoplanets by all survey. The green dots are the nine TOIs within the Desert already confirmed and known planets. The blue shaded region is the Desert defined in this work. The grey dashed lines are the upper and lower boundaries of the Hot Neptune Desert as calculated by \protect\cite{Mazeh_2016}.}
    \label{fig:9 candidates in the Desert}
\end{figure*}

In Fig. \ref{fig:9 candidates in the Desert} we show the whole sample of 18 robust candidates obtained from our vetting procedure, along with the confirmed TESS exoplanets in the (P,R) plane. 
TIC 63898957.01 and TIC 358070912.01 with an estimated $R=(2.35 \pm 0.11) R_\oplus$ and $R=(2.65 \pm 0.18) R_\oplus$ respectively, are hot sub-Neptunes outside our Desert which lie on the lower boundary of the Hot Neptune Desert calculated by \cite{Mazeh_2016}. TIC 365733349.01 with $R=(5.08 \pm 0.11) R_\oplus$ is just slight out of our Desert but still within the $3R_\oplus<R<10R_\oplus$ range.
TIC 372172128.01, TIC 73540072.01, TIC 124235800.01, TIC 153078576.01, TIC 146523262.01 and TIC 439456714.01 are instead found within the Desert boundaries.

\begin{table*}
\caption{Best fit planet parameters with $68\%$ confidence interval of the new nine likely hot Neptunes with \texttt{juliet}.}
\label{tab:transit_fit}
\begin{tabular}{cc|ccc|cccc}
\hline

\multicolumn{1}{c}{TIC ID} & \multicolumn{1}{c|}{TOI} & \multicolumn{3}{c|}{Posterior Parameters}                            & \multicolumn{4}{c}{Derived parameters}                                                      \\[0.2cm]
\multicolumn{1}{c}{}       & \multicolumn{1}{c|}{}    & Period (days)        & $T_0$ (BJD-2457000)    & $\rho_{*} (kg/m^3)$  & R ($R_\oplus)$  & $a$ (AU)            & $b$                       & $i (^{\circ})$          \\[0.2cm] \hline 
63898957                   & 261.01                   & $3.3636\pm 0.0017$   & $1383.9184 \pm 0.0059$ & $1523_{-85}^{+74}$   & $2.35 \pm 0.11$ & $0.0518 \pm 0.0014$ & $0.086_{-0.056}^{+0.078}$ & $89.5_{-0.5}^{+0.3}$    \\[0.2cm]
439456714                  & 277.01                   & $3.9940\pm 0.0004$   & $2119.8264 \pm 0.0017$ & $4838_{-368}^{+361}$ & $2.65\pm 0.18$  & $0.0269\pm 0.0011$  & $0.36_{-0.14}^{+0.09}$    & $88.9_{-0.4}^{+0.5}$ \\[0.2cm]
365733349                  & 1288.01                  & $2.69998\pm 0.00015$ & $1712.3589\pm 0.0007$  & $2098_{-61}^{+56}$   & $5.08\pm 0.11$  & $0.0429\pm 0.0006$  & $0.072_{-0.049}^{+0.065}$ & $89.5_{-0.4}^{+0.3}$    \\[0.2cm]
73540072                   & 1853.01                  & $1.2437\pm 0.0002$   & $2670.8431\pm 0.0016$  & $1499\pm 110$        & $3.52\pm 0.17$  & $0.0186\pm 0.0006$  & $0.63_{-0.05}^{+0.04}$    & $82.6_{-0.6}^{+0.7}$    \\[0.2cm]
372172128                  & 2196.01                  & $1.19544\pm 0.00036$ & $2036.5083\pm 0.0038$  & $939\pm 80$          & $ 3.82\pm 0.38$ & $0.0201\pm 0.0009$  & $0.72_{-0.09}^{+0.05}$    & $80.0_{-0.9}^{+1.4}$    \\[0.2cm]
153078576                  & 2407.01                  & $2.70302\pm 0.00022$ & $2116.620\pm 0.001$    & $5872_{-328}^{+312}$ & $4.14\pm 0.16$  & $0.0349\pm 0.0011$  & $0.1_{-0.4}^{0.09}$       & $89.6_{-0.4}^{+0.3}$    \\[0.2cm]
146523262                  & 2465.01                  & $3.759\pm 0.001$     & $2174.8386\pm 0.0042$  & $2060_{-120}^{+138}$ & $3.88\pm 0.17$  & $0.064 \pm 0.002$  & $0.09_{-0.06}^{+0.08}$    & $89.5_{-0.4}^{+0.3}$    \\[0.2cm]
358070912                  & 3261.01                  & $0.88299\pm 0.00019$ & $2112.3973\pm 0.0003$  & $1343\pm 115$        & $4.33\pm 0.37$  & $0.01624\pm 0.0008$ & $0.48\pm 0.12$            & $82.7_{-2.2}^{+1.9}$    \\[0.2cm]
124235800                  & 4898.01                  & $2.76307\pm 0.00049$ & $2310.6620\pm0.0018$   & $8812_{-799}^{+711}$ & $3.90\pm 0.21$  & $0.049\pm 0.002$    & $0.24\pm 0.15$            & $89.0_{-0.6}^{+0.5}$    \\ [0.2cm]\hline
\end{tabular}
\end{table*}

\section{Follow-up observations}
\label{sec:followup}
Without using any follow-up observation we have already statistically validated two out the nine candidate Hot Neptunes in our sample. 
In this section we summarize the publicly available follow-up observations obtained by the TFOP Working Groups that have been used in this work.

\subsection{High-resolution Imaging}

TESS has a large pixel scale, about $21''$/pixel, with a focus-limited PSF. 
As a result, the flux measured in a single pixel may be contaminated by the contribution of nearby or background and foreground sources.
So, even if there is no centroid offset in the TESS image as revealed by the \textit{centroids} analysis performed with \texttt{DAVE}, there might be  additional sources within the $21''$-wide pixel  contaminating the transit event or being the origin of the transit-like signal.
In the first case the light contamination due to the unresolved sources could bias the depth of the transit, resulting in underestimated exoplanetary radii \citep{2015ApJ...805...16C,2017AJ....153...71F}. 
As a consequence, the mean density of the planet will be poorly constrained or incorrect even if the the mass of the planet is determined with high accuracy. Moreover, the transiting object may not be a planet at all, but a brown dwarf or a small stellar binary.

In order to discard stellar companions and foreground/background stars unresolved scenario, we used the adaptive optics and speckle imaging observations of our targets as obtained by the TFOPWG by means of several facilities all over the world.
Each of the nine TOIs in our sample has been observed within the TESS-EXOFOP program.
TFOP comprises a large working group of astronomical observers supporting the TESS Mission yield by providing different follow-up measurements. These observations are summarized in Table \ref{tab:followup_table} and the corresponding high-contrast curves are shown in Figure \ref{fig:high-contrast curves}.

\subsubsection{Southern Astrophysical Research Telescope - Cerro Pachón}
The $4.1$ m Southern Astrophysical Research (SOAR) Telescope observations of TIC 63898957, TIC 439456714, TIC 73540072, TIC 372172128, TIC 153078576, TIC 146523262, TIC 358070912 and TIC 124235800 were carried with speckle interferometric imaging HRCam. The contrast curves extracted from these observations are shown in Figure \ref{fig:high-contrast curves}.
For more information, we refer the reader to \cite{2020AJ....159...19Z,2021AJ....162..192Z}.
\subsubsection{Gemini North Telescope - Hawaii \& Gemini South Telescope -  Cerro Pachón}
TIC 439456714 was observed using speckle interferometric imaging of 'Alopeke and Zorro mounted upon the 8 m Gemini North and South telescopes respectively. In this work we will not use the Zorro data due to the poor seeing at the moment of observation. Moreover this TIC has also been observed performing adaptive optics (AO) observations taken by  the Near‐Infrared Imager (NIRI) instrument mounted on the Gemini North telescope.
Furthermore, 'Alopeke also observed TIC 63898957, TIC 365733349 and TIC 73540072.
We also used AO images of TIC 365733349 obtained by NIRI. The contrast curves extracted from these observations are shown in Figure \ref{fig:high-contrast curves}.
For more information, we refer the reader to \cite{10.1117/12.2311539} and \cite{2021FrASS...8..138S}.
\subsubsection{Keck Telescope - Hawaii}
We used the AO observations of TIC 73540072 obtained with the the near-infrared imager NIRC2 instrument, placed on Keck-II. The contrast curves extracted from these observations are shown in Figure \ref{fig:high-contrast curves}.
\subsubsection{Shane Telescope - Lick Observatory}
We observed TIC 63898957 on UT 2019 September 14 using the ShARCS camera on the Shane 3-meter telescope at Lick Observatory \citep{2012SPIE.8447E..3GK, 2014SPIE.9148E..05G, 2014SPIE.9148E..3AM}. The observation was taken with the Shane adaptive optics system in natural guide star mode. The final images were constructed using sequences of images taken in a 4-point dither pattern with a separation of 4$\arcsec$ between each dither position. Two image sequences were taken of this star: one with a $Ks$ filter ($\lambda_0 = 2.150$ $\mu$m, $\Delta \lambda = 0.320$ $\mu$m) and one with a $J$ filter ($\lambda_0 = 1.238$ $\mu$m, $\Delta \lambda = 0.271$ $\mu$m). A more detailed description of the observing strategy and reduction prodecure can be found in \cite{2020AJ....160..287S}. The contrast curves extracted from these observations are shown in Figure \ref{fig:high-contrast curves}. With the $Ks$ filter, we achieve contrasts of 4.0 at $0\farcs5$ and 5.7 at $1\farcs0$. With the $J$ filter, we achieve contrasts of 4.0 at $0\farcs5$ and 5.6 at $1\farcs0$. We find no nearby stellar companions within our detection limits.
\subsubsection{WIYN - Kitt Peak}
\begin{figure*}
    \centering
    \includegraphics[width=\textwidth]{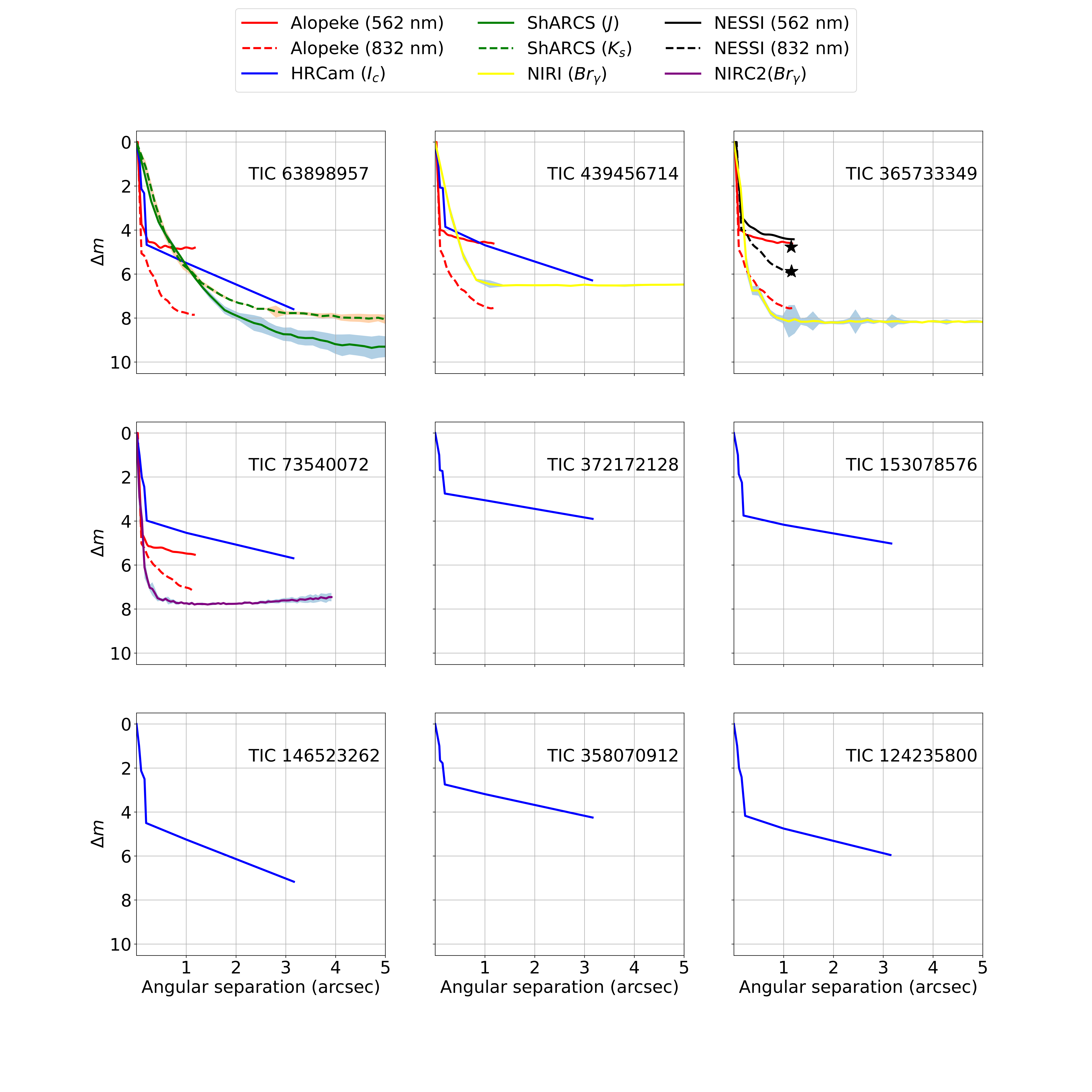}
    \caption{High-contrast curves extracted from the imaging observations summarized in Table \ref{tab:followup_table}. TIC 365733349 have  $<2''$ companions represented by two black star. Curves with shading (from adaptive optics imaging) were created by calculating the mean and rms error of the contrasts in circular annuli around the target star. These curves are incorporated into the \texttt{TRICERATOPS} analysis in order to better constraint its results as described in Section \ref{sec:results}.} 
    \label{fig:high-contrast curves}
\end{figure*}

TIC 365733349 was observed with the NN-EXPLORE Exoplanet Stellar Speckle Imager (NESSI) mounted on the $3.5$ m WIYN telescope located at Kitt Peak.  The contrast curves extracted from these observations are shown in Figure \ref{fig:high-contrast curves}. For more information, we refer the reader to \cite{2018PASP..130e4502S}.
\begin{table*}
\caption{Follow-up observations of the likely Hot Neptunes.}
\label{table:followup_table}
	\centering
	\label{tab:followup_table}
	\begin{tabular}{lllcccccr} % four columns, alignment for each
		\hline
		TIC ID & Telescope & Instrument & Observation
date (UT) & Filter & Image Type & Companion\\
		\hline
		63898957& Gemini-N ($8$ m)  &'Alopeke& 2019-10-13 & $562$ nm & Speckle & No \\
		& Gemini-N ($8$ m)  &'Alopeke& 2019-10-13 & $832$ nm & Speckle & No \\
		& Shane ($3$ m) &ShARCS& 2019-09-14& \textit{J} & AO & No \\
		& Shane ($3$ m) &ShARCS&2019-09-14 & $\textit{K}_s$ & AO & No \\
		& SOAR ($4.1$ m) &HRCam& 2019-07-14& $\textit{I}_c$ & Speckle & No \\
		\hline
		439456714& Gemini-N ($8$ m)  &'Alopeke&2020-08-10 & $562$ nm & Speckle & No \\
		& Gemini-N ($8$ m)  &'Alopeke&2020-08-10 & $832$ nm & Speckle & No \\
		& Gemini-N ($8$ m)  &NIRI& 2019-06-20&$\text{Br}_\gamma$ & AO & No \\
		& Gemini-S ($8$ m)  &Zorro&2019-07-17& $562$ nm & Speckle & No \\
		& Gemini-S ($8$ m)  &Zorro&2019-07-17 & $832$ nm & Speckle & No \\
		& SOAR ($4.1$ m) &HRCam& 	2019-08-12& $\textit{I}_c$ & Speckle & No \\
		\hline
		365733349 & Gemini-N ($8$ m)  &'Alopeke&2021-06-24 & $562$ nm & Speckle & No \\
		& Gemini-N ($8$ m)  &'Alopeke& 2021-06-24& $832$ nm & Speckle & No \\
		& Gemini-N ($8$ m)  &NIRI&2019-11-08	 &$\text{Br}_\gamma$ & AO & Yes \\
		& WIYN ($3.5$ m)  &NESSI&2019-11-17	 & $562$ nm & Speckle & No \\
		& WIYN ($3.5$ m)  &NESSI&2019-11-17	 & $832$ nm & Speckle & No \\
		\hline
		73540072 & Gemini-N ($8$ m)  &'Alopeke&2020-06-10 & $562$ nm & Speckle & No \\
		& Gemini-N ($8$ m)  &'Alopeke&2020-06-10 & $832$ nm & Speckle & No \\
		& SOAR ($4.1$ m) &HRCam&2021-02-27 & $\textit{I}_c$ & Speckle & No \\
		& Keck ($10$ m) &NIRC2&2020-05-28 & $\text{Br}_\gamma$ & AO & No \\
        \hline
        372172128 & SOAR ($4.1$ m) &HRCam&2020-10-31& $\textit{I}_c$ & Speckle & No \\
        \hline
        153078576 & SOAR ($4.1$ m) &HRCam&2020-12-03	 & $\textit{I}_c$ & Speckle & No \\
        \hline
        146523262 &  SOAR ($4.1$ m) &HRCam&2021-02-27 & $\textit{I}_c$ & Speckle & No \\
        \hline
        358070912 &  SOAR ($4.1$ m) &HRCam&2021-07-14 & $\textit{I}_c$ & Speckle & No \\
       \hline
        124235800&  SOAR ($4.1$ m) &HRCam&2022-03-20	 & $\textit{I}_c$ & Speckle & No \\
		\hline
		
	\end{tabular}
	
\end{table*}
\subsection{Reconnaissance Spectroscopy}
We used  the publicly available reconnaissance spectra of our TOIs in order to rule out a false positive scenario caused by a spectroscopic binary system that would not necessarily be detected by high-contrast imaging observationsWe stress that those stars without any available reconnaissance spectra can not be statistically validated even if they passed the \texttt{TRICERATOPS} test.
The Tillinghast Reflector Echelle Spectrograph (TRES) mounted on the 1.50 m Tillinghast Telescope located at the Fred Lawrence Whipple Observatory (FLWO) in southern Arizona acquired the reconnaissance spectra of TIC 63898957, TIC 439456714, TIC 365733349, TIC 73540072 and TIC 146523262. In addition, we also used the reconnaissance spectra of TIC 365733349 acquired by the cross-dispersed high-resolution FIbre-fed Echelle Spectrograph (FIES; \citealt{2014AN....335...41T}) mounted on the 2.6 m Nordic Optical Telescope (NOT) on La Palma, Spain. The left stars of this work  are all southern targets with a declination too low for both TRES and FIES.\newline
The reconnaissance spectra obtained by TRES and FIES were reduced using the Stellar Parameter Classification (SPC, \citealt{2012Natur.486..375B}) pipeline. SPC cross matches an observed spectrum against a library grid of synthetic template spectra based on Kurucz atmospheric models \citep{1993KurCD..18.....K}. SPC firstly fits for the stellar parameters $T_{eff}$, $\log g$, $[m/H]$ and $v_{\star}\sin i$ then computes the cross correlation function normalized peak height using a three dimensional third-order polynomial fit. Yonsei-Yale
isochrones \citep{2001ApJS..136..417Y} are used to set priors on the surface gravity which otherwise would not be well constrained by observed spectra. Due to SPC was designed for sun-like stars, it works less robustly with cooler, lower mass stars as well as hotter and fast rotating stars. Hence, TIC 439456714 is too cool for reliable SPC parameters. Table \ref{tab:spc} summarizes the spectroscopic follow-up observations as well as the SPC extracted parameters for the aforementioned TICs. We emphasize that the majority of the parameters generated by SPC agree with those obtained by \texttt{ARIADNE}. We visually examined these spectra and found no evidence of a composite spectrum, which could have occurred if there was a stellar companion (for more details see Appendix \ref{appendix:tres_spectra}).
\begin{table*}
\caption{SPC outcomes based on TRES/FIES spectra.}
\label{table:spc}
	\centering
	\label{tab:spc}
	\begin{tabular}{lllcccccr} % four columns, alignment for each
		\hline
		TIC ID & Telescope & Instrument & Spectra resolution & Number of spectra used & $T_{eff}$ (K) & $\log g$&$[m/H]$\\
		\hline
		63898957 & FLWO ($1.50$ m)  &TRES& 44000 & 2 & 5953  & 4.24 & -0.11  \\
		\hline
		365733349 & NOT ($2.6$ m) & FIES & 67000 & 13 & 5368 & 4.40 & 0.24\\
		& FLWO ($1.50$ m)  &TRES& 44000 & 3 & 5402 & 4.44 & 0.24  \\
		\hline
		73540072 & FLWO ($1.50$ m)  &TRES& 44000 & 1 & 5175   & 4.63 & 0.28\\  
		\hline
		146523262& FLWO ($1.50$ m)  &TRES& 44000 & 2 & 6159    & 4.35& 0.33\\
		\hline
	\end{tabular}
	
\end{table*}

\section{Results}
\label{sec:results}

In this Section we analyze in detail each of the nine planet candidates vetted in Sect. \ref{sec:vetting}, shown in Table \ref{table:10new}, using the additional follow-up measurements aforementioned to better constrain the FPP and NFPP values.
Due to intrinsic statistical scattering in the calculation, for each TOI at this step we run the code $30$ times in order to increase the robustness of our results, taking the mean value along with the standard deviation for both FPP and NFPP values.

However, as we discuss in the following, despite  a candidate with FPP $ \lesssim 0.05$ and NFPP $<0.001$
is not considered to be statistically validated,
we argue that there is a relatively high chance of being a true planet, and so it should be considered a high priority target for  follow-up observations.

In the following section, for each target, we estimate the radial velocity semi-amplitude $K$ that these planets may produce in a spectroscopic follow-up observation.
To do so, we predicted the masses of each planet by using the \texttt{forecaster} tool \citep{2017ApJ...834...17C}. Due to the big uncertainties in the predictions by \texttt{forecaster}, by simply propagating the errors we obtain large error bars for the $K$ estimates. 
Hence one should look at these values just to get an idea of the order of magnitude and whether it is within the capability of modern spectrographs.
Eventually, for each planet we also calculated the Transmission Spectroscopy Metric (TSM), \citep{2018PASP..130k4401K}, 
a quantity proportional to the expected SNR of the transmission spectroscopy based on the strength of the spectral features along with the brightness of the host star in a cloud-free atmosphere. 
\cite{2018PASP..130k4401K} argued that TSM nearly reproduces the expected SNR over a 10-hour observation in the NIRISS bandpass, \cite{2012SPIE.8442E..2RD}, in order to check whether these TOIs could be amenable targets for future JWST observations.
In this work the TSM values might be slightly optimistic since the planets this close to their stars are likely to be denser than typical planets of their size and/or have higher mean molecular weight atmospheres $\mu$ due to photoevaporation \citep{2018MNRAS.479.5012O}. 
In order to perform a more refined calculation, one should re-scale the Equation (1) in \cite{2018PASP..130k4401K} taking into account the mean molecular weight $\mu$ of a typical Hot Neptune.

\subsection{TIC 63898957.01}
TIC 63898957.01, or TOI 261.01, is a planet candidate with $R = (2.35\pm 0.11) R_\oplus$ orbiting a bright G0V/G1V star ($V=9.56, d= 114.1_{-0.2}^{+0.8}\,\text{pc}$) in $\approx 3.3$ days.It has been observed by TESS in sector 3 with cadences of both $2$ and $30$ minutes and in sector 30 with cadences of both $2$ and $10$ minutes. A transiting planet candidate has been detected by both the SPOC \citep{2016SPIE.9913E..3EJ} and the QLP \citep{2020RNAAS...4..204H} pipelines. In both sectors the transit was clear (SNR $>300$) and well above the noisy level in the \textit{Modelshift} module of \texttt{DAVE}. The transit is box-shaped consistent with a true planet. Moreover TFOPWG has already found and confirmed TOI 261.02 which is $2.88 R_\oplus$ planet orbiting the host star in $\approx 13$ days. In light of this TOI 261.01 would make a multiple system composed by a Neptune and sub-Neptune orbiting a Sun-like star.
The Lomb-Scargle periodogram of the raw lightcurve finds a maximum peak of $3\cdot 10^{-4}$ suggesting that the star is quiet. There is no clear offset of the photocenter. In both sectors, the difference images show that the variation of light is spread over $3$ pixels but the highest loss of flux occurs in correspondence of the target.
We also inspected whether the background lightcurve at the time of transit could have altered the signal and in both sectors we found no anomalies. 
The SPOC data validation report draws attention to a slight centroid offset at sub-pixel level ($\approx 12$ arcsec) for this TOI when observed in sector 3. However, the sub-pixel offset may be because the fourth transit, in sector 3, occurred in correspondence of a TESS momentum dump.
This target has been observed by 'Alopeke (Gemini-N), ShARCS (Shane) and HRCam (SOAR) but neither AO nor speckle images found any companions nearby the target.

The \texttt{TRICERATOPS} analysis of this TIC returns FPP$= 0.14 \pm 0.04$ and NFPP $=0$. This candidate planet does not satisfy the requirements needed to be considered statistically validated. Assuming this is a real planet, using the \texttt{forecaster} tool, we computed its expected mass $M$ and its RV semiamplitude $K$. We obtain $M=6.17_{-2.54}^{+4.73} \, M_{\oplus}$ and
$K = \, 1.79_{-0.72}^{1.31} \, {\rm m \, s}^{-1}$.
The TSM calculated for this TOI is $58.1$.

\subsection{TIC 439456714.01}
TIC 439456714.01, or TOI 277.01, is a planet candidate with $R=(2.65 \pm 0.18) R_\oplus$ orbiting a high proper motion $(\mu_\alpha=-115.333\,\text{mas}\cdot\text{yr}^{-1}, \mu_\delta=-249.417\,\text{mas}\cdot\text{yr}^{-1})$ K6V star ($V=13.63, d= 65.17_{-0.09}^{+0.18}\,\text{pc}$) in $\approx 4$ days. 
It has been observed by TESS in sector 3 with cadences of both $2$ and $30$ minutes and in sector 30 with cadences of both $2$ and $10$ minutes. Both the SPOC and the QLP pipelines detected a transiting-like feature in both sectors. With a SNR of $338$, the transit is clear and well above the noisy level in the \textit{Modelshift} module of \texttt{DAVE}. Moreover, the shape of the transit is consistent with that of a transiting planet. The Lomb-Scargle periodogram of the raw lightcurve finds a maximum peak of $3\cdot 10^{-4}$ suggesting that the star is quiet.
No anomalies in the background flux have been detected at the time of transit in both sectors.
The difference image does not show any no clear offset of the light photocenter.
The SPOC data validation report draws attention to a slight centroid offset at sub-pixel level ($\approx 11$ arcsec) for this TOI.
The follow-up observations consist of both speckle and AO images provided by Zorro, Alopeke, HRCam for the former and NIRI for the latter.  No evidence of stellar companion was found by these measurements within detection limits.

The \texttt{TRICERATOPS} analysis of this TIC returns FPP$=(2.2\pm 3.9)\cdot 10^{-4}$ and NFPP $=0$.
This TIC achieves the requirements needed to be considered statistically validated.
We hereafter refer to this validated planet as TOI-277 b.
By using the \texttt{forecaster} tool, we estimated its mass $M$ and its RV semiamplitude $K$, in particular $M=7.58_{-3.12}^{+5.69}\,M_{\oplus}$.
$K=10.37_{-4.29}^{+7.79}\,m\cdot s^{-1}$.
The TSM calculated for this TOI is $116.7$.

\subsection{TIC 365733349.01}
TIC 365733349.01, or TOI 1288.01, is a planet candidate with $R=(5.08\pm 0.11) R_\oplus$ orbiting a G6V/G7V star ($V=10.44, d= 115.06_{-0.15}^{+0.40}\,\text{pc}$) in $\approx 2.7$ days. It has been observed by TESS in sectors $15,16,17,18$ and $24$ with cadences of both $2$ and $30$ minutes. In all sectors both SPOC and QLP pipelines detected a transiting planet candidate. With a SNR $>900$, the transit is clear and well above the noisy level in the \textit{Modelshift} module of \texttt{DAVE}. The difference image produced in sector 18 shows a negative net flux suggesting some problematic at the time of transit (UC: Unreliable Centroids).
We investigated this issue by plotting the lightcurve at pixel level and we detected a modulation of the flux in the nearby pixels at $(1793.4\pm 1.5) $ TJD which may have caused the negative net flux observed.
Nevertheless, when observed in other sectors, the analysis of centroid does not display any shift of the light photocenter as well as no more anomalies.  The shape of the transit is consistent with that of a transiting planet. The Lomb-Scargle periodogram of the raw lightcurve finds a maximum peak of $4\cdot 10^{-4}$ suggesting that the star is quiet. 
The SPOC data validation report does not show any anomalies for this TOI. The follow-up observations consist of both speckle and AO images provided by Alopeke, NIRI and NESSI instruments.  AO observations of NIRI detected 2 visual companions for this TIC at $1.152''$ and $1.579''$, which are $4.77$ and $5.88$ K-mag fainter respectively. The relative large distances suggest these may not be bound companions, in addition they are much fainter than the target to significantly contaminate the lightcurve.

The \texttt{TRICERATOPS} analysis of this TIC returns FPP$=(1.8\pm 7.8)\cdot 10^{-4}$ and NFPP $=0$.
This TIC satisfies the requirements needed to be considered statistically validated.
We hereafter refer to this validated planet as TOI-1288 b.
By using the \texttt{forecaster} tool, we estimated its mass $M$ and its RV semiamplitude $K$, in particular $M=22.62_{-9.67}^{+18.25}\,M_{\oplus}$.
$K=10.70_{-4.58}^{+8.64}\,$ m s$^{-1}.$
The TSM calculated for this TOI is $110.1$.

\subsection{TIC 73540072.01}
TIC 73540072.01, or TOI 1853.01, is a planet candidate with $R=(3.52\pm 0.17) R_\oplus$ orbiting a G8V-G9V star ($V=12.18, d=166.8_{-0.5}^{+0.9}\,\text{pc}$) in $\approx 1.2$ days. It has been observed by TESS in only sector 23 with a cadence of $10$ minutes and detected by the QLP pipeline.
The lightcurve is much noisy, SNR $=1.3$, and the transit is just above the noisy level. The shape of the transit is slightly V-shaped, but we did not find any statistical significant odd-even difference.
As far as the pixel analysis is concerned, the \textit{centroids} module does not show any centroid offset but we would not trust the measurements because the whole $13\times 13$ pixels image shows a flux variation at the time of the transit.
We inspected this anomaly by plotting the background lightcurve which shows a negative flux at the time of many transits ($1932.1,1933.4,1934.6,1935.8$ TJD) in the sector 23, due to unknown systematics.
This target has recently been observed in TESS sector 50 with $2$ minutes cadence and processed by the SPOC pipeline.We run \texttt{DAVE} with respect to this sector. In sector 50 the lightcurve exhibits a clear box-shaped transit and the \textit{centroids} module now generates a clear difference image with the overall light photocenter in correspondence of the target position.
The Lomb-Scargle periodogram of the raw lightcurve finds a maximum peak of $4\cdot 10^{-4}$ suggesting that the star is quiet.
The follow-up observations consist of both speckle and AO images provided by Alopeke, HRCam and NIRC2 instruments. No evidence of stellar companion was found by these measurements.

The \texttt{TRICERATOPS} analysis of this TIC returns FPP$=(1.68 \pm 0.02)\cdot 10^{-2} $ and NFPP $=0$.
This TIC does not achieve the requirements needed to be considered statistically validated.
Despite the fact that we do not consider this TOI statistically validated, its FPP value is low enough to consider this target amenable of more spectroscopic follow-up measurements.
By using the \texttt{forecaster} tool, we estimated its possible mass $M$ and RV semiamplitude $K$, in particular $M=11.82_{-4.91}^{+9.27}\,M_{\oplus}$ and
$K=8.5_{-3.9}^{+6.8}\,$ m s$^{-1}.$
The TSM calculated for this TOI is $77.3$.

\subsection{TIC 372172128.01}
TIC 372172128.01, or TOI 2196.01, is a planet candidate with $R=(3.82\pm 0.38) R_\oplus$ orbiting a G3V star ($V=12.30, d= 263.5_{-0.5}^{+3.0}\,\text{pc}$) in $\approx 1.2$ days. It has been observed by TESS in sector 13 with a cadence of $30$ minutes and in sector 27 with cadences of both $2$ and $10$ minutes. In both sectors the SPOC and QLP pipelines recognised a transiting-like feature in the lightcurve.
The lightcurve has a SNR $>100$ so the transit is clear and well above the noisy level in the \textit{Modelshift} module. The difference does not show any centroid offset.
Moreover we found no systematics in the background flux at the time of each transit in the sector. 
The Lomb-Scargle periodogram of the raw lightcurve finds a maximum peak of $4\cdot 10^{-4}$ suggesting that the star is quiet.
The SPOC data validation report does not show any anomalies for this TOI. 
The follow-up observations consist of speckle images provided by the HRCam that did not detect any companion.

The \texttt{TRICERATOPS} analysis of this TIC returns FPP$=(3.03\pm 0.05)\cdot 10^{-1}$ and NFPP $=0$.
This TIC does not achieve the requirements needed to be considered statistically validated.
Assuming it is a real planet, using the \texttt{forecaster} tool, we estimated its mass $M$ and its RV semiamplitude $K$, in particular $M=12.98_{-6.00}^{+10.06}\,M_{\oplus}$ and $K=7.81_{-3.62}^{+6.05} \, {\rm m s}^{-1}$.
The TSM calculated for this TOI is $57.4$.

\subsection{TIC 153078576.01}
TIC 153078576.01, or TOI 2407.01, is a planet candidate with $R=(4.14\pm 0.16) R_\oplus$ orbiting a K9V-M0V ($V=14.68, d= 92.23_{-0.12}^{+0.20}\,\text{pc}$) in $\approx 2.7$ days. It has been observed by TESS in sectors 3,4 with a cadence of $10$ minutes, in sectors 30 and 31 with cadences of both $2$ and $10$ minutes. Both the SPOC and the QLP pipelines detected a transiting-like feature in all four sector.
The transit is clear (SNR $>22$) and well above the noise level in the \textit{Modelshift} module. The difference image obtained in the \textit{centroids} module does not show any centroid offset.
The Lomb-Scargle periodogram of the raw lightcurve finds a maximum peak of $3 \, \cdot \,10^{-2}$ 
in correspondence of three days suggesting that the star is slightly variable.
No anomalies in the background flux have been detected.
The SPOC data validation report does not discuss any anomalies for this TOI.
The follow-up observations consist of just speckle images provided by the HRCam instrument which did not find any companion.

The \texttt{TRICERATOPS} analysis of this TIC returns FPP $=(5\pm 3)\cdot 10^{-3}$ and NFPP $=0$.
This TIC does not achieve the requirements needed to be considered statistically validated. We note that if we had used the less conservative constraint FPP$<10^{-2}$ and had reconnaissance spectroscopy of the target star, this candidate would have been statistically validated.
Assuming it is a real planet, using the \texttt{forecaster} tool, we estimated its mass $M$ and its RV semiamplitude $K$, in particular
$M=16.18_{-6.84}^{+12.29}\,M_{\oplus}$ and
$K=13.12_{-5.54}^{+9.97}\,m\cdot s^{-1}$.
The TSM calculated for this TOI is $66.5$.

\subsection{TIC 146523262.01}
TIC 146523262.01, or TOI 2465.01, is a planet candidate with $R=(3.88\pm 0.17) R_\oplus$ orbiting an F9V star ($V=10.78, d= 217.7_{-0.9}^{+1.7}\,\text{pc}$) in $\approx 3.8$ days. It has been observed by TESS only in sector 32 with cadences of both $2$ and $10$ minutes and detected by both the SPOC and QLP pipelines.
With a SNR $>350$ the transit is clear and well above the noisy level in the \textit{Modelshift} module. The \textit{centroids} analysis shows no offset of the light photocenter. The Lomb-Scargle periodogram of the raw lightcurve finds a maximum peak of $6\cdot 10^{-5}$ suggesting that the star is very quiet.
Moreover, we found no anomalies in the background lightcurve at the time of each transits in the whole sector.
The SPOC data validation report does not show any anomalies for this TOI.
The follow-up observations consist of just speckle images provided by the HRCam instrument that did not find any companion.

The \texttt{TRICERATOPS} analysis of this TIC returns FPP$=(2.8\pm 0.2)\cdot 10^{-2}$ and NFPP $=0$.
This TIC does not achieve the requirements needed to be considered statistically validated.
As done for TIC 73540072, the FPP value is low enough to consider this target amenable of more spectroscopic follow-up measurements. Assuming it is a real planet, using the \texttt{forecaster} tool, we estimated its mass $M$ and its RV semiamplitude $K$, in particular $M=13.62_{-5.83}^{+10.59}\,M_{\oplus}$ and
$K=4.82_{-2.06}^{+3.74}\,m\cdot s^{-1}$.
The TSM calculated for this TOI is $52.4$.

\subsection{TIC 358070912.01}
TIC 358070912.01, or TOI 3261.01, is a planet candidate with $R=(4.33\pm 0.37) R_\oplus$ orbiting a G8V star ($V=13.26, d=301.6_{-1.0}^{+1.5}\,\text{pc}$) in $\approx 0.88$ days. It has been observed by TESS for a total of 7 sectors from 2018 and 2021. In particular it has been observed in sectors 2,6 and 13 with a cadence of $30$ minutes and in sectors 27,28 and 29 with a cadence of $10$ minutes. In all sectors the QLP pipeline identified a transiting planet candidate.We here stress that due to the lack of $2$-minute cadence data, regardless of \texttt{TRICERATOPS} outcomes, this candidate can not be validated by our procedure.
The SNR $\approx 20$ is high enough to see a quite clear transit in the phase-folded lightcurve in the \textit{Modelshift} module. 
The \textit{centroids} outcome shows no offset of the light photocenter with respect to the target position.
The transit is slightly V-shaped and we noticed a statistical significant odd-even difference only in sector 6.
We found no clear issues in the background flux at the time of transit for all sectors.
Despite some suspicious oddities, we had no enough red flags to rule out this TIC from our analysis.
The follow-up observations consist of just speckle images provided by the HRCam instrument that did not detect any companion.

The \texttt{TRICERATOPS} analysis of this TIC returns FPP$=(6.3\pm 0.2)\cdot 10^{-1}$ and NFPP $=(2.4\pm 0.8)\cdot 10^{-4}$.
This TIC does not achieve the requirements needed to be considered statistically validated.
This target offers an important food for thought about the workflow adopted in this work. Despite we were not such confident in ranking it as a PC due to some red flags, we also could not rule out as a FP. Probably due to statistical fluctuations, it passed the first run of \texttt{TRICERATOPS} with FPP $<0.5$. When we re-run the code augmenting the robustness of the analysis along with the follow-up measurements we obtained an averaged FPP $>0.5$ confirming the goodness of our preliminary doubts.
Assuming it is a real planet, using the \texttt{forecaster} tool, we estimated its mass $M$ and its RV semi-amplitude $K$, in particular $M=15.59_{-6.56}^{+11.43}\, M_{\oplus}$ and
$K=12.95_{-5.62}^{+9.49} \, {\rm m s}^{-1}$.
The TSM calculated for this TOI is 53.8.

\subsection{TIC 124235800.01}
TIC 124235800.01, or TOI 4898.01, is a planet candidate with $R=(3.90\pm 0.21) R_\oplus$ orbiting a K7V-K8V ($V=13.62, d= 99.2_{-0.2}^{+0.3}\,\text{pc}$) in $\approx 2.76$ days. It has been observed by TESS in sector 10 with a cadence of $30$ minutes and in sector 37 with a cadence of $10$ minutes. A transiting planet candidate has been catched by the QLP pipeline.
The target star is nearby ($\approx 4'$) two brighter known stars (TIC 124235806 and  124235788) which falsify the \textit{centroids} module outcome whose light photocenter measurements are unrealiable. 
Nevertheless, we found no anomalies in the background flux at the time of each transit in both sectors.
The follow-up observations consist of just speckle images provided by the HRCam instrument that did not find any companion.

The \texttt{TRICERATOPS} analysis of this TIC returns FPP$=(1.2 \pm 3.5) \cdot 10^{-5}$ and NFPP $=0$.
This TIC meets the requirements for statistical validation, but due to the lack of 2-minute cadence data and the absence of reconnaissance spectroscopy, we consider it a good priority target for further investigation.
Assuming it is a real planet, using the \texttt{forecaster} tool, we estimated its mass $M$ and its RV semiamplitude $K$, in particular, $M=16.29_{-7.00}^{+11.93}\,M_{\oplus}$ and $K=11.26_{-4.37}^{+8.24}\,m\cdot s^{-1}$.
The TSM calculated for this TOI is $65.3$. 

\begin{table*}
\caption{Vetting results for the nine likely hot Neptunes using follow-up observations.}
\label{table:fin_table_disp}
	\centering
	\label{tab:fin_table_disp}
	\begin{tabular}{llrlccr} % four columns, alignment for each
		\hline
		TIC ID & TOI &TSM & DAVE & FPP & NFPP&Validated\\
		\hline
		124235800.01& 4898.01& $65.3$ & UC, long cadence data  & $(1.2\pm 3.5)\cdot 10^{-5}$  & $0$ & N\\
		365733349.01& 1288.01 & $110.1$ &  UC in sector 18 & $(1.8\pm 7.8)\cdot 10^{-4}$ & $0$ & Y \\
		439456714.01&277.01& $116.7$ &  Clear & $(2.2\pm 3.9)\cdot 10^{-4}$ & $0$ & Y \\
		153078576.01& 2407.01 & $66.5$ & Slight variable star  & $(5\pm 3)\cdot 10^{-3}$  & $0$ & N\\
		73540072.01& 1853.01& $77.3$ & UC, Low SNR in sector 23 &$(1.68\pm 0.02)\cdot 10^{-2}$  & 0 & N \\
		146523262.01& 2465.01 & $52.4$ & Clear & $(2.8\pm 0.2)\cdot 10^{-2}$  & $0$ &N\\
		63898957.01& 261.01 & $58.1$ & Clear & $(1.3\pm 0.1)\cdot 10^{-1}$ & 0 & N\\
		372172128.01& 2196.01& $57.4$ & Clear &$(3.03\pm 0.05)\cdot 10^{-1}$ & 0 & N \\
		358070912.01& 3261.01& $65.3$ & slightly V-shaped transit, OED in sect.6, long cadence data  & $(6.3 \pm 0.2)\cdot 10^{-1}$  & $(2.4\pm 0.8)\cdot 10^{-4}$& N\\
		\hline
		
	\end{tabular}
	
\end{table*}

\section{Potential for Atmospheric Characterization}
\label{sec:discussion}

Characterization of the Hot Neptunes atmosphere is a crucial step to address the riddle of their 
formation and evolution. Transit spectroscopy with JWST will allow to shed light on whether or not 
Hot Neptunes are loosing their envelope along with its composition. 

\begin{figure}
    \centering
    \includegraphics[width=0.5\textwidth]{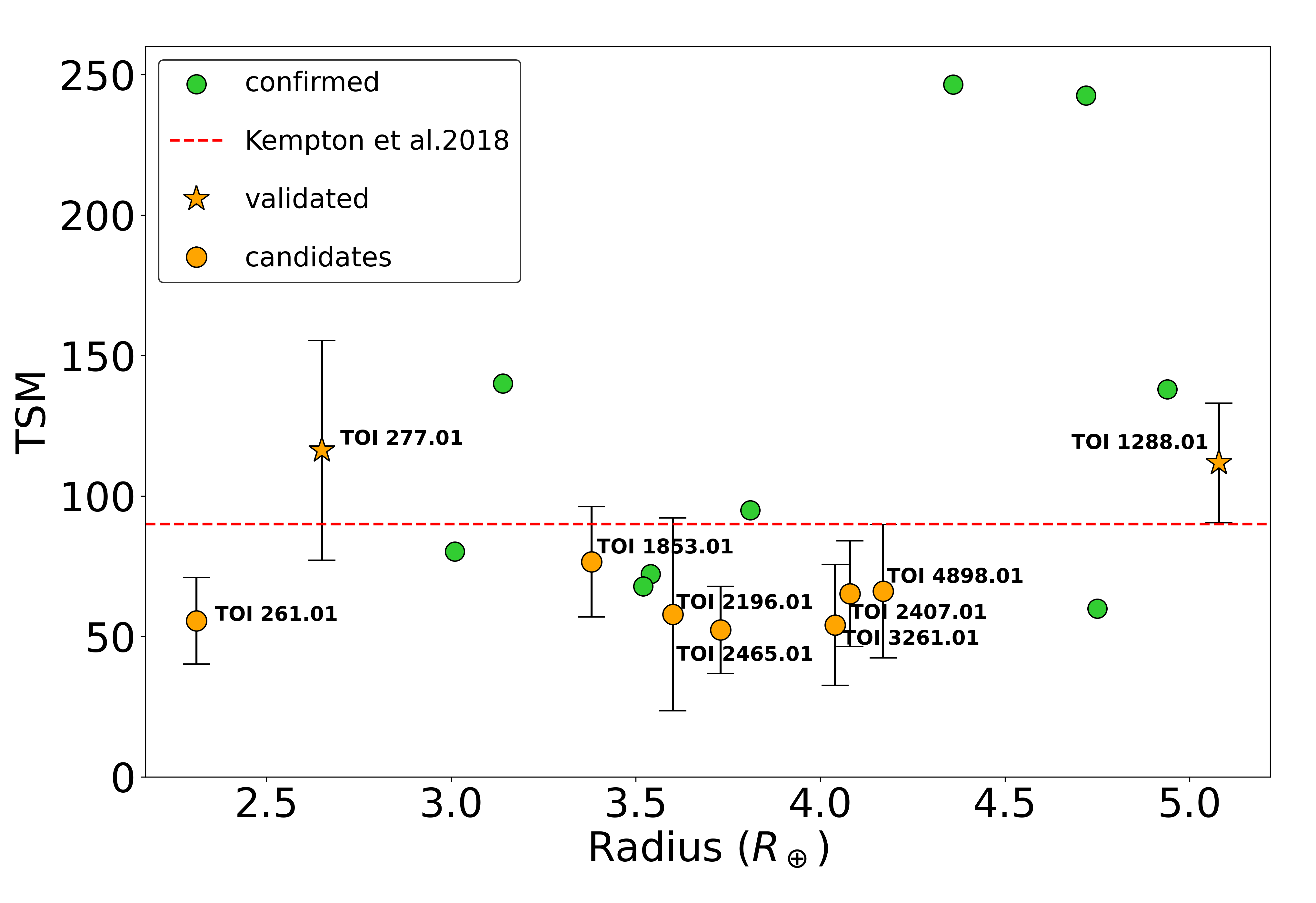}
    \caption{The planet candidates (orange circles) and the validated Hot Neptunes (stars) along with the nine already confirmed and known planets (green circles) in the (R, TSM) plane obtained in this work. The red dashed line represents the threshold value adopted by \protect\cite{2018PASP..130k4401K} above which a $1.5 \, R_\oplus < R < 10 \, R_\oplus$ can be selected as high quality atmospheric characterization targets.}
    \label{fig:tsm plot}
\end{figure}

In Sect. \ref{sec:results} we calculated the TSM index for each target of the final sample using the stellar and planetary parameters obtained in this work. The TSM is a valuable tool to rank exoplanetary targets for spectroscopic atmospheric characterization.
For planets with $1.5 \, R_\oplus < R < 10 \, R_\oplus$, \cite{2018PASP..130k4401K} recommend to select targets with TSM larger than 90. 
In Fig. \ref{fig:tsm plot} we show the distribution of the nine TOIs in the (R, TSM) plane along with the nine confirmed and known exoplanets discussed in Sect. \ref{sec:vetting}. The large error bars originate from the big uncertainties in the planetary masses predicted by \texttt{forecaster}.
Two of our validated exoplanets (TOI-277 b and TOI-1288 b) are well above the mentioned threshold
and would be highest priority targets for atmospheric characterization. However, we remark that 
the TSM parameter is an approximate index for
transmission spectroscopy follow-up measurements,  to be refined with accurate mass measurement of the planets. 
Exoplanets TOI 1853.01 and TOI 2196.01, if confirmed as true planet by follow-up observations, 
would be interesting targets as well. 
TOI-4898 b is just below the limit for effective atmospheric characterization. 
All other exoplanets from our sample lie slightly below the the threshold TSM=90. 
We put the planets discussed in this work into context by comparing their TSM values 
with the same quantity for other Hot Neptunes from literature.

\section{Conclusions}
\label{sec:end}

We presented a homogeneous and complete vetting analysis of all the 250 TOIs with $P<4$ days and $3R_\oplus<R<5R_\oplus$, inside the so called Hot Neptune Desert \citep{Mazeh_2016}. Our vetting technique is based on two main steps. First, we visually inspected each TOI at both the pixel and flux levels using \texttt{DAVE}, and for each of them we gave a disposition: Planetary Candidate (PC), probable False Positive (pFP) and False Positive (FP).
Secondly, we run \texttt{TRICERATOPS}, only on those TOIs ranked as PC in the previous step.

We obtained a sample of 18 planet candidates with FPP $<50\%$ and NFPP $<0.1\%$ which were identified as likely Hot Neptunes. The chosen threshold values for 
the parameters FPP and NFPP 
roughly correspond to a a probability larger than $50\%$ of turning out to be real planets.

The false-positive rate is about $75\%$, much higher with respect to that obtained by  \cite{2022MNRAS.513..102C}. 
This might be the consequence of dealing with intrinsically rare objects, see \cite{2019AJ....158..109H}.

We performed a detailed  analysis of the nine TOIs in the selected sample which were not already confirmed by the TFOPWG or that are not known exoplanets. 
At this stage, we used  the available follow-up high-contrast imaging observations collected by the TFOPWG
and  publicly available on ExoFOP and re-run \texttt{TRICERATOPS}. 
Results are summarized in Table \ref{tab:fin_table_disp}: two TOIs have been statistically validated 
(that is, we obtain FPP $<0.3\%$, NFPP $<0.1\%$) in this further step.
We summarize the properties of the validated exoplanets:
\begin{itemize}
    \item TOI-277 b is a $2.65 \, R_\oplus$ hot sub-Neptune orbiting a high proper motion K star $65$ pc away in $3.99$ days;
    \item TOI-1288 b is a $5.08 \, R_\oplus$ hot Neptune orbiting a late G star $115$ pc away in $2.69$ days;
\end{itemize}

Out of 3888 confirmed transiting exoplanets, just 39 have $P<4$ days and $3R_\oplus<R< 5 R_\oplus$, 
while $71$ populated the extended region up to $R<10R_\oplus$.
Our sample of planet candidates, if confirmed by further analyses in the future, will enlarge the mass/radius vs period distribution within the Neptune Desert by adding new hot Neptunes. 
It will be useful for demographic studies which will shed light on the origins of the observed dearth of close-in Neptune/sub-Jovian planets by looking for statistical correlations among stellar and planetary parameters. 
An important contribution in solving the riddle of the origin of Neptunian Desert  might come from planetary population synthesis tools \citep{Mordasini2018} which take into account all the physical processes occurring during the protoplantery disk stage up when the planetary system is stable and formed.

There is great prospect for JWST spectroscopy, as well Ariel ESA’s M4 mission, to test the hypothesis that the hot Neptunes are giant planets observed in the evolutionary stage in which they are
loosing their gas envelope. 
Moreoever, the planetary atmospheric characterization performed by JWST and Ariel, along with the mass estimates that will be obtained with the high-accuracy spectrographs, will be of paramount importance to understand whether these objects form {\it in situ} or they have migrated from further out in  the system.

Both JWST and Ariel spectrometers cover wide spectral intervals, spanning from the visual to infrared wavelengths (0.6 – 28.3 $\mu$m and 0.5 – 7.8 $\mu$ m, for JWST and Ariel respectively), allowing to investigate spectral features related to both the main volatile atmospheric gases (e.g. H2O, CO2, CH4, NH3, HCN, H2S) and to more exotic refractory elements (e.g. metal oxydes like TiO, CaO, VO, SiO). While the presence of refractory elements would be compatible with an in situ formation scenario (see Sect. 1), in the migration hypothesis the molecular hydrogen and helium envelope could still be abundant and detectable through the observation of collision induced absorption spectral signatures and enhanced Rayleigh scattering in the visual/near infrared.

\section*{Acknowledgements}

This research has made use of the NASA Exoplanet Archive, which
is operated by the California Institute of Technology, under contract
with the National Aeronautics and Space Administration under the
Exoplanet Exploration Program.

We acknowledge the use of the Exoplanet Follow-up Observation Program website, which is operated by the California Institute of Technology, under contract with the National Aeronautics and Space Administration under the Exoplanet Exploration Program. The TFOP is led by the Smithsonian Astrophysical Observatory (SAO), in coordination with MIT, as part of the TESS Science Office.

This work has made use of data from the European Space Agency (ESA) mission
{\it Gaia} (\url{https://www.cosmos.esa.int/gaia}), processed by the {\it Gaia}
Data Processing and Analysis Consortium (DPAC,
\url{https://www.cosmos.esa.int/web/gaia/dpac/consortium}). Funding for the DPAC
has been provided by national institutions, in particular the institutions
participating in the {\it Gaia} Multilateral Agreement.
JSJ greatfully acknowledges support by FONDECYT grant 1201371 and from the ANID BASAL projects ACE210002 and FB210003.
JIV acknowledges support of CONICYT-PFCHA/Doctorado Nacional-21191829.

\textit{Software}: \texttt{DAVE} \citep{kostov2019discovery}, \texttt{TRICERATOPS} \citep{2021AJ....161...24G}, \texttt{forecaster} \citep{2017ApJ...834...17C}.

\section*{Data Availability}
The data underlying this article will be shared on reasonable request
to the corresponding author

% The inclusion of a Data Availability Statement is a requirement for articles published in MNRAS. Data Availability Statements provide a standardised format for readers to understand the availability of data underlying the research results described in the article. The statement may refer to original data generated in the course of the study or to third-party data analysed in the article. The statement should describe and provide means of access, where possible, by linking to the data or providing the required accession numbers for the relevant databases or DOIs.



\bibliographystyle{mnras}
\bibliography{biblio}

%\begin{thebibliography}{99}
%\end{thebibliography}

%%%%%%%%%%%%%%%%%%%%%%%%%%%%%%%%%%%%%%%%%%%%%%%%%%

%%%%%%%%%%%%%%%%% APPENDICES %%%%%%%%%%%%%%%%%%%%%

% \appendix
\appendix
\section{DAVE results} 
\label{appendix: dave_results}
In this Appendix we show the main \textit{DAVE} outcomes for the nine TOIs analyzed in Sect.\ref{sec:results}. For each figure below, the top panel corresponds to the \textit{Modelshift} analysis while the bottom panel corresponds to the \textit{centroids} analysis.
%%%%%%%%%%%%%%%%%%%%%%%%%%%%%%%%%%%%%%%%%%%%%%%%%%
\begin{figure}
     \centering
     \begin{subfigure}[b]{0.45\textwidth}
         \centering
         \includegraphics[width=\textwidth]{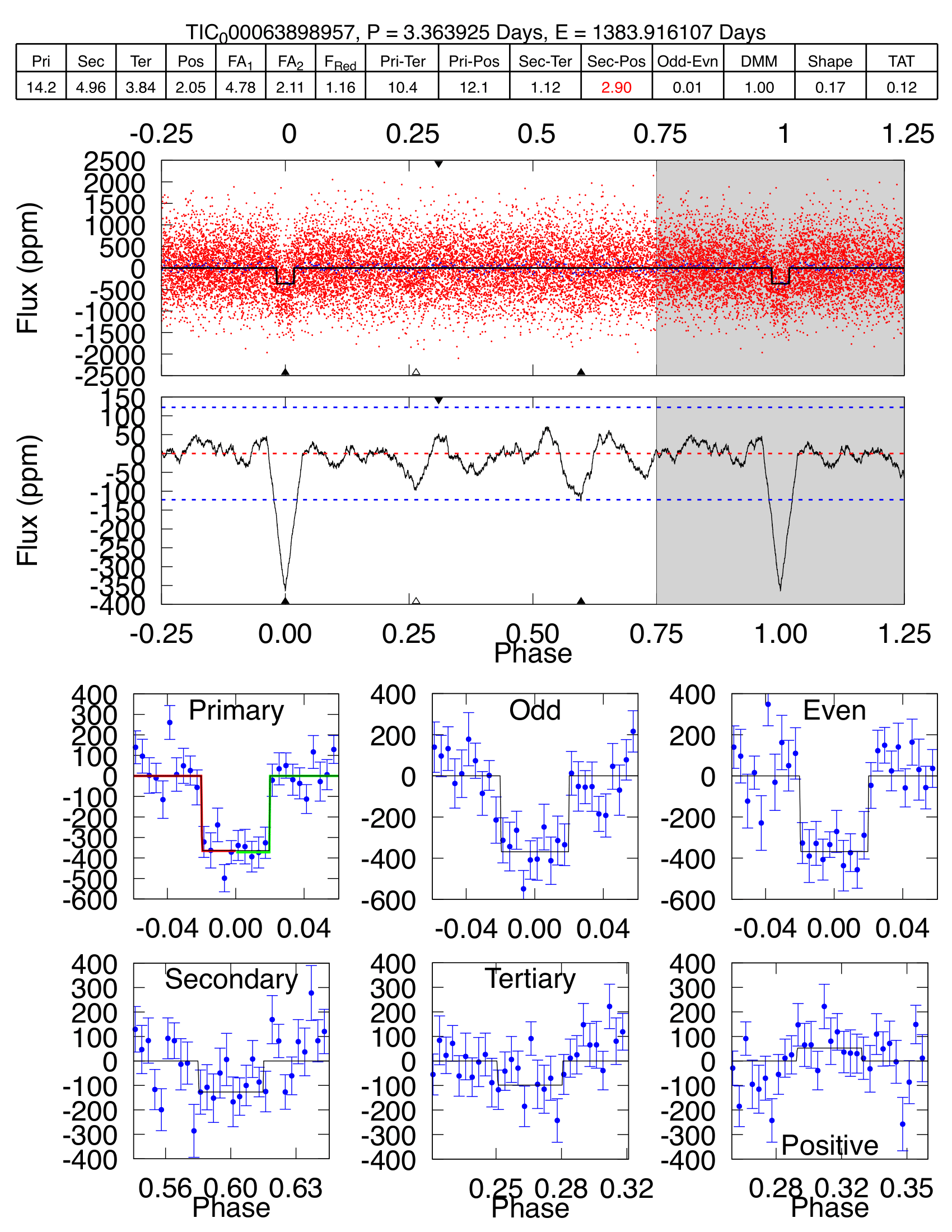}
         \caption{\textit{Modelshift} for TIC 63898957.}
         \label{fig:Modshift 63898957}
     \end{subfigure}
     \hfill
    \begin{subfigure}[b]{0.45\textwidth}
         \centering
         \includegraphics[width=\textwidth]{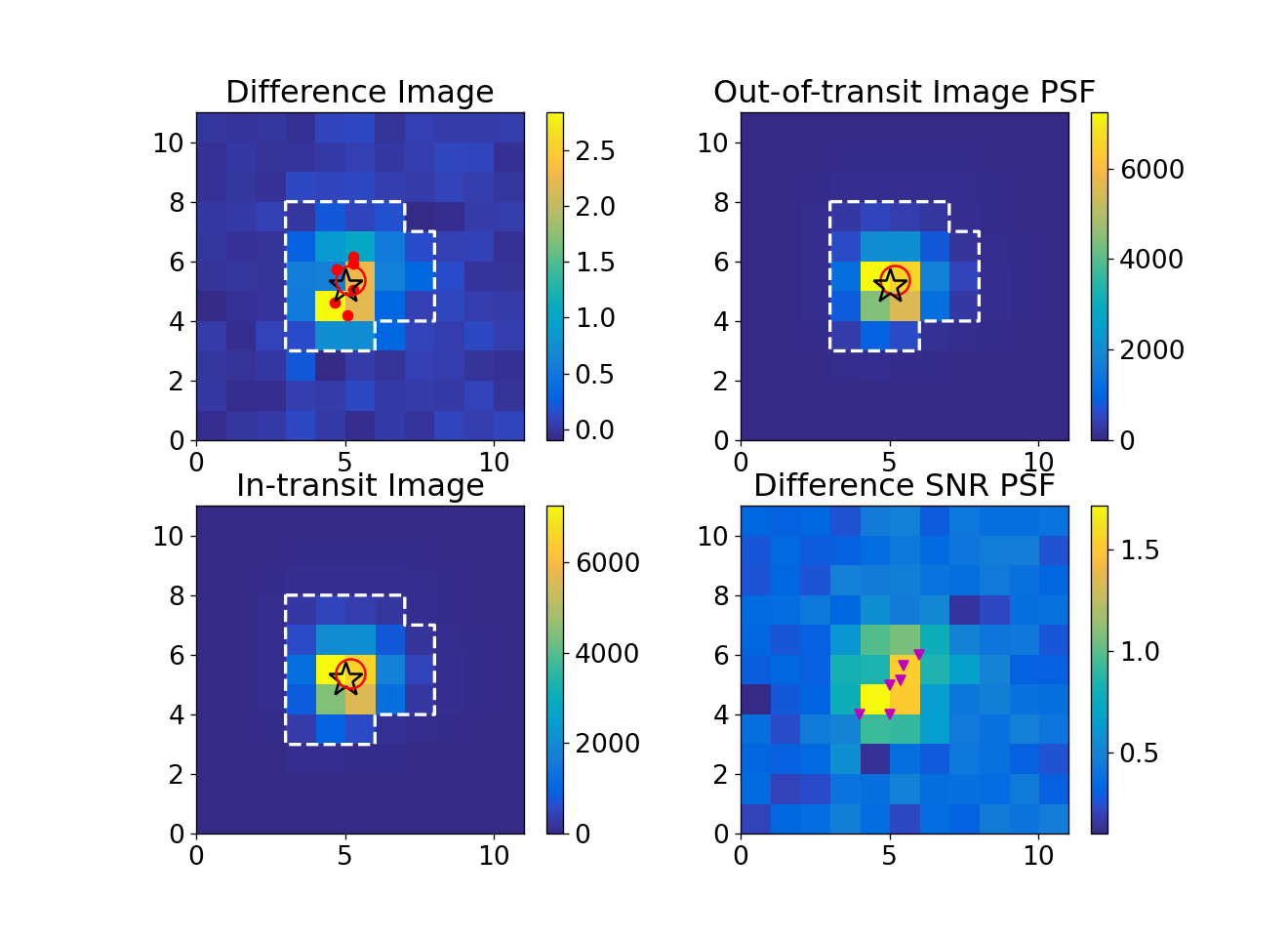}
         \caption{\textit{centroids} analysis for TIC 63898957.}
         \label{fig:centroids 63898957}
     \end{subfigure}
        \caption{\texttt{DAVE} outcomes for TIC 63898957 based on TESS sector 3 data.}
        \label{fig:DAVE 63898957}
\end{figure}
%%%%%%%%%%%%%%%%%%%%%%%%%%%%%%%%%%%%%%%%%%%%%%%%%%
\begin{figure}
     \centering
     \begin{subfigure}[b]{0.4\textwidth}
         \centering
         \includegraphics[width=\textwidth]{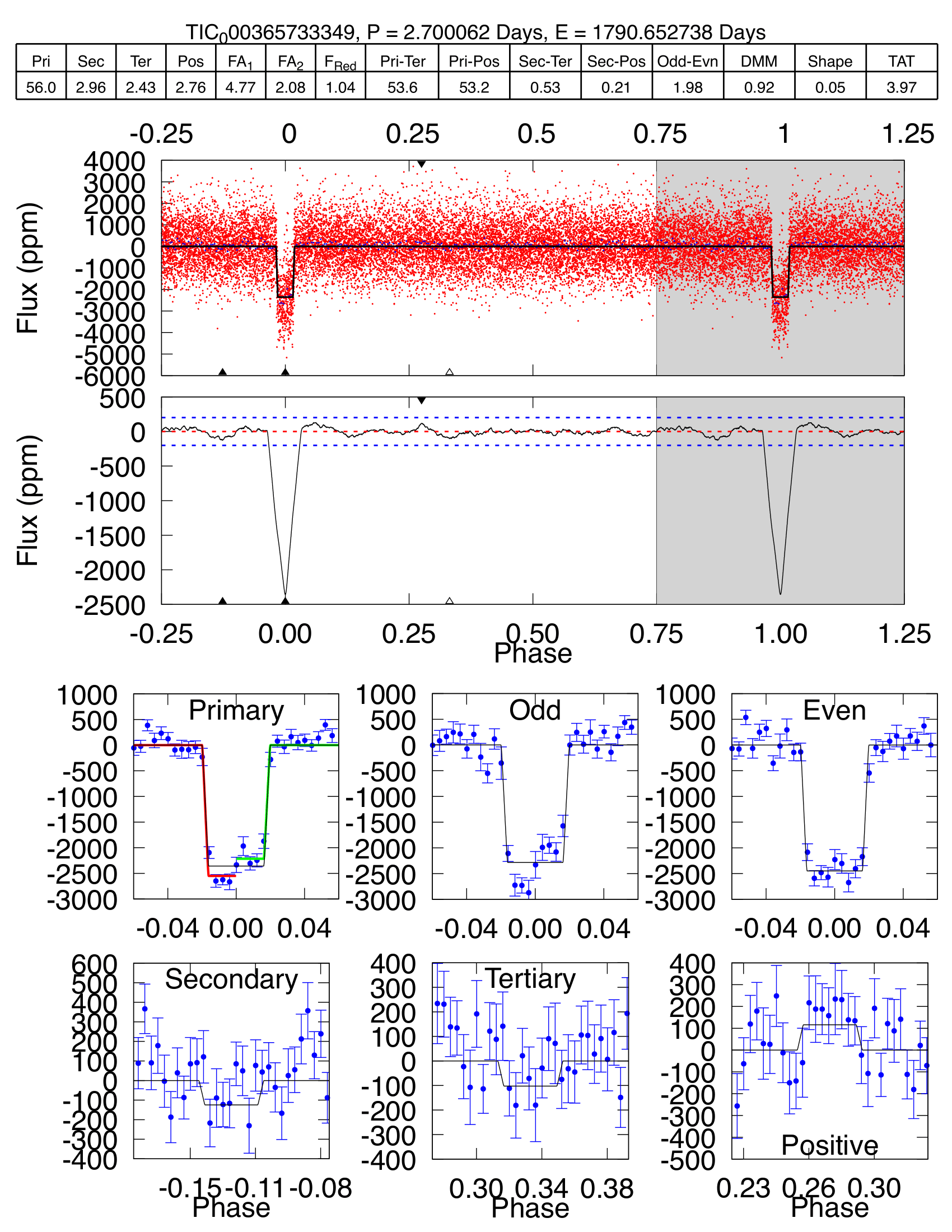}
         \caption{\textit{Modelshift} for TIC 365733349.}
         \label{fig:Modshift 365733349}
     \end{subfigure}
     \hfill
    \begin{subfigure}[b]{0.4\textwidth}
         \centering
         \includegraphics[width=\textwidth]{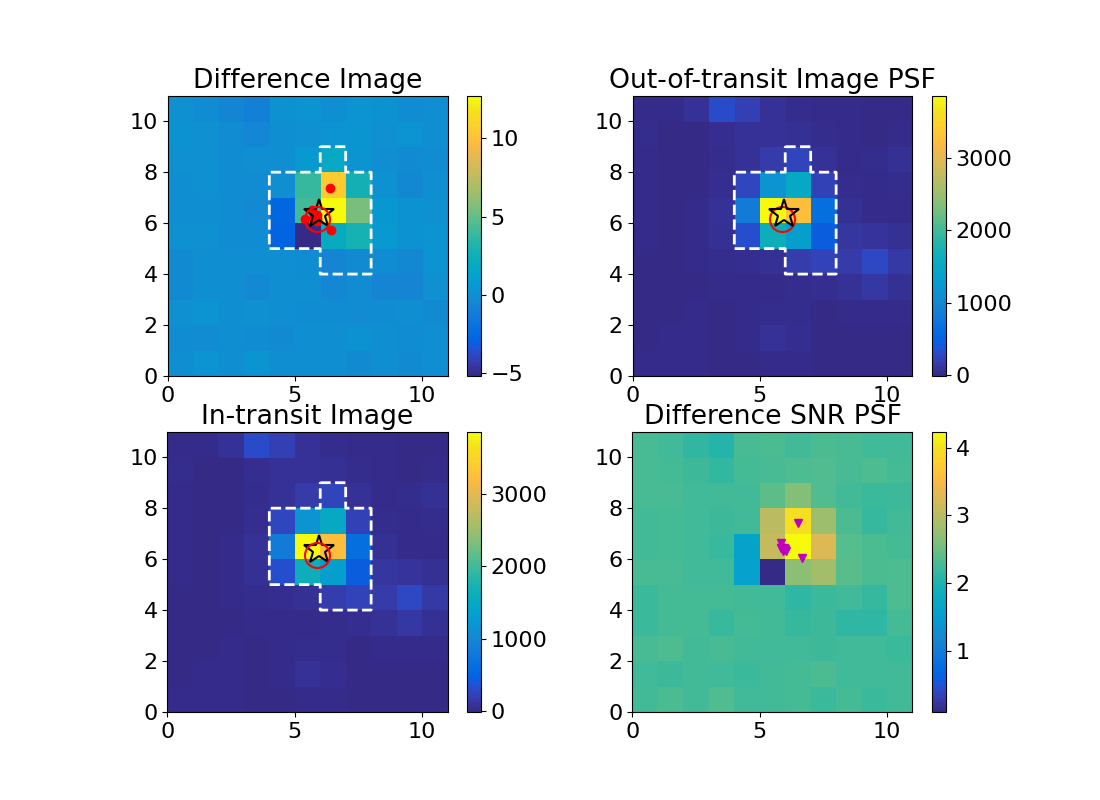}
         \caption{\textit{centroids} analysis for TIC 365733349.}
         \label{fig:centroids 365733349}
     \end{subfigure}
        \caption{\texttt{DAVE} outcomes for TIC 365733349 based on TESS sector 18 data.}
        \label{fig:DAVE 365733349}
\end{figure}
%%%%%%%%%%%%%%%%%%%%%%%%%%%%%%%%%%%%%%%%%%%%%%%%%%
\begin{figure}
     \centering
     \begin{subfigure}[b]{0.4\textwidth}
         \centering
         \includegraphics[width=\textwidth]{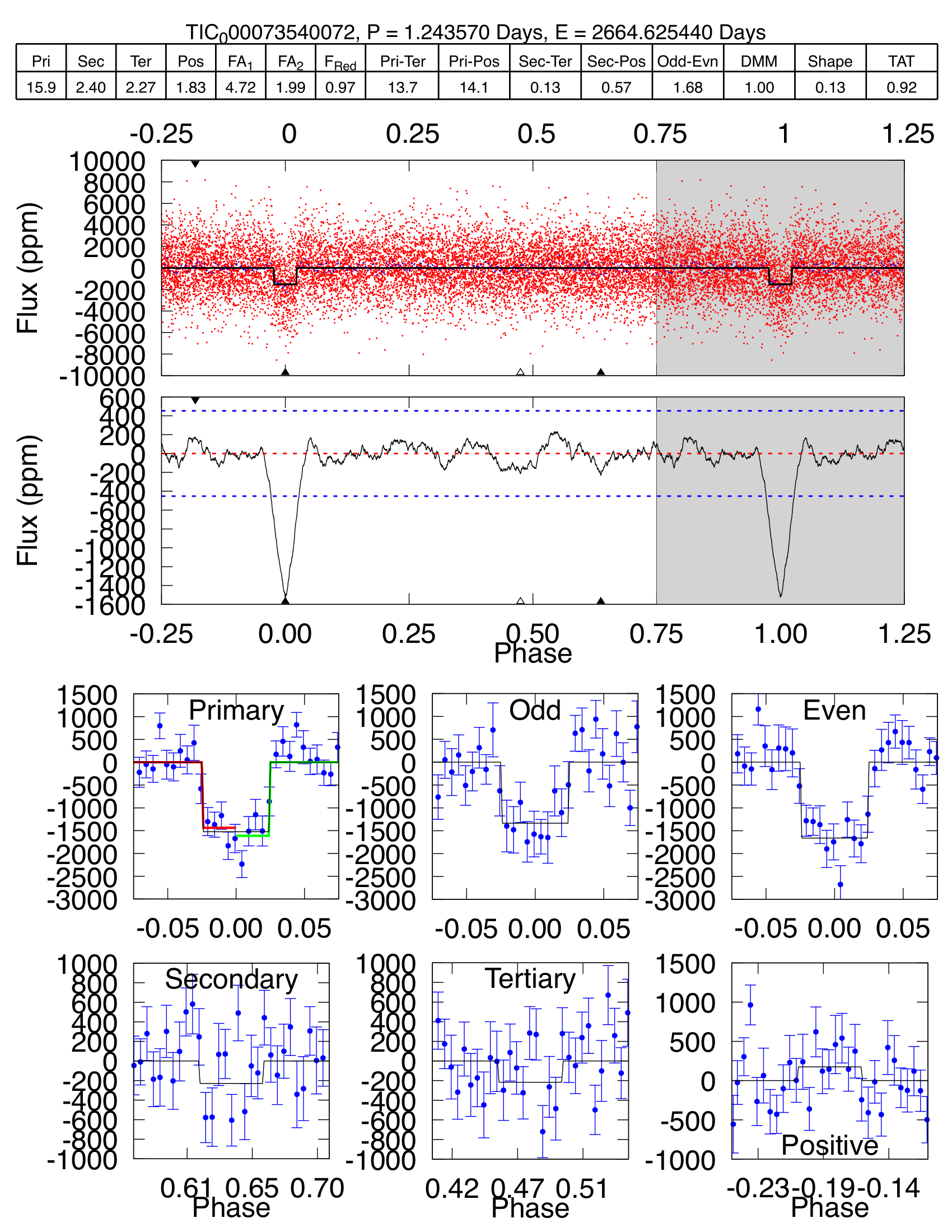}
         \caption{\textit{Modelshift} for TIC 73540072.}
         \label{fig:Modshift 73540072}
     \end{subfigure}
     \hfill
    \begin{subfigure}[b]{0.4\textwidth}
         \centering
         \includegraphics[width=\textwidth]{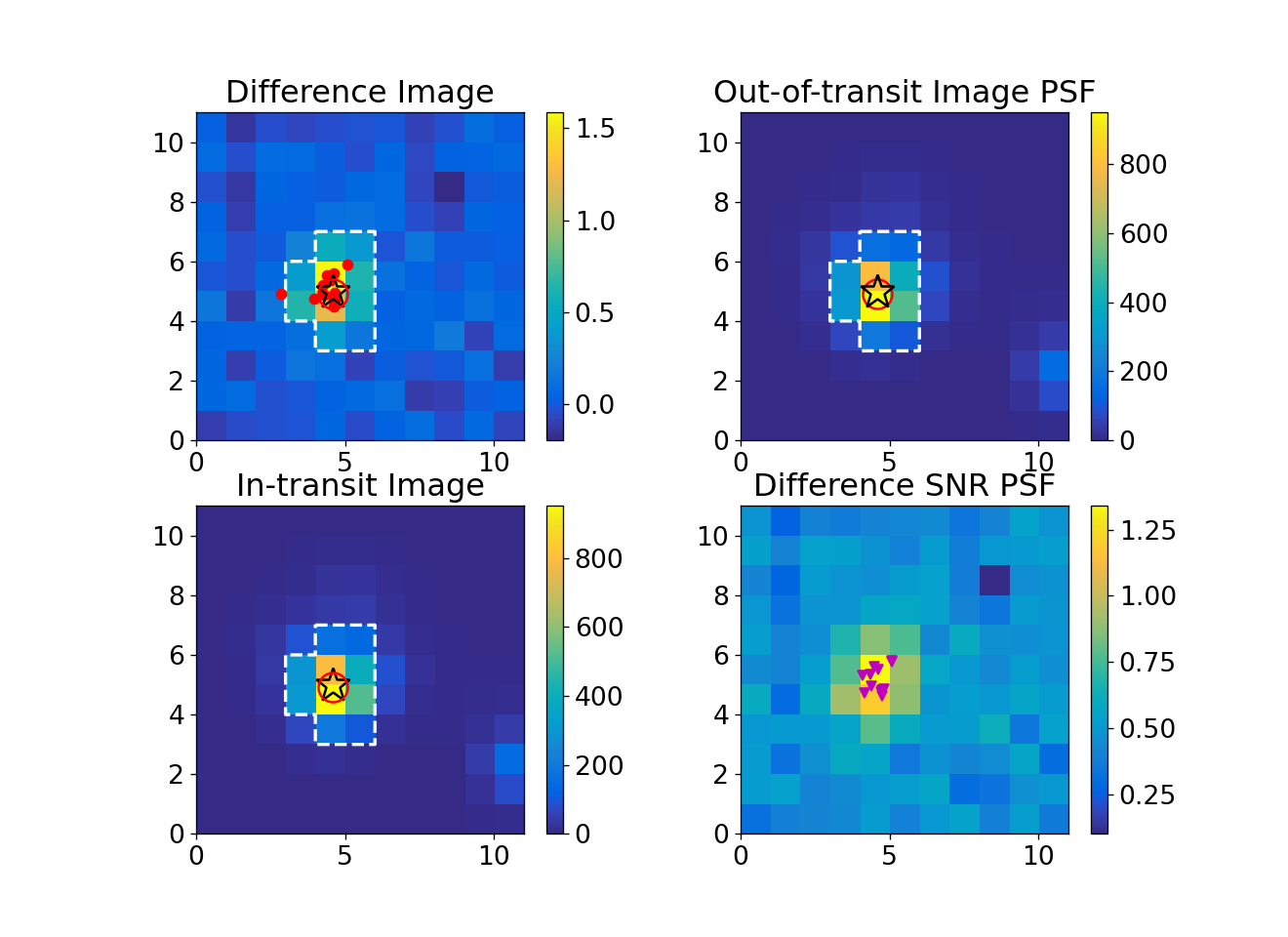}
         \caption{\textit{centroids} analysis for TIC 73540072.}
         \label{fig:centroids 73540072}
     \end{subfigure}
        \caption{\texttt{DAVE} outcomes for TIC 73540072 based on TESS sector 50 data.}
        \label{fig:DAVE 73540072}
\end{figure}
%%%%%%%%%%%%%%%%%%%%%%%%%%%%%%%%%%%%%%%%%%%%%%%%%%
\begin{figure}
     \centering
     \begin{subfigure}[b]{0.4\textwidth}
         \centering
         \includegraphics[width=\textwidth]{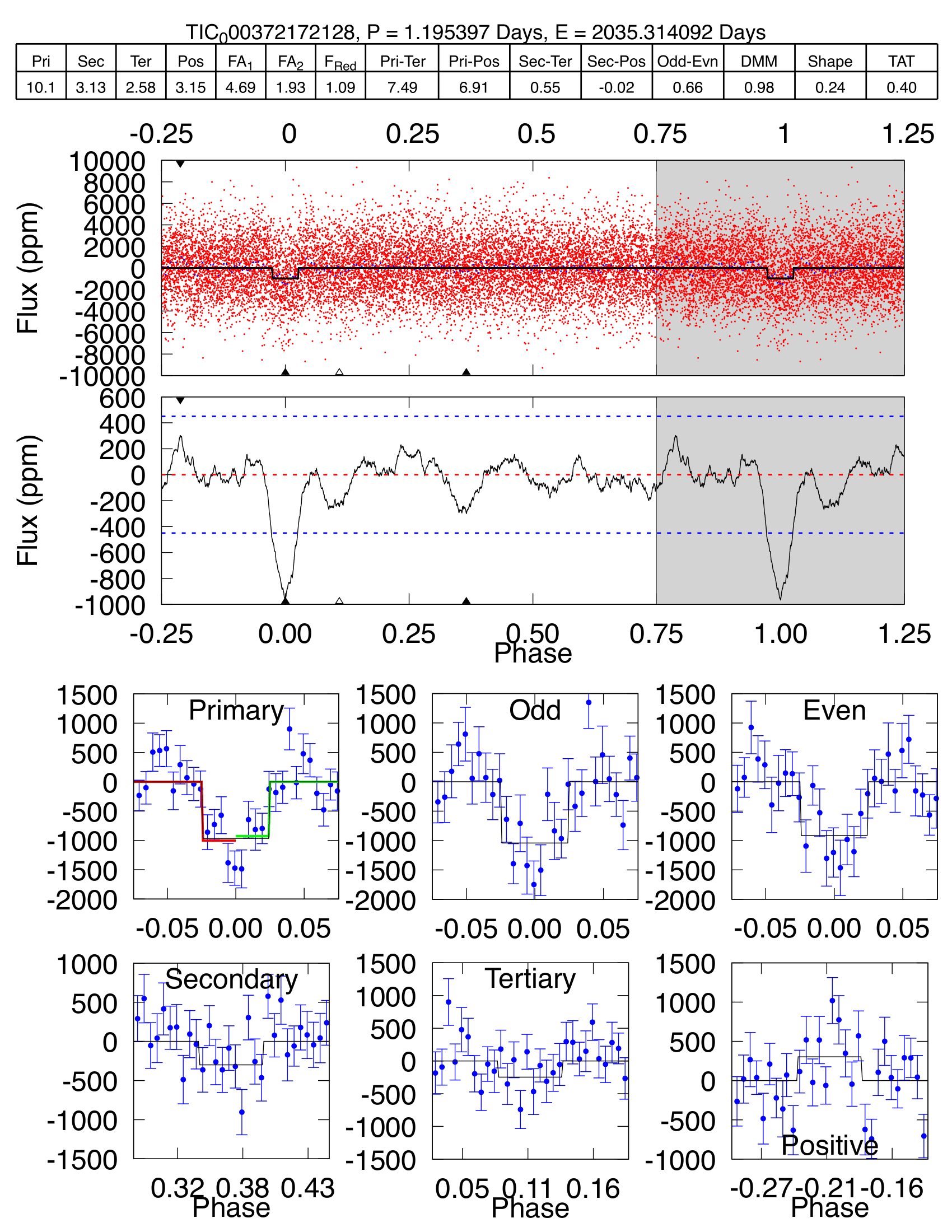}
         \caption{\textit{Modelshift} for TIC 372172128.}
         \label{fig:Modshift 372172128}
     \end{subfigure}
     \hfill
    \begin{subfigure}[b]{0.4\textwidth}
         \centering
         \includegraphics[width=\textwidth]{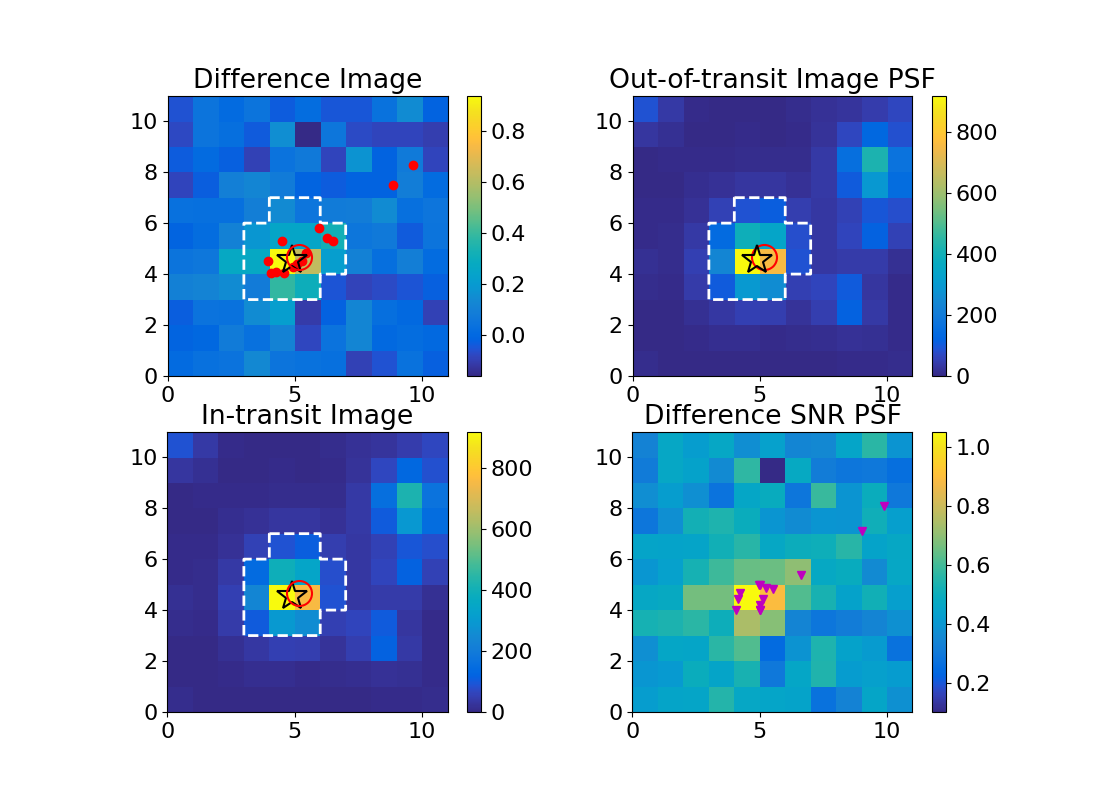}
         \caption{\textit{centroids} analysis for TIC 372172128.}
         \label{fig:centroids 372172128}
     \end{subfigure}
        \caption{\texttt{DAVE} outcomes for TIC 372172128 based on TESS sector 27 data.}
        \label{fig:DAVE 372172128}
\end{figure}
%%%%%%%%%%%%%%%%%%%%%%%%%%%%%%%%%%%%%%%%%%%%%%%%%%
\begin{figure}
     \centering
     \begin{subfigure}[b]{0.4\textwidth}
         \centering
         \includegraphics[width=\textwidth]{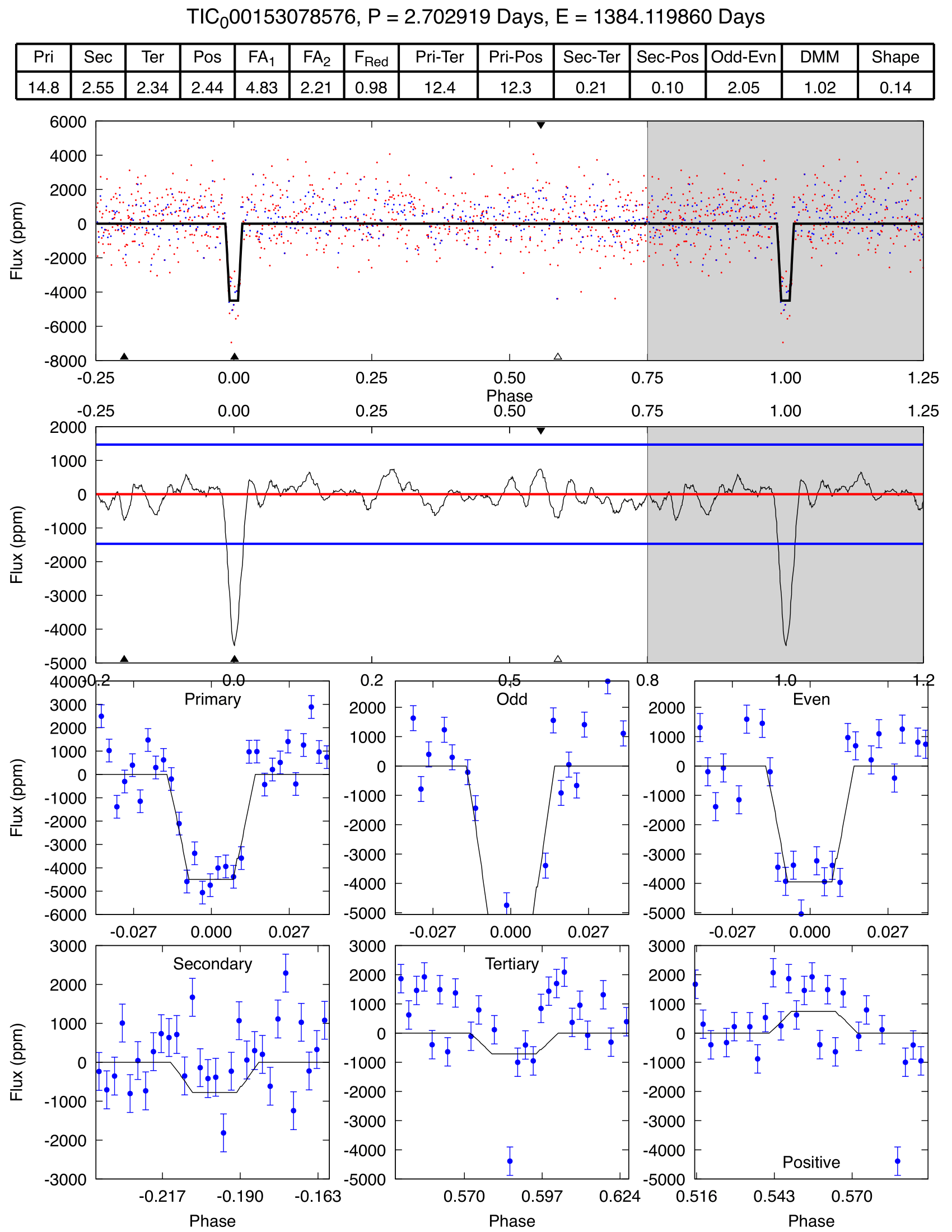}
         \caption{\textit{Modelshift} for TIC 153078576.}
         \label{fig:Modshift 153078576}
     \end{subfigure}
     \hfill
    \begin{subfigure}[b]{0.4\textwidth}
         \centering
         \includegraphics[width=\textwidth]{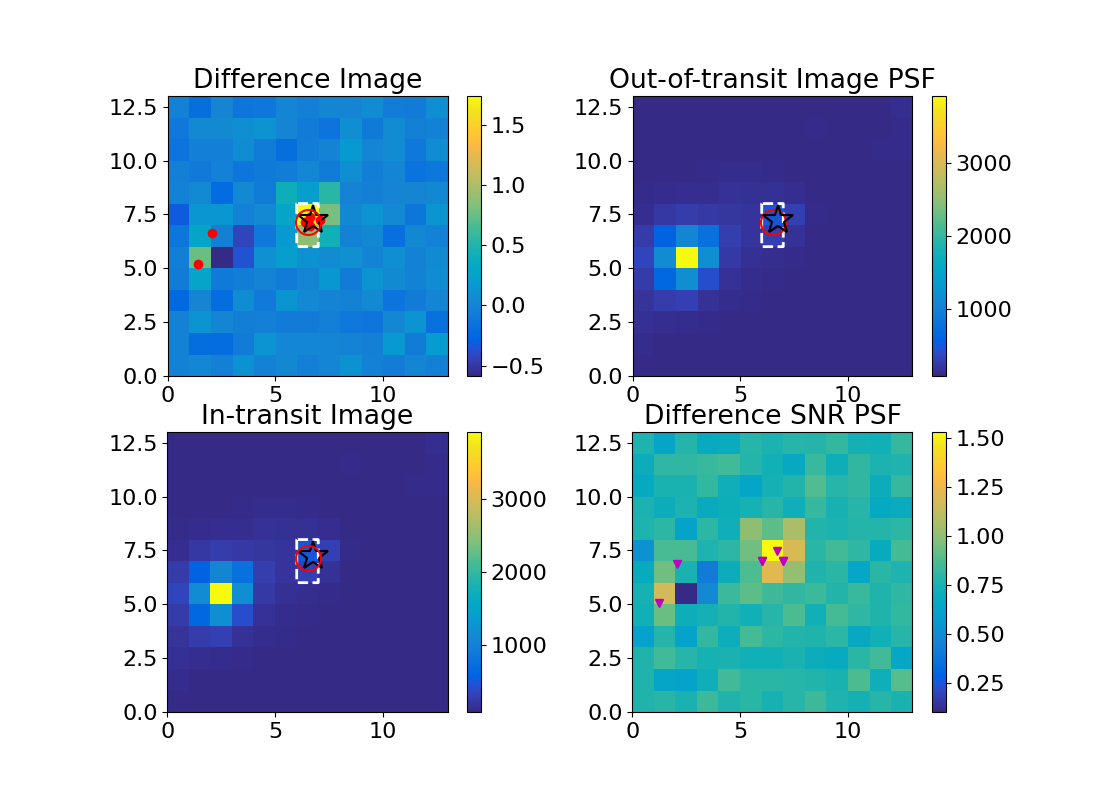}
         \caption{\textit{centroids} analysis for TIC 153078576.}
         \label{fig:centroids 153078576}
     \end{subfigure}
        \caption{\texttt{DAVE} outcomes for TIC 153078576 based on TESS sector 3 data.}
        \label{fig:DAVE 153078576}
\end{figure}
%%%%%%%%%%%%%%%%%%%%%%%%%%%%%%%%%%%%%%%%%%%%%%%%%%
\begin{figure}
     \centering
     \begin{subfigure}[b]{0.4\textwidth}
         \centering
         \includegraphics[width=\textwidth]{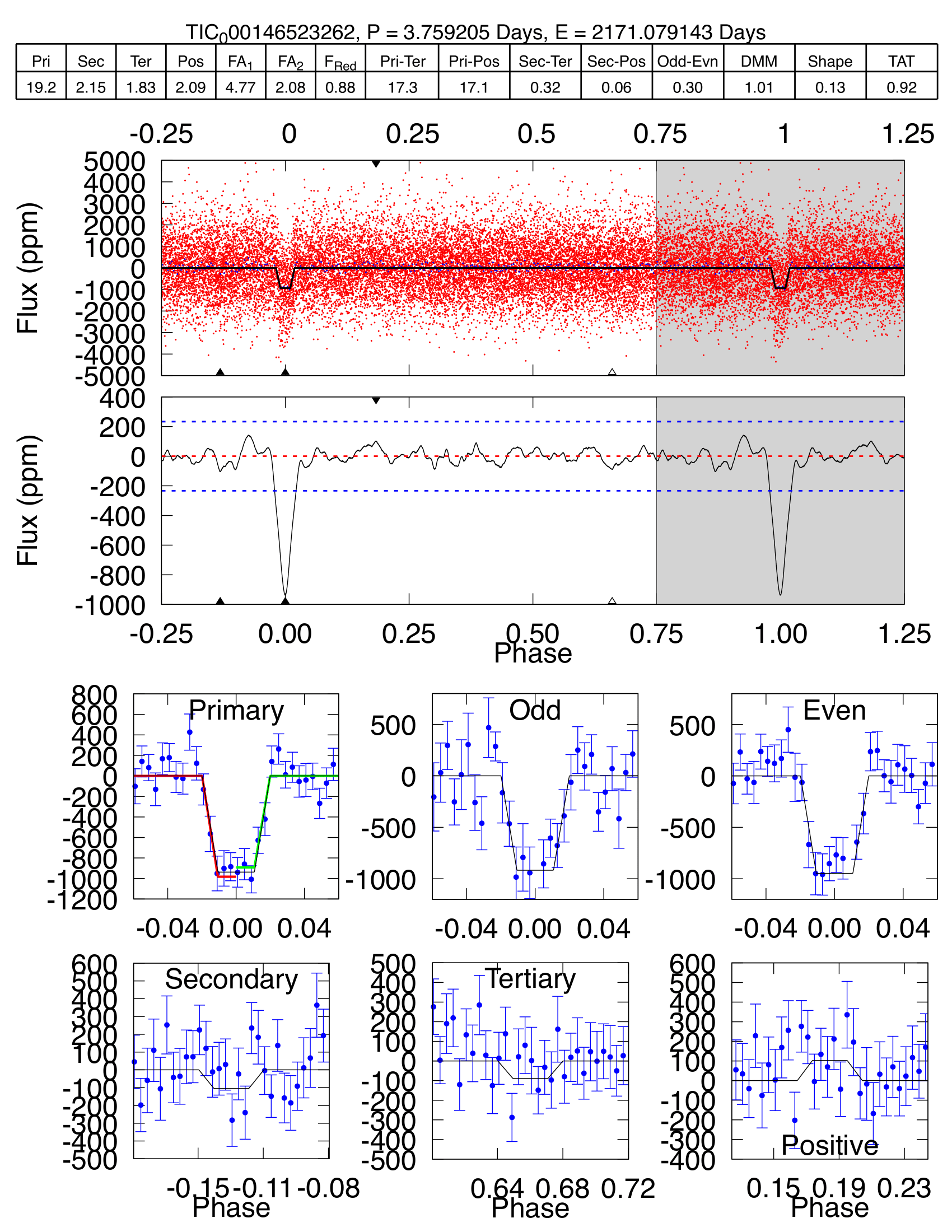}
         \caption{\textit{Modelshift} for TIC 146523262.}
         \label{fig:Modshift 146523262}
     \end{subfigure}
     \hfill
    \begin{subfigure}[b]{0.4\textwidth}
         \centering
         \includegraphics[width=\textwidth]{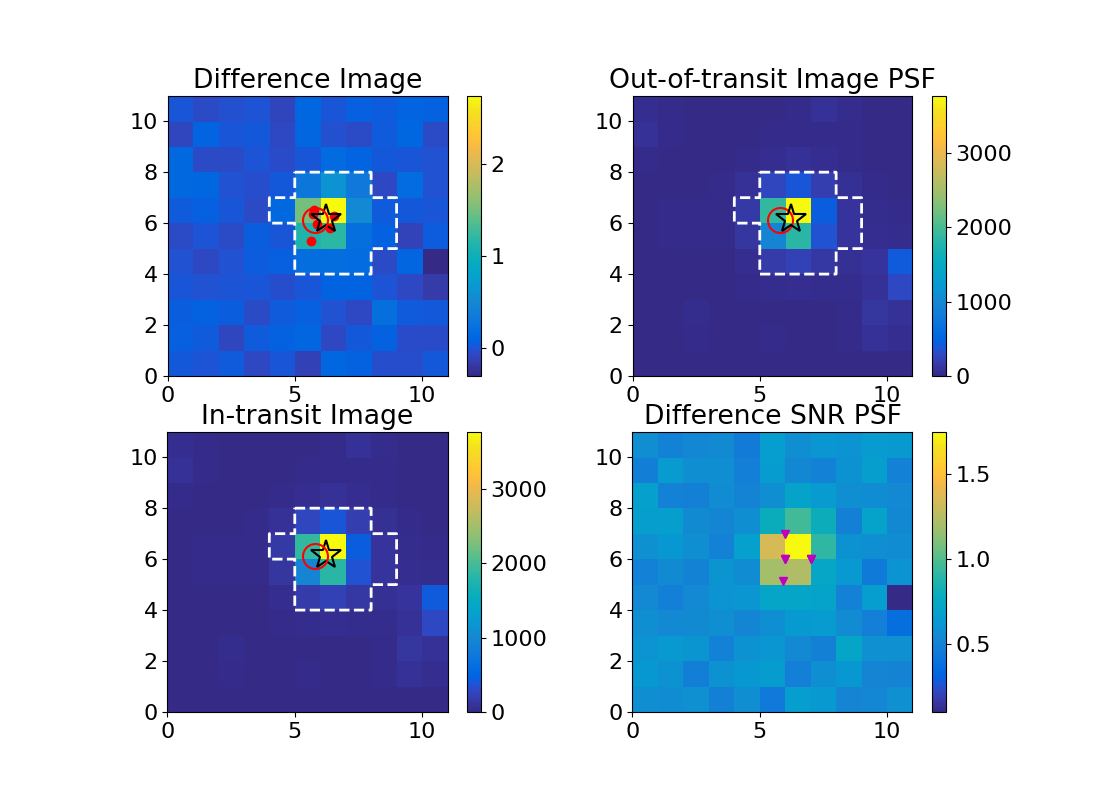}
         \caption{\textit{centroids} analysis for TIC 146523262.}
         \label{fig:centroids 146523262}
     \end{subfigure}
        \caption{\texttt{DAVE} outcomes for TIC 146523262 based on TESS sector 32 data.}
        \label{fig:DAVE 146523262}
\end{figure}
%%%%%%%%%%%%%%%%%%%%%%%%%%%%%%%%%%%%%%%%%%%%%%%%%%
\begin{figure}
     \centering
     \begin{subfigure}[b]{0.4\textwidth}
         \centering
         \includegraphics[width=\textwidth]{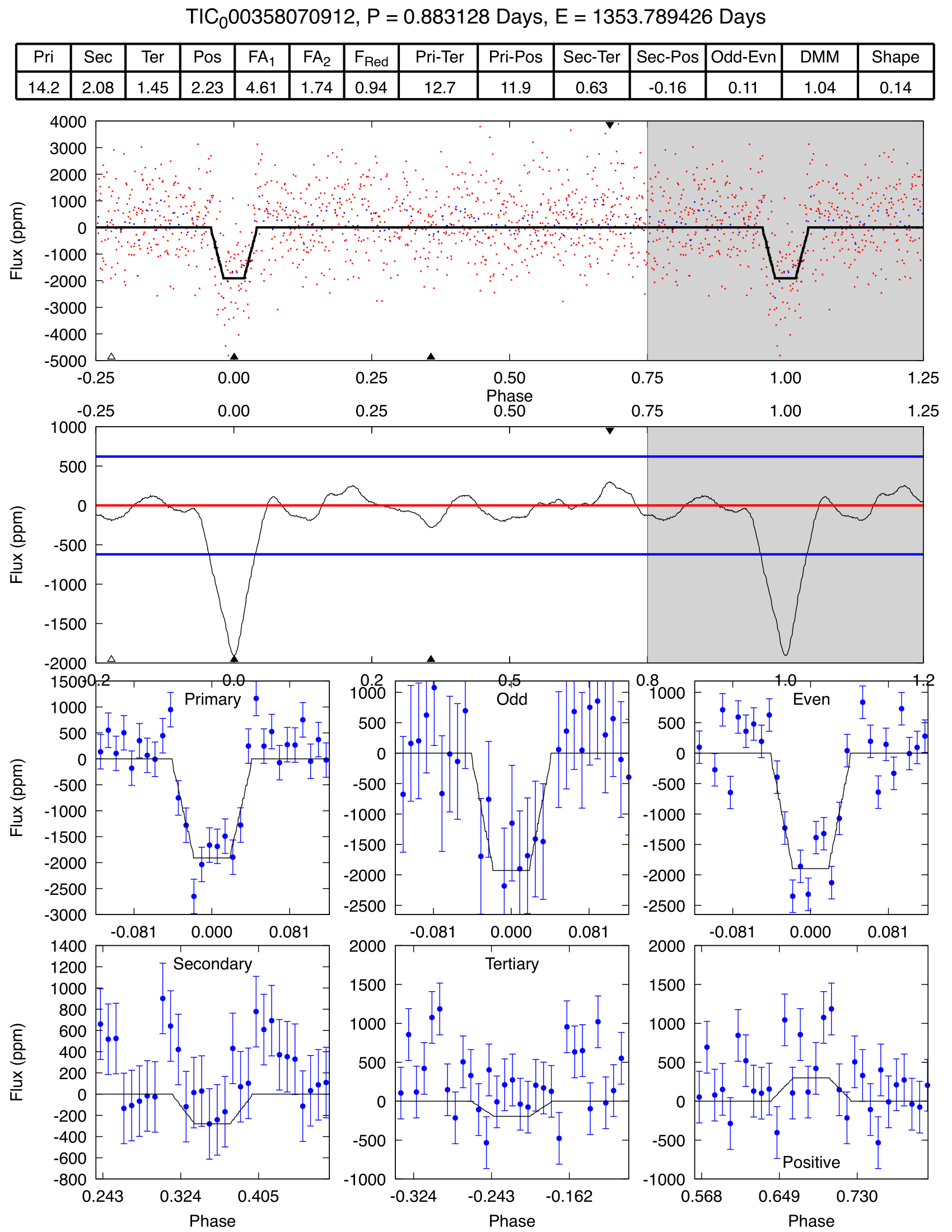}
         \caption{\textit{Modelshift} for TIC 358070912.}
         \label{fig:Modshift 358070912}
     \end{subfigure}
     \hfill
    \begin{subfigure}[b]{0.4\textwidth}
         \centering
         \includegraphics[width=\textwidth]{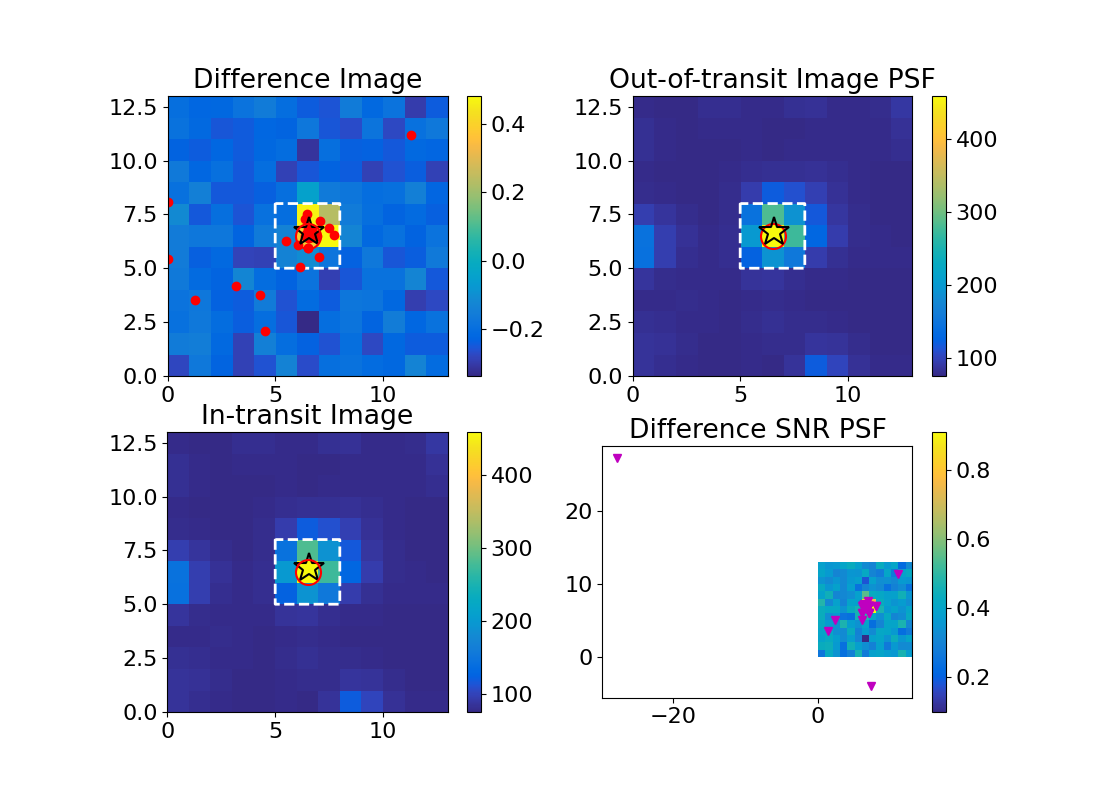}
         \caption{\textit{centroids} analysis for TIC 358070912.}
         \label{fig:centroids 358070912}
     \end{subfigure}
        \caption{\texttt{DAVE} outcomes for TIC 358070912 based on TESS sector 2 data.}
        \label{fig:DAVE 358070912}
\end{figure}
%%%%%%%%%%%%%%%%%%%%%%%%%%%%%%%%%%%%%%%%%%%%%%%%%%
\begin{figure}
     \centering
     \begin{subfigure}[b]{0.4\textwidth}
         \centering
         \includegraphics[width=\textwidth]{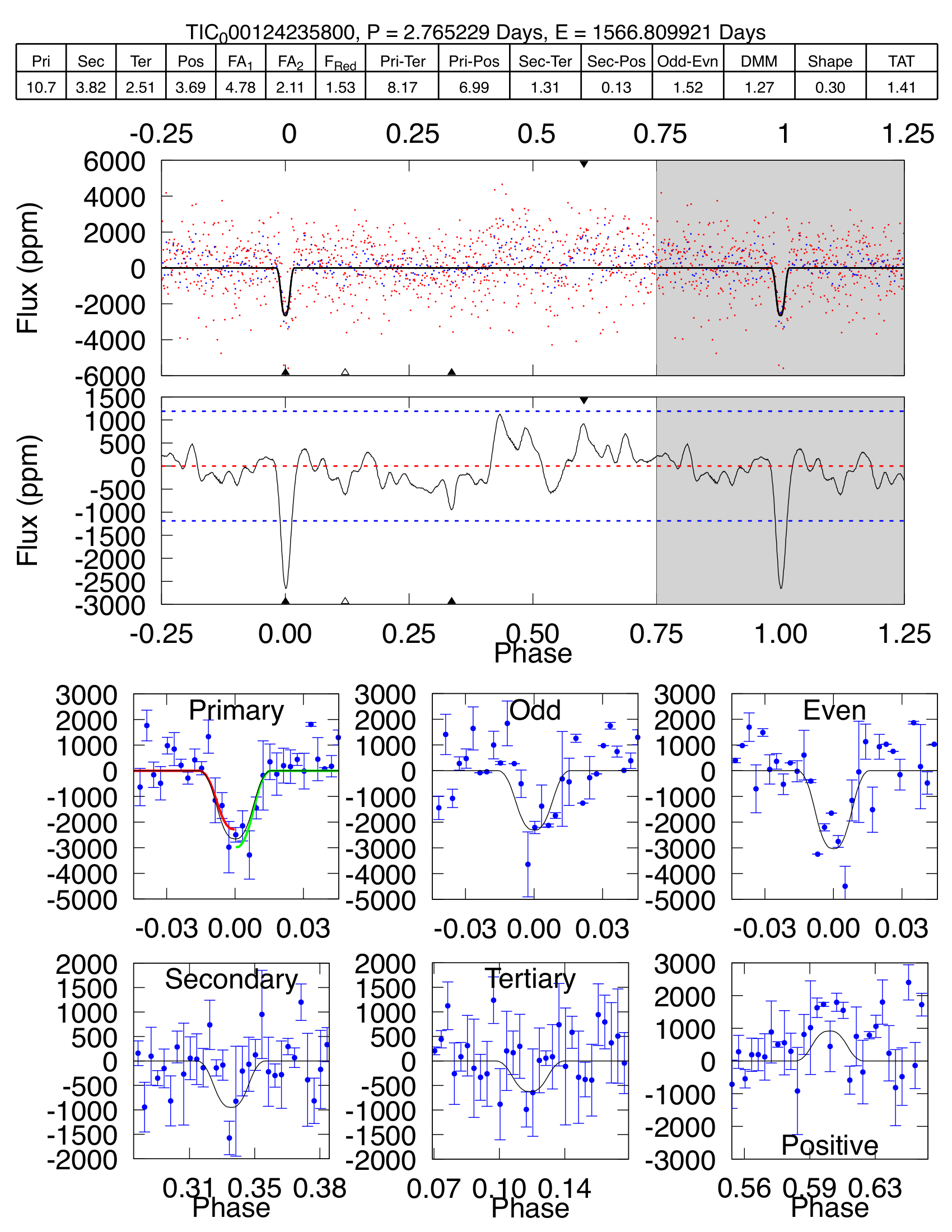}
         \caption{\textit{Modelshift} for TIC 124235800.}
         \label{fig:Modshift 124235800}
     \end{subfigure}
     \hfill
    \begin{subfigure}[b]{0.4\textwidth}
         \centering
         \includegraphics[width=\textwidth]{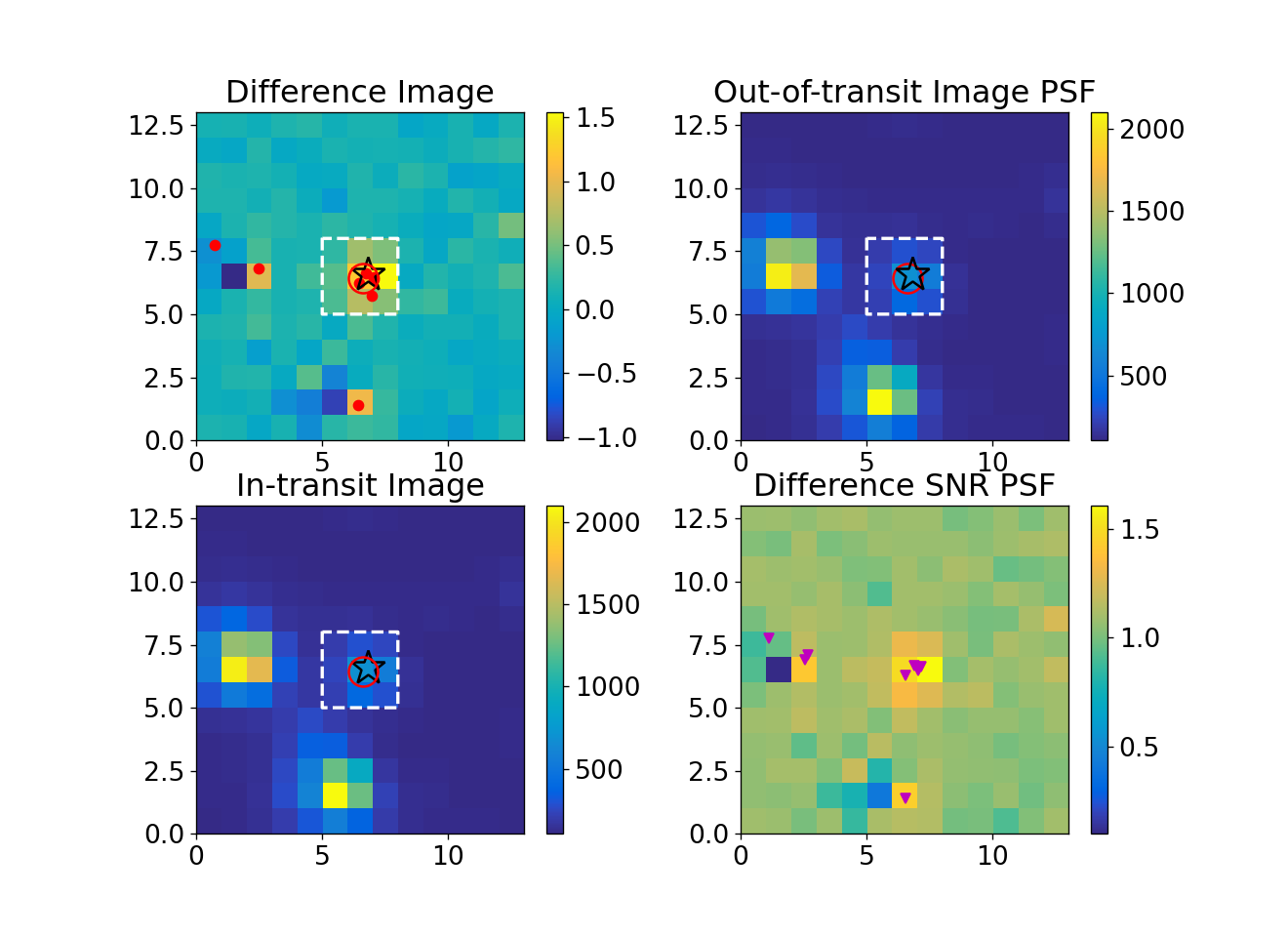}
         \caption{\textit{centroids} analysis for TIC 124235800.}
         \label{fig:centroids 124235800}
     \end{subfigure}
        \caption{\texttt{DAVE} outcomes for TIC 124235800 based on TESS sector 10 data.}
        \label{fig:DAVE 124235800}
\end{figure}
%%%%%%%%%%%%%%%%%%%%%%%%%%%%%%%%%%%%%%%%%%%%%%%%%%%%%%%
\begin{figure}
     \centering
     \begin{subfigure}[b]{0.4\textwidth}
         \centering
         \includegraphics[width=\textwidth]{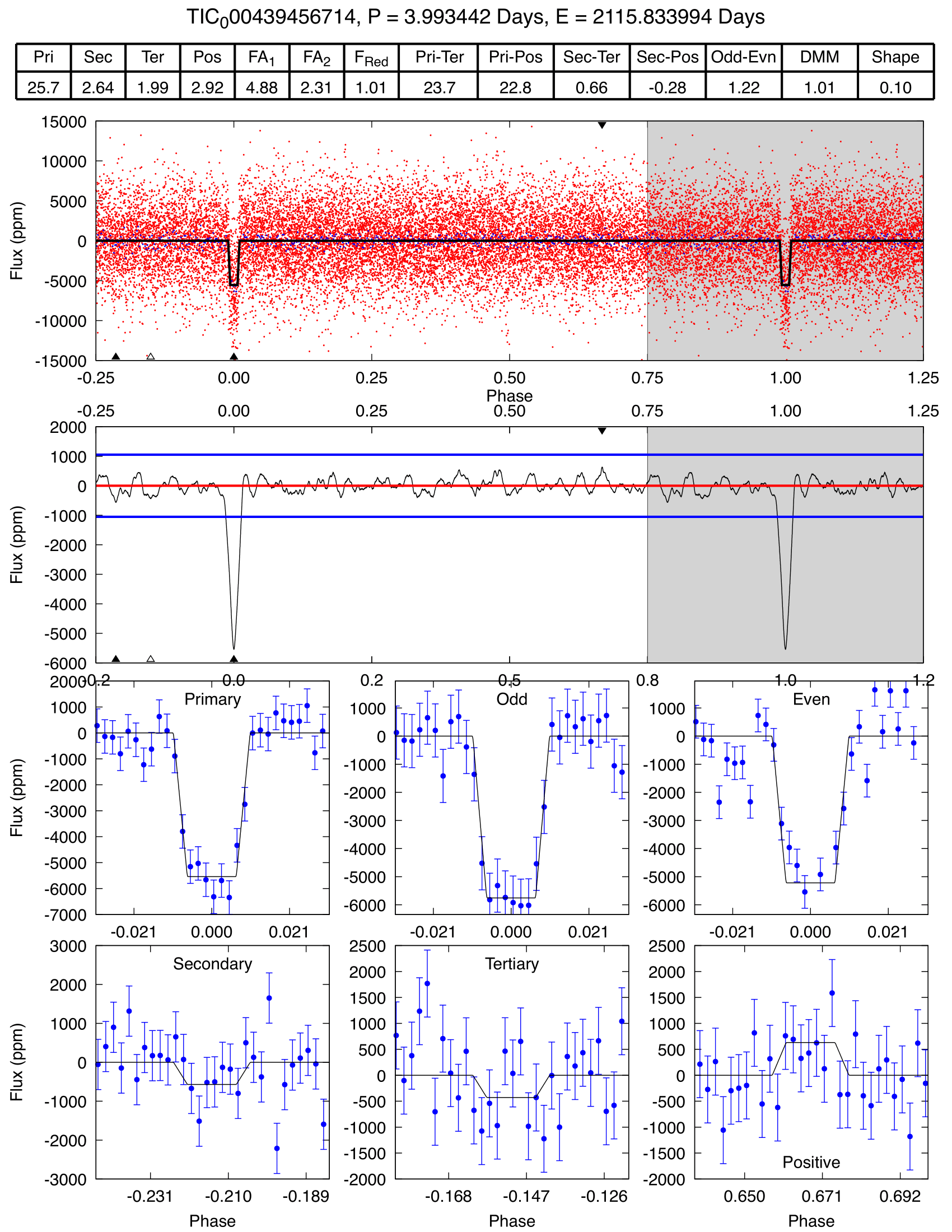}
         \caption{\textit{Modelshift} for TIC 439456714.}
         \label{fig:Modshift 439456714}
     \end{subfigure}
     \hfill
    \begin{subfigure}[b]{0.4\textwidth}
         \centering
         \includegraphics[width=\textwidth]{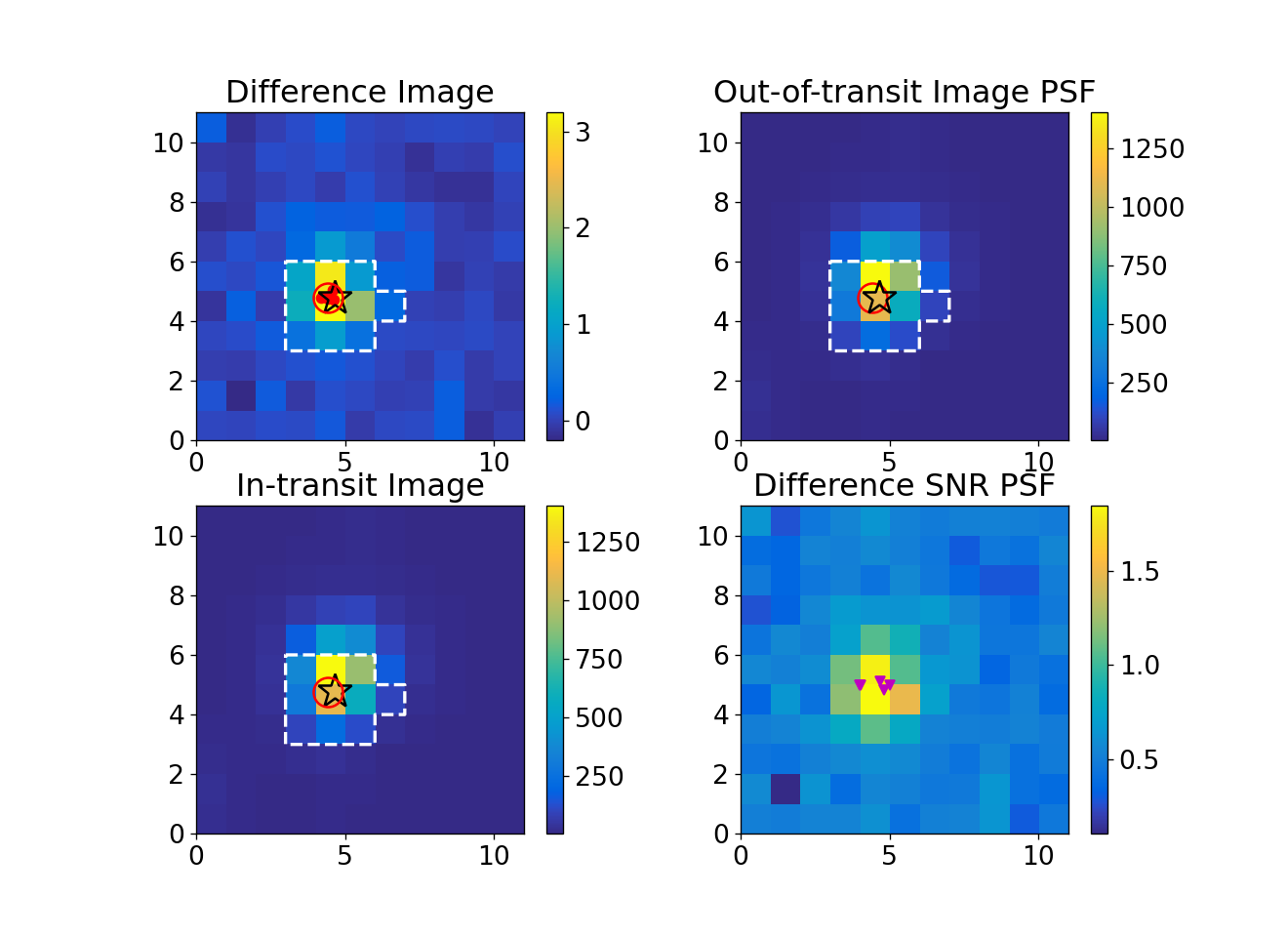}
         \caption{\textit{centroids} analysis for TIC 439456714.}
         \label{fig:centroids 439456714}
     \end{subfigure}
        \caption{\texttt{DAVE} outcomes for TIC 124235800 based on TESS sector 30 data.}
        \label{fig:DAVE 439456714}
\end{figure}
%%%%%%%%%%%%%%%%%%%%%%%%%%%%%%%%%%%%%%%%%%%%%%%%%%%%%%%
\section{TRES extracted spectra}
\label{appendix:tres_spectra}
In this Appendix we show the TRES extracted spectra of TIC 63898957, TIC 365733349, TIC 73540072 and TIC 146523262. The upper panel shows the observed spectrum (blue line) along with the best-fit model (red line). The lower panel shows two CCF peak fits, one using an observed rotating template and a second with a non-rotating template.
\begin{figure*}
     \centering
     \begin{subfigure}[b]{0.75\textwidth}
         \centering
         \includegraphics[width=\textwidth]{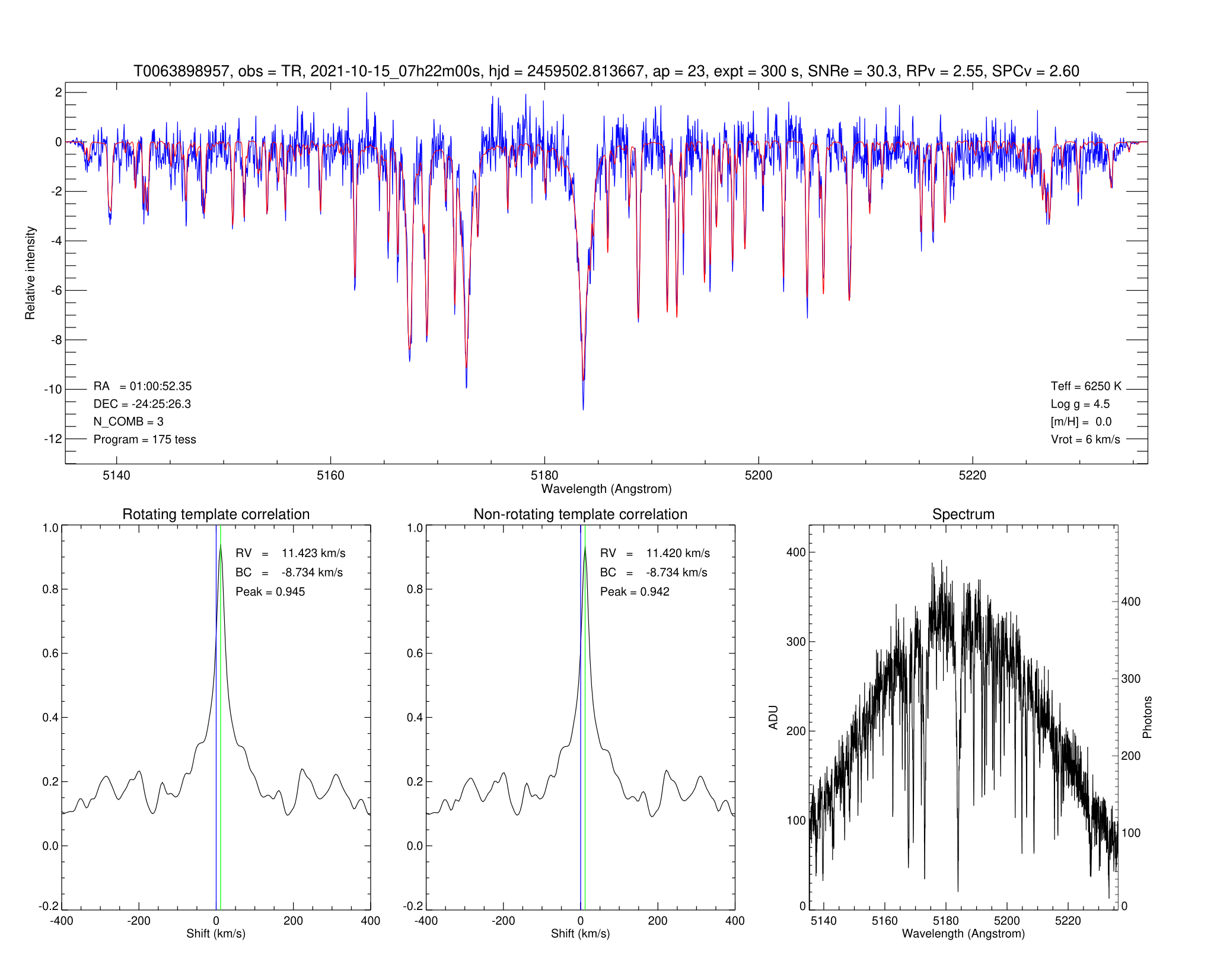}
         %\label{fig:}
     \end{subfigure}
     \\
    \begin{subfigure}[b]{0.75\textwidth}
         \centering
         \includegraphics[width=\textwidth]{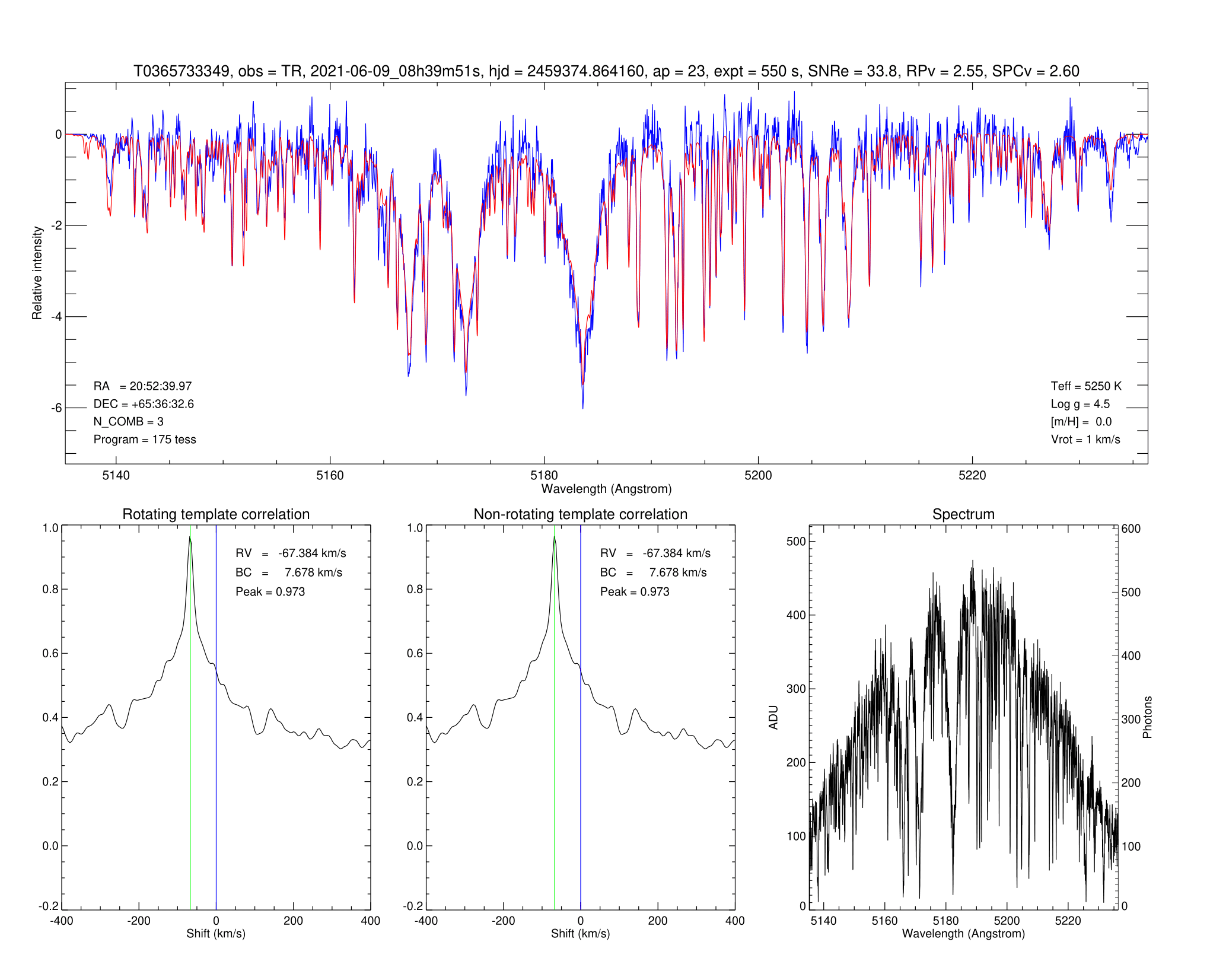}
         %\label{fig:}
     \end{subfigure}
     \label{fig:tres_spectra_1}
     \caption{TRES extracted spectra of TIC 63898957 (upper) and TIC 365733349 (lower).}
\end{figure*}
\begin{figure*}
        \begin{subfigure}[b]{0.75\textwidth}
         \centering
         \includegraphics[width=\textwidth]{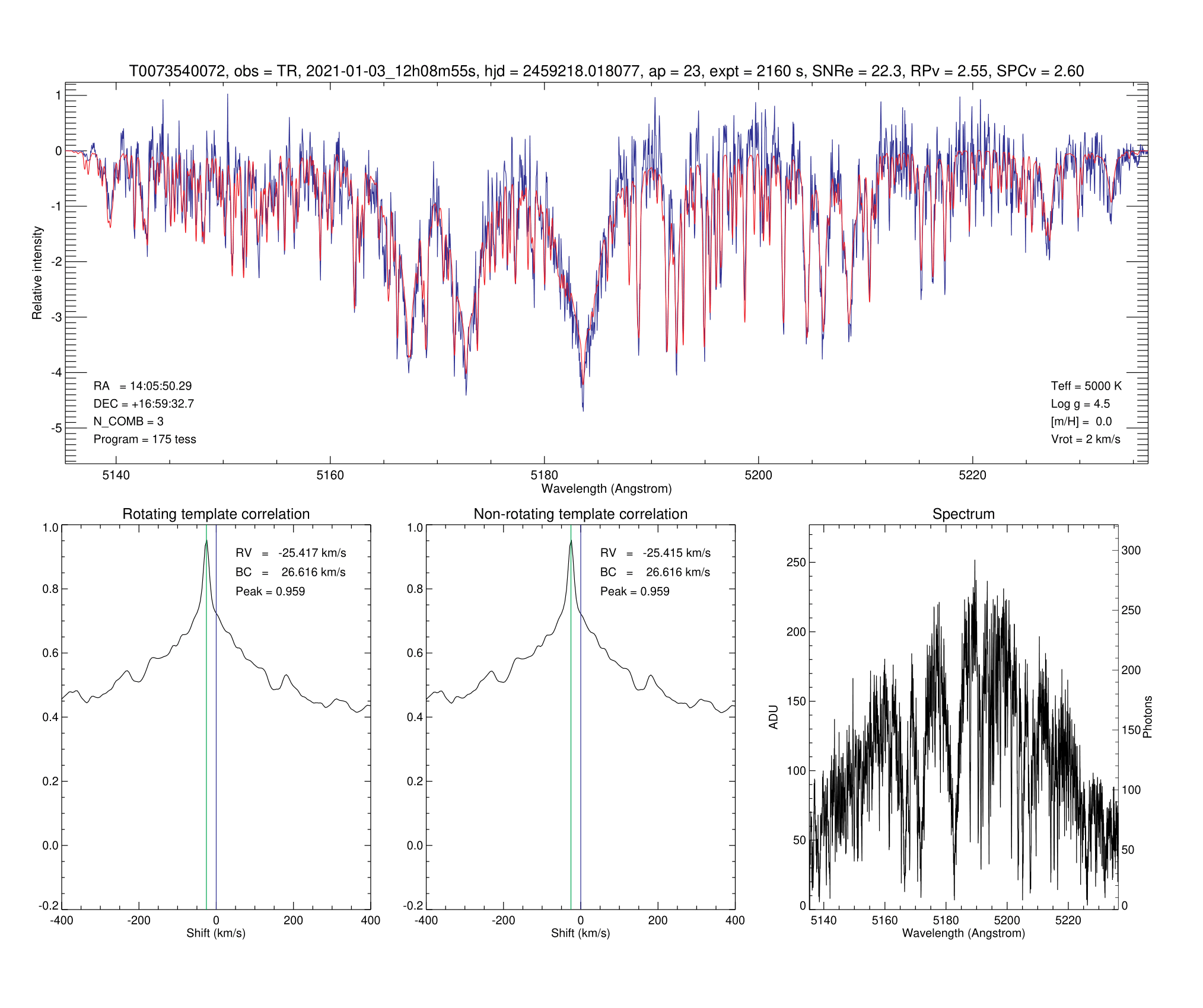}
         
         %\label{fig:}
     \end{subfigure}
     \\
    \begin{subfigure}[b]{0.75\textwidth}
         \centering
         \includegraphics[width=\textwidth]{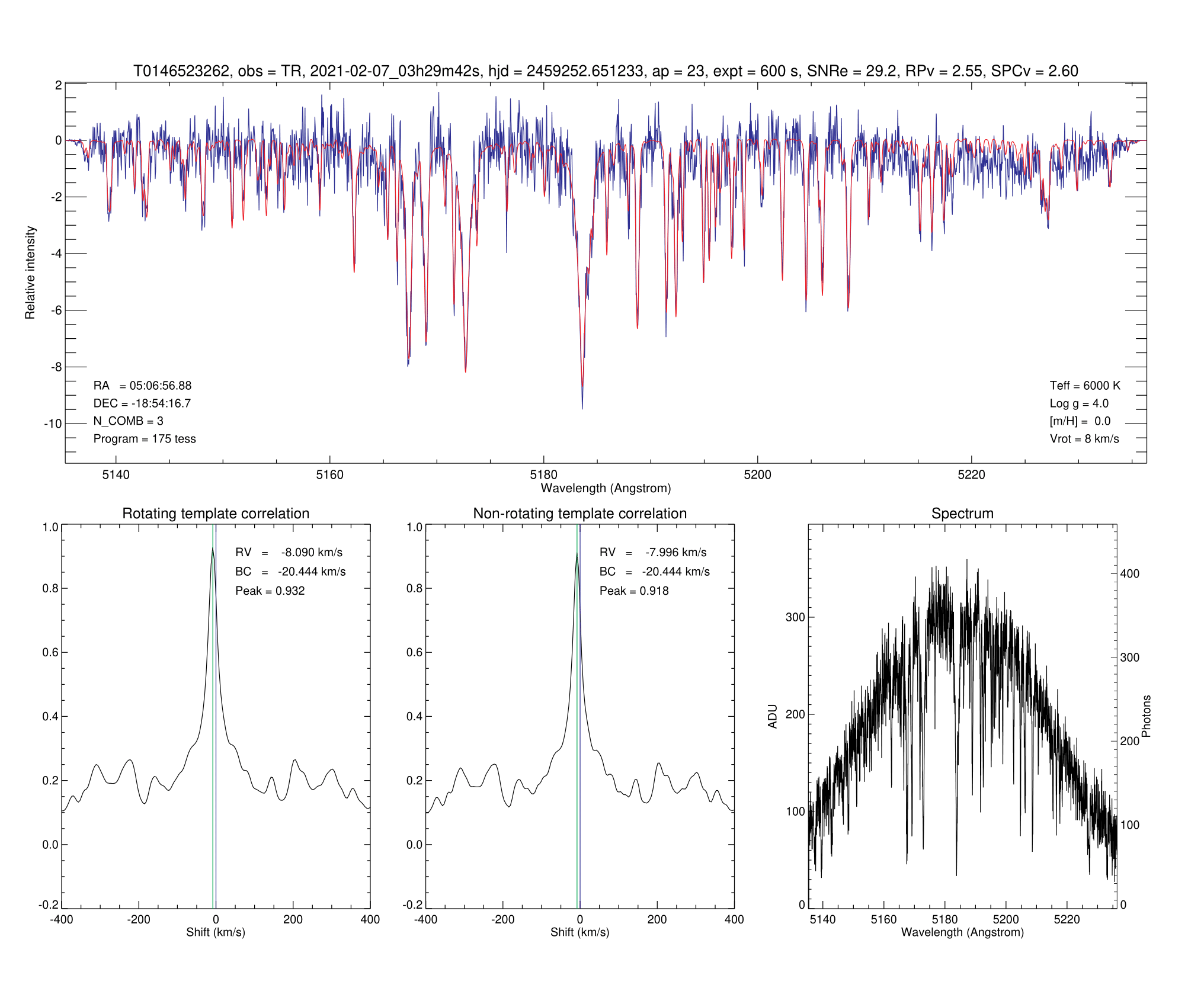}
         
         %\label{fig:}
     \end{subfigure}
        \label{fig:tres_spectra_2}
        \caption{TRES extracted spectra of TIC 73540072 (upper) and TIC 146523262 (lower).}
    \end{figure*}
\begin{figure*}
        \begin{subfigure}[b]{0.8\textwidth}
         \centering
         \includegraphics[width=\textwidth]{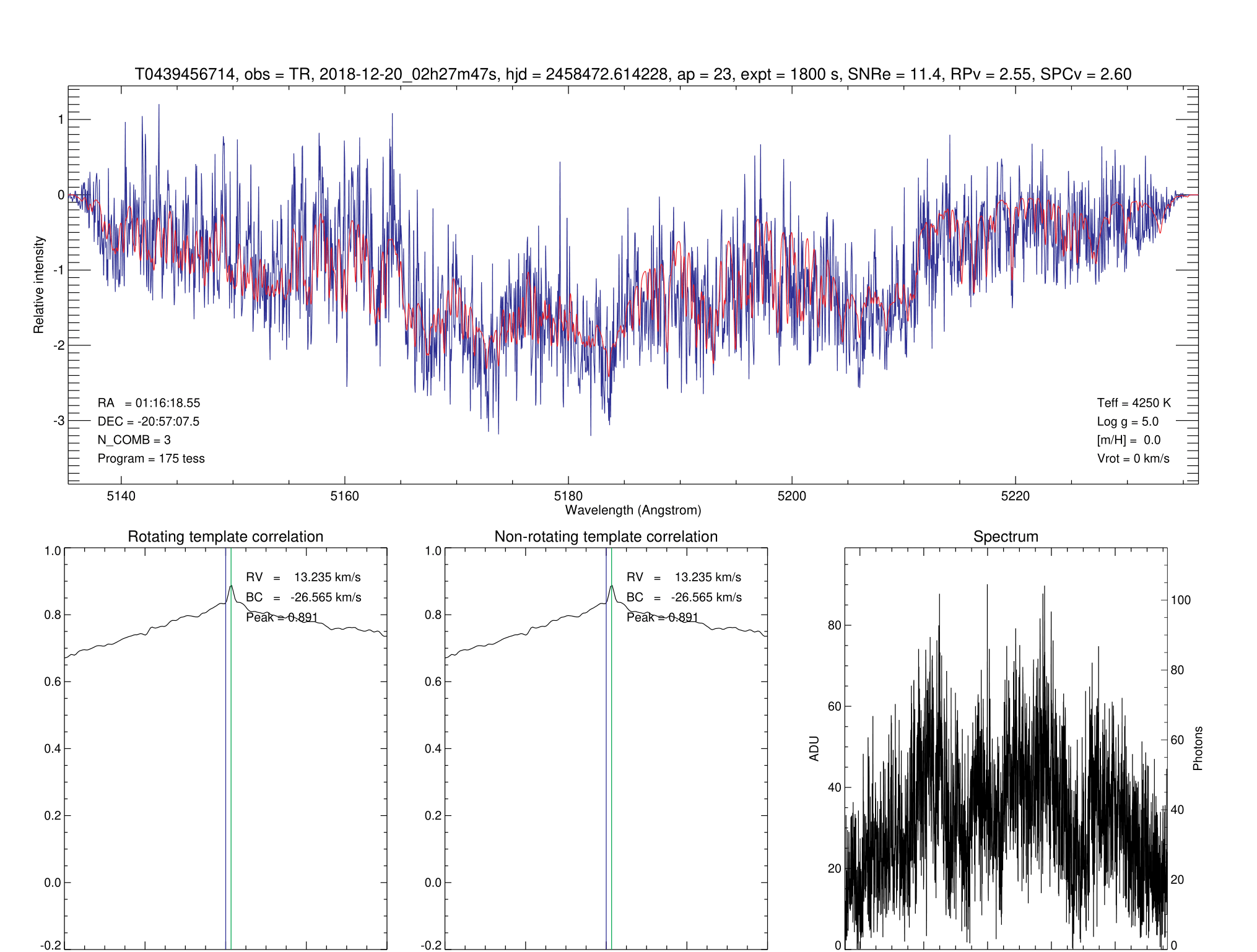}
         
         %\label{fig:}
     \end{subfigure}
      \label{fig:tres_spectra_3}
        \caption{TRES extracted spectra of TIC 439456714.}
    \end{figure*}

%If you want to present additional material which would interrupt the flow of the main paper, it can be placed in an Appendix which appears after the list of references.

%%%%%%%%%%%%%%%%%%%%%%%%%%%%%%%%%%%%%%%%%%%%%%%%%%

% Don't change these lines
\bsp	% typesetting comment
\label{lastpage}
\end{document}